\documentclass[%
superscriptaddress,
preprint,
preprintnumbers,
amsmath,amssymb,
aps,
prd,
floatfix,
notitlepage,
showpacs,
]{revtex4-2}
\usepackage{graphicx}
\usepackage{adjustbox}
\usepackage{hyperref}
\usepackage{xcolor}
\usepackage{subcaption}
\usepackage[justification=raggedright,singlelinecheck=false]{caption}
\usepackage{booktabs} 
\usepackage{multirow}
\usepackage{float} 
\usepackage{placeins} 
\usepackage{siunitx}
\usepackage{subcaption}
\usepackage[skip=0pt]{caption} 
\usepackage{amsmath} 
\newcommand{\appendixsetup}{
    \renewcommand{\thesection}{Appendix \Alph{section}} 
    \renewcommand{\theequation}{A\arabic{equation}} 
    \renewcommand{\thetable}{A\arabic{table}} 
    \setcounter{table}{0}\setcounter{equation}{0}
}
\usepackage{hyperref} 
\usepackage{titlesec} 
\usepackage{chngcntr} 

\usepackage{parskip}
\newcommand{\tr}{{\rm tr}}

\newcommand{\Id}{\mbox{1\hspace{-1.2mm}I}}

\newcommand{\bea}{\begin{eqnarray}}
\newcommand{\eea}{\end{eqnarray}}
\newcommand{\BAN}{\begin{eqnarray*}}
\newcommand{\EAN}{\end{eqnarray*}}

\newcommand{\blue}{\color{blue}}

\newcommand{\LB}{$<325$}
\newcommand{\BB}{$<650$}

\usepackage{titlesec}
\titleformat{\paragraph}
{\normalfont\normalsize\bfseries}{\theparagraph}{1em}{}
\titlespacing*{\paragraph}
{0pt}{3.25ex plus 1ex minus .2ex}{1.5ex plus .2ex}
\titlespacing*{\section}{0pt}{5pt}{5pt} 

\renewcommand\thesection{\arabic{section}}

\renewcommand{\thetable}{\arabic{table}}
\usepackage{titlesec}
\titlespacing*{\section}{0pt}{12pt}{8pt}
\titlespacing*{\subsection}{0pt}{10pt}{6pt}
\titlespacing*{\subsubsection}{0pt}{8pt}{4pt}

\begin{document}

\newcommand{\ASIOP}
{Institute of Physics, Academia Sinica, Taipei, Taiwan~11529}

\newcommand{\NTU}
{Department of Physics, National Taiwan University, \\ 
Taipei, Taiwan~10617}

\newcommand{\NTNU}
{Department of Physics, National Taiwan Normal University, \\  
Taipei, Taiwan~11677}

\newcommand{\CTP}
{Center for Theoretical Physics, Department of Physics, National Taiwan University, \\ 
Taipei, Taiwan~10617}

\newcommand{\NCTS}
{Physics Division, National Center for Theoretical Sciences,  \\
Taipei, Taiwan~10617}
\newcommand{\LBNL}
{Nuclear Science Division, Lawrence Berkeley National Laboratory, Berkeley, CA 94720, USA}

\title{Symmetries in High-Temperature Lattice QCD with $(u, d, s, c, b)$ Optimal Domain-Wall Quarks}

\author{Ting-Wai~Chiu}
\email{twchiu@phys.ntu.edu.tw}
\affiliation{\NTNU}
\affiliation{\ASIOP}
\affiliation{\NTU}
\affiliation{\NCTS}
\affiliation{\LBNL}

\begin{abstract}

We investigate the spatial $ z $-correlators of meson operators in $ N_f=2+1+1+1 $ lattice QCD with 
optimal domain-wall quarks across eight temperatures ranging from 325 to 3250 MeV. 
The meson operators include a complete set of Dirac bilinears 
for ten flavor combinations. 
%
Our findings reveal a hierarchical restoration of chiral symmetry in QCD with $ (u, d, s, c, b) $ quarks, 
progressing sequentially from $ SU(2)_L \times SU(2)_R \times U(1)_A $ to 
$ SU(3)_L \times SU(3)_R \times U(1)_A $, then to $ SU(4)_L \times SU(4)_R \times U(1)_A $, 
and finally to $ SU(5)_L \times SU(5)_R \times U(1)_A $ as the temperature increases.  
Additionally, we explore the emergence of the $ SU(2)_{CS} $ chiral-spin symmetry 
and compare the temperature windows for all flavor combinations. 
Our results indicate that the temperature windows for the emergent $SU(2)_{CS}$ symmetry 
are primarily dominated by the $ \bar{u} b $ and $ \bar{s} b $ sectors.

\end{abstract}

\maketitle

\section{Introduction}
\label{intro}  

Understanding the symmetries of high-temperature QCD is a crucial first step in 
determining the properties and dynamics of matter under extreme conditions. 
These studies are essential for gaining insight into the mechanisms governing matter creation
in the early universe and for interpreting the results of relativistic heavy-ion collision
experiments, such as those conducted at the LHC and RHIC, as well as future 
electron-ion collision experiments at planned electron-ion colliders.  
Lattice QCD provides a nonperturbative framework to explore the symmetries of 
high-temperature QCD from first principles. 
Since 1987 \cite{Detar:1987kae,Detar:1987hib}, numerous lattice studies have utilized 
the screening masses of meson $z$-correlators to investigate the effective 
restoration of $U(1)_A$ and $ SU(2)_L \times SU(2)_R $ chiral symmetries 
of $u$ and $d$ quarks in high-temperature QCD (see, 
e.g., ref.~\cite{Bazavov:2019www} and references therein).  
However, the hierarchical restoration of chiral symmetry in high-temperature QCD 
has not been discussed or studied in the literature, 
with the exception of refs.~\cite{Chiu:2023hnm,Chiu:2024jyz}.


In $ N_f=2+1+1+1 $ QCD with nonzero quark masses, the theory does not exhibit 
$ SU(N)_L \times SU(N)_R \times U(1)_A $ chiral symmetry for any integer $ N $ from 2 to 5,
due to the explicit symmetry breaking induced by the quark masses. However, 
as the temperature $ T $ increases, each quark gains thermal energy on the order 
of $ \pi T $, and eventually, its rest mass energy becomes negligible 
when $ \pi T \gg m_q $.  
Given that quark masses span from a few MeV to several GeV, 
chiral symmetry is restored in a hierarchical manner as the temperature 
rises: first, the $ SU(2)_L \times SU(2)_R \times U(1)_A $ symmetry 
of $ (u, d) $ quarks is restored, followed by 
the $ SU(3)_L \times SU(3)_R \times U(1)_A $ symmetry of $ (u, d, s) $ quarks, 
then the $ SU(4)_L \times SU(4)_R \times U(1)_A $ symmetry of $ (u, d, s, c) $ quarks, 
and finally the $ SU(5)_L \times SU(5)_R \times U(1)_A $ symmetry 
of $ (u, d, s, c, b) $ quarks. 
This hierarchical pattern was first pointed out in ref.~\cite{Chiu:2023hnm}.  
It is important to note that the top quark can be neglected in QCD, 
as it is extremely short-lived, decaying into a $W$-boson and 
a $b$-quark (most frequently), or an $s$- or $d$-quark (the rarest) before it can interact with gluons. 
Furthermore, since the QCD action with nonzero quark masses does not possess 
exact chiral symmetries, the term ``hierarchical restoration of chiral symmetry" 
should be regarded as ``hierarchical emergence of chiral symmetry".

In ref.~\cite{Chiu:2024jyz}, the hierarchical restoration of chiral symmetry was first 
observed in $ N_f=2+1+1 $ lattice QCD with $ (u,d,s,c) $ domain-wall quarks at the physical point. 
The restoration progresses sequentially from $ SU(2)_L \times SU(2)_R \times U(1)_A $ to 
$ SU(3)_L \times SU(3)_R \times U(1)_A $, and subsequently to $ SU(4)_L \times SU(4)_R \times U(1)_A $, 
as the temperature increases from 190 MeV to 1540 MeV.  
While this observation provides strong evidence supporting the hierarchical restoration 
of chiral symmetry in QCD, it remains incomplete, 
as the emergence of $ SU(5)_L \times SU(5)_R \times U(1)_A $ symmetry 
for $ (u,d,s,c,b) $ quarks has not yet been verified. 
This limitation motivates the present study, 
which aims to complete the picture of hierarchical chiral symmetry restoration 
in $ N_f=2+1+1+1 $ lattice QCD.

However, simulating $ N_f=2+1+1+1 $ lattice QCD with $ (u,d,s,c,b) $ quarks 
at the physical point remains a grand challenge, as discussed in ref.~\cite{Chiu:2020tml}. 
To control both discretization and finite volume errors, 
the constraints $ a \lesssim 0.03 $~fm and $ M_\pi L > 4 $ must be satisfied, 
which necessitate a lattice size larger than $ 180^3 \times N_t $, 
exceeding the capabilities of current lattice computations.  

Since our primary objective is to observe the emergence 
of $ SU(5)_L \times SU(5)_R \times U(1)_A $ symmetry in QCD with $ (u,d,s,c,b) $ 
quarks at temperatures $ T \ge T_{c1}^{\bar b b} > T_{c1}^{\bar c c} $ 
(see the definition of $T_{c1}^{\bar q Q}$ in (\ref{eq:Tc1_qQ})) 
after the restoration of $ SU(4)_L \times SU(4)_R \times U(1)_A $ symmetry 
for $ (u,d,s,c) $ quarks at the lower temperature $ T_{c1}^{\bar c c} $, 
this problem can be qualitatively addressed in lattice QCD with physical $ (s,c,b) $ 
quarks but unphysically heavy $ u/d $ quarks (e.g., with $ M_\pi \sim 700 $~MeV). 
Under these conditions, simulations can be conducted on 
$ 40^3 \times (64, 20, 16, 12, 10, 8, 6, 4, 2) $ lattices using a modest GPU cluster. 
The ``zero" temperature ensemble for the $ 40^3 \times 64 $ lattice has already been 
generated in ref.~\cite{Chiu:2020tml}, along with the basic physical properties of mesons 
with flavor contents $ \bar b b$, $\bar b c$, $\bar b s$, and $\bar c c$.  
In this exploratory study, we generate eight ensembles at finite temperatures, 
summarized in Table~\ref{tab:8_ensembles}. 
Details of the simulation algorithms, the determination of the lattice spacing $a$,  
the $(s, c, b)$ physical quark masses, and the residual masses of $(u/d, s, c, b)$ quarks 
have been given in ref.~\cite{Chiu:2020tml} and the references therein.  
It is important to note that any results derived from these ensembles 
are subject to systematic uncertainties arising from 
unphysically heavy $ u/d $ quarks, as well as discretization 
and finite volume effects. These uncertainties cannot be quantified 
within the present study, as the gauge ensembles include only 
a single unphysical $ u/d $ quark mass, one spatial volume, and a single lattice spacing.  
{\it Our goal here is not to provide a precise determination of the temperatures   
for the hierarchical restoration of chiral symmetry in $ N_f=2+1+1+1 $ lattice QCD, 
but rather to offer a qualitative picture of the hierarchical restoration of chiral symmetry 
in this system.} This work represents a first step toward more precise determinations 
of $ T_{c1}^{\bar q Q} $ with controlled systematics in future lattice studies, 
which will require simulations at the physical point and sufficiently large spatial volumes, 
(i.e., $> 180^3 \times N_t$).

\begin{table}[h!]
\caption{\small The lattice parameters and statistics of the eight gauge ensembles 
for computing the meson correlators. The HMC simulations are performed with 
the Wilson plaquette gauge action \cite{Wilson:1974sk} at $\beta = 6/g^2 = 6.70$, the two-flavor
optimal domain-wall fermion action for $u/d$ quarks \cite{Chiu:2002ir,Chiu:2011bm}, 
and the exact one-flavor optimal domain-wall fermion action for $s$, $c$, and $b$ quarks 
\cite{Chen:2014hyy,Chiu:2015sea}.
The lattice spacing $a = 0.0303(2)$~fm is determined by 
Wilson flow \cite{Narayanan:2006rf,Luscher:2010iy} 
with the condition $ \left< t^2 E(t) \right>|_{t=t_0} = 0.3$ 
and input $\sqrt{t_0} = 0.1416(8)$~fm \cite{Bazavov:2015yea}.
The bare quark masses are $(m_{u/d}, m_s, m_c, m_b) a = (0.010, 0.015, 0.200, 0.850)$, where 
$m_s$, $m_c$ and $m_b$ are at the physical point, while the $u/d$ quarks are at the unphysical point
with $ M_\pi \sim 700 $~MeV.   
The last four columns are the residual masses \cite{Chen:2012jya} of $u/d$, $s$, $c$, and $b$ quarks.}
\setlength{\tabcolsep}{3pt}
\centering
\small
\begin{tabular}{|cccccccc|}
\hline 
    $ N_x $
  & $ N_t $
  & $T$[MeV]
  & $N_{\rm confs}$
  & $ (m_{u/d} a)_{\rm res} $
  & $ (m_{s} a)_{\rm res} $
  & $ (m_{c} a)_{\rm res} $
  & $ (m_{b} a)_{\rm res} $
\\
\hline
\hline
40 & 20 & 
325 & 306 &
$ 8.21(33) \times 10^{-7}$ & $8.22(33) \times 10^{-7}$ & $8.70(33) \times 10^{-7}$ & 
i$9.74(36) \times 10^{-7}$ \\
40 & 16 & 
406 & 382 & 
$9.35(28) \times 10^{-7}$ & $9.42(28) \times 10^{-7}$ & $9.82(28) \times 10^{-7}$ & 
$1.09(31) \times 10^{-6}$ \\
40 & 12 & 
524 & 380 & 
$9.46(28) \times 10^{-7}$ & $9.46(28) \times 10^{-7}$ & $9.74(28) \times 10^{-7}$ & 
$1.08(31) \times 10^{-7}$\\
40 & 10 & 
650 & 260 & 
$9.12(33) \times 10^{-7}$ & $9.12(33) \times 10^{-7}$ & $9.25(34) \times 10^{-7}$ & 
$1.01(36) \times 10^{-8}$ \\
40 & 8 & 
813 & 337 & 
$1.03(3) \times 10^{-6}$ & $1.02(3) \times 10^{-6}$ & $1.04(3) \times 10^{-6}$ & $1.12(3) \times 10^{-6}$ \\
40 & 6 & 
1084 & 411 &
$1.03(3) \times 10^{-6}$ & $1.03(3) \times 10^{-6}$ & $1.02(3) \times 10^{-6}$ & $1.07(3) \times 10^{-6}$ \\
40 & 4 & 
1626 & 337 &
$1.13(4) \times 10^{-6}$ & $1.13(4) \times 10^{-6}$ & $1.12(4) \times 10^{-6}$ & $1.13(4) \times 10^{-6}$ \\
40 & 2 & 
3252 & 727 & 
$1.7(5) \times 10^{-7}$ & $1.7(5) \times 10^{-7}$ & $ 1.9(5) \times 10^{-7} $ & $2.3(5) \times 10^{-7}$\\
\hline
\end{tabular}
\label{tab:8_ensembles}
\end{table}

In this study, we adopt the same strategy as in refs.~\cite{Chiu:2024jyz,Chiu:2023hnm} 
to examine the hierarchical restoration of chiral symmetry in high-temperature QCD 
by analyzing the splittings of meson $ z $-correlators within symmetry multiplets.  
In general, the meson $ z $-correlator, $ C_\Gamma(zT) $, 
of the meson interpolator $ \bar{q} \Gamma Q $ is expressed 
as a function of the dimensionless variable:  
\bea
zT = \frac{n_z a}{N_t a} = \frac{n_z}{N_t},
\label{eq:zT}
\eea
where $ T = \frac{1}{N_t a} $ is the temperature.  
For the classification of meson operators, along with their names and notations, 
we refer to Table II in ref.~\cite{Chiu:2024jyz}. Additionally, 
we adopt the symmetry breaking parameters as defined in ref.~\cite{Chiu:2024jyz}, 
following the same conventions and notations used therein.  
For the convenience of the reader, we summarize our conventions and notations 
in \ref{app:A}.

We also recall the following notation introduced in ref.~\cite{Chiu:2024jyz}:  
\begin{equation}  
\label{eq:Tc1_qQ}  
T_{c1}^{\bar{q} Q} \equiv \max( T_c^{\bar{q} Q}, T_1^{\bar{q} Q} ),  
\end{equation}  
where $ T_c^{\bar{q} Q} $ and $ T_1^{\bar{q} Q} $ represent the temperatures 
at which the restoration of $ SU(2)_L \times SU(2)_R $ and $ U(1)_A $ chiral symmetries 
occurs, respectively, as determined via meson $ z $-correlators 
with flavor content $ \bar{q} Q $.  
For $ T > T_{c1}^{\bar{q} Q} $, the theory exhibits 
the $ SU(2)_L \times SU(2)_R \times U(1)_A $ chiral symmetry 
in the $ \bar{q} Q $ sector.

Besides the hierarchical restoration of chiral symmetry, we are also interested in 
the emergence of symmetries that are not inherent to the full QCD action 
but apply only to specific components of it. 
One such example is the $ SU(2)_{CS} $ chiral-spin symmetry 
(with $ U(1)_A $ as a subgroup) \cite{Glozman:2014mka,Glozman:2015qva}, 
which is a symmetry of the chromoelectric part of the quark-gluon interaction 
and the color charge. Since free fermions and the chromomagnetic part 
of the quark-gluon interaction do not possess $ SU(2)_{CS} $ symmetry, 
its emergence in high-temperature QCD suggests the possible existence of hadronlike objects 
predominantly bound by chromoelectric interactions.  
The first indication of approximate $ SU(2)_{CS} $ symmetry was observed 
in the multiplets of $ z $-correlators of vector mesons at temperatures 
$ T \sim 220-500 $ MeV in $ N_f=2 $ lattice QCD with 
domain-wall fermions \cite{Rohrhofer:2019qwq}. In ref.~\cite{Chiu:2023hnm}, 
we investigated the emergence of $ SU(2)_{CS} $ symmetry in $ N_f=2+1+1 $ 
lattice QCD with optimal domain-wall quarks at the physical point. 
Our findings indicated that $ SU(2)_{CS} $ symmetry breaking in 
the $ \bar{u} d $ sector of $ N_f=2+1+1 $ lattice QCD is larger than 
that in $ N_f=2 $ lattice QCD at the same temperature, for both $ z $-correlators 
and $ t $-correlators of vector mesons composed of $ u $ and $ d $ quarks.  
In ref.~\cite{Chiu:2024jyz}, our study was extended to all flavor combinations 
($ \bar{u} d $, $ \bar{u} s $, $ \bar{s} s $, $ \bar{u} c $, $ \bar{s} c $, $ \bar{c} c $), 
revealing that the temperature windows for the emergence 
of $ SU(2)_{CS} $ symmetry are predominantly dominated by 
$ \bar{u} c$ and $\bar{s} c$ sectors. 
In this work, we further extend our investigation to $ N_f=2+1+1+1 $ lattice QCD 
with physical $ (s,c,b) $ quarks but unphysically heavy $ u/d $ quarks,  
with $ M_\pi \sim 700 $ MeV.

The outline of this paper is as follows.  
In Section~\ref{HRCS}, we present the hierarchical restoration of chiral symmetry 
in $ N_f=2+1+1+1 $ QCD, progressing from $ SU(2)_L \times SU(2)_R \times U(1)_A $ 
to $ SU(3)_L \times SU(3)_R \times U(1)_A $, then to 
$ SU(4)_L \times SU(4)_R \times U(1)_A $, 
and finally to $ SU(5)_L \times SU(5)_R \times U(1)_A $.  
In Section~\ref{SU2_CS}, we estimate the approximate temperature windows for the  
of emergent $ SU(2)_{CS} $ symmetry across ten flavor combinations. 
Our findings indicate that the $ SU(2)_{CS} $ symmetry is predominantly 
governed by the $ \bar{u} b$ and $ \bar{s} b $ sectors.  
In Section~\ref{conclusions}, we summarize our findings and provide concluding remarks.
The appendices contain supplementary details.   
\ref{app:A} summarizes the notations and conventions used in this paper.   
\ref{app:B} estimates the variation of $\sqrt{t_0}$ as $M_\pi$ changes 
from 700 MeV to 140 MeV (the physical point).  
\ref{app:C} tabulates the numerical values of 
$\kappa_{VA}$, $\kappa_{TX}$, $\kappa$, and $\kappa_{CS}$ for $zT = 1$, 2, and 3 
in each flavor sector.  
\ref{app:D} provides the corresponding numerical values for 
$N_f=2+1+1$ lattice QCD at the physical point \cite{Chiu:2024jyz} at $zT = 0.5$, 1, and 2.

\section{Hierarchical restoration of chiral symmetry}
\label{HRCS}

First, we recall the general features of symmetry breaking parameters
as discussed in ref.~\cite{Chiu:2024jyz}.

In general, the degeneracy of any two meson $z$-correlators
$C_{\Gamma_A}(zT)$ and $C_{\Gamma_B}(zT)$ with flavor content $\bar q Q$
can be measured by the symmetry breaking parameter
\bea
\label{eq:kappa_AB}
\kappa_{AB}(zT) = \frac{\left| C_{\Gamma_A}(zT)-C_{\Gamma_B}(zT) \right|}
                       {C_{\Gamma_A}(zT)+C_{\Gamma_B}(zT)}, \hspace{4mm}  z > 0.
\eea
If $ C_{\Gamma_A} $ and $ C_{\Gamma_B} $ are exactly degenerate at $T$,
then $\kappa_{AB} = 0 $ for any $z$, and the symmetry is effectively restored at $T$.
On the other hand, if there is any discrepancy between $C_A$ and $C_B$ at any $z$,
then $\kappa_{AB} $ is nonzero at this $z$, and the symmetry is not exactly restored at $T$.
Here the denominator of (\ref{eq:kappa_AB}) serves as (re)normalization and
the value of $\kappa_{AB}$ is bounded between zero and one.
Obviously, this criterion is more stringent than the equality of the
ground-state screening masses, $m_A^{scr} = m_B^{scr}$,
which are extracted from $C_{\Gamma_A}$ and $C_{\Gamma_B}$ at large $z$.

For example, the effective restoration of $ SU(2)_L \times SU(2)_R $ chiral symmetry
for any $ \bar{q} Q $ implies that the correlators of the vector and axial-vector mesons
are identical at all distances, i.e., $C_{V_k}(zT) = C_{A_k}(zT), (k=1,2,4)$ for any $z$ at fixed $T$.
Since each correlator consists of contributions from both the ground state and excited states,
the equality of these correlators implies that the screening masses of the vector and axial-vector mesons
are identical for the ground state as well as for each excited state.
Similarly, the effective restoration of $U(1)_A$ symmetry for any $ \bar{q} Q $ implies that
the correlators of the pseudoscalar and scalar mesons are equal at all distances,
$ C_P(zT) = C_S(zT) $ for any $z$ at fixed $T$. As with the vector and axial-vector correlators,
this equality indicates that the screening masses of the pseudoscalar and scalar mesons are degenerate
for both the ground state and each excited state.

{\it
Therefore, examining the degeneracy of the correlators of symmetry partners at any $z < N_z/2 $
(accounting for the periodic boundary condition in the $z$ direction)
provides a more rigorous test of symmetry restoration than focusing solely on the degeneracy of
the ground-state screening masses at large distances. Consequently, the symmetry-breaking parameters
presented in this work offer more reliable insights into chiral symmetry breaking and restoration
compared to approaches that rely only on the degeneracy of ground-state screening masses
of symmetry partners.
}


The $SU(2)_L \times SU(2)_R $ symmetry breaking for any $\bar q Q$ sector can be measured by
\bea
\label{eq:k_VA_z}
\kappa_{VA}(zT) = \frac{\left| C_{V_k}(zT) - C_{A_k}(zT) \right| }{C_{V_k}(zT) + C_{A_k}(zT)},
\hspace{4mm} z > 0, \hspace{4mm} (k=1,2,4).
\eea
In principle, any component of (\ref{eq:k_VA_z}) can serve as the $SU(2)_L \times SU(2)_R$ 
symmetry-breaking parameter. Due to the $S_2$ symmetry of the $z$-correlators, 
the $k=1$ and $k=2$ components are identical. To enhance the statistics,
we average over $k=1$ and $k=2$ components to measure
the $SU(2)_L \times SU(2)_R$ symmetry breaking.  

In general, to determine to what extent the
$SU(2)_L \times SU(2)_R$ chiral symmetry is restored, 
it is necessary to examine whether $\kappa_{VA}$ is sufficiently small.
To this end, we use the following criterion for the manifestation of
$SU(2)_L \times SU(2)_R$ chiral symmetry at $T$ for a fixed $zT$, 
\bea
\label{eq:SU2_crit_z}
\kappa_{VA}(zT) \le \epsilon_{VA},
\eea
where $ \epsilon_{VA} $ is a small parameter which defines the precision of the chiral symmetry.
For fixed $zT$ and $\epsilon_{VA}$, the temperature $T_c$ 
is the lowest temperature satisfying (\ref{eq:SU2_crit_z}), i.e.,
\bea
\label{eq:Tc_epsilon}
\kappa_{VA}(zT) < \epsilon_{VA} \ \text{for} \ T > T_c.
\eea


The $U(1)_A$ symmetry breaking for any $\bar q Q$ sector can be measured
by the $z$-correlators in the pseudoscalar and scalar channels, with
\bea
\label{eq:k_PS_z}
\kappa_{PS}(zT) = \frac{\left| C_P(zT)-C_S(zT) \right|}{C_P(zT)+C_S(zT)}, \hspace{4mm}  z > 0,
\eea
as well as in the tensor vector and axial-tensor vector channels, with
\bea
\label{eq:k_TX_z}
\kappa_{TX}(zT) = \frac{\left| C_{T_k}(zT) - C_{X_k}(zT) \right| }{C_{T_k}(zT) + C_{X_k}(zT)},
\hspace{4mm} z > 0, \hspace{4mm} (k=1,2,4).
\eea
In principle, (\ref{eq:k_PS_z}) and any component of (\ref{eq:k_TX_z}) can serve as
the $U(1)_A$ symmetry-breaking parameter.
In the following, we use (\ref{eq:k_TX_z}) with $k=4$ to measure the $U(1)_A$ symmetry breaking.
The reason for choosing the $k=4$ component is for consistency since
the $k=4$ component of (\ref{eq:k_TX_z}) is also needed to measure the $U(1)_A$ symmetry breaking
in the multiplet $\{C_{A_1}, C_{T_4}, C_{X_4}\}$ of the $SU(2)_{CS}$ chiral-spin symmetry which
contains $U(1)_A$ as a subgroup. (See the discussion
in Section~\ref{SU2_CS} and our notatations and conventions in refs.~\cite{Chiu:2023hnm,Chiu:2024jyz}.)

Similar to (\ref{eq:SU2_crit_z}), we use the following criterion for the manifestation of
$U(1)_A$ symmetry at $T$ for a fixed $zT$
\bea
\kappa_{TX}(zT) \le \epsilon_{TX},
\label{eq:U1_TX_crit_z}
\eea
where $\epsilon_{TX}$ is a small parameter which defines the precision of $U(1)_A$ symmetry.
For fixed $zT$ and $\epsilon_{TX}$, the temperature $T_1$
is the lowest temperature satisfying (\ref{eq:U1_TX_crit_z}), i.e.,
\bea
\label{eq:T1_epsilon}
\kappa_{TX}(zT) < \epsilon_{TX} \ \text{for} \ T > T_1.
\eea


Next, consider QCD with $N_f$ quarks $(q_1, q_2, \cdots, q_{N_f})$ where $N_f \ge 2$.
As discussed in ref.~\cite{Chiu:2024jyz},
upon neglecting the disconnected diagrams in the meson $z$-correlators,
the $SU(N)_L \times SU(N)_R$ chiral symmetry of $N$ ($2 \le N \le N_f$) quarks
is manifested by the degeneracies of
meson $z$-correlators in the vector and axial-vector channels,
$C_{V_k}^{\bar q_i q_j}(z) = C_{A_k}^{\bar q_i q_j}(z)$, $(k=1,2,4)$,
for {\it all} flavor combinations of $N$ quarks ($\bar q_i q_j$, $i,j=1, \cdots, N$).
Thus, to determine the temperature $T_c$ for the manifestation of the
$SU(N)_L \times SU(N)_R$ chiral symmetry of $N$ quarks,
it needs to measure $\kappa_{VA}^{\bar q_i q_j}$ for {\it all} flavor combinations of $N$ quarks,
and check whether they {\it all} satisfy (\ref{eq:Tc_epsilon})
for fixed $zT$ and $\epsilon_{VA}$.
This amounts to finding the largest $T_c^{\bar q_i q_j}$ satisfying (\ref{eq:Tc_epsilon})
among all flavor combinations of $N$ quarks, i.e.,
\bea
\label{eq:Tc_N}
T_c^N = \max( T_c^{\bar q_i q_j}, i,j=1, 2, \cdots, N).
\eea

Similarly, about the $U(1)_A$ symmetry of $N$ ($2 \le N \le N_f$) quarks,
upon neglecting the disconnected diagrams in the meson $z$-correlators,
it is manifested by the degeneracies of meson $z$-correlators 
in the pseudoscalar and scalar channels,
$C_{P}^{\bar q_i q_j}(z) = C_{S}^{\bar q_i q_j}(z)$,
as well as in the tensor vector and axial-tensor vector channels,
$C_{T_k}^{\bar q_i q_j}(z) = C_{X_{k}}^{\bar q_i q_j}(z)$, $(k=1,2,4)$,
for {\it all} flavor combinations of $N$ quarks ($\bar q_i q_j$, $i,j=1, \cdots, N$).
To determine the temperature $T_1$ for the manifestation of 
$U(1)_A $ symmetry via the tensor vector and axial-tensor vector channels,
it needs to measure $\kappa_{TX}^{\bar q_i q_j}$ for {\it all} flavor combinations of $N$ quarks,
and check whether they {\it all} satisfy (\ref{eq:T1_epsilon}) for fixed $zT$ and $\epsilon_{TX}$.
This amounts to finding the largest $T_1^{\bar q_i q_j}$ satisfying (\ref{eq:T1_epsilon})
among all flavor combinations of $N$ quarks, i.e.,
\bea
\label{eq:T1_N}
T_1^N = \max( T_1^{\bar q_i q_j}, i,j=1, 2, \cdots, N).
\eea 
Then, for the $\epsilon_{VA}$ and $\epsilon_{TX}$ specified
in (\ref{eq:Tc_epsilon}) and (\ref{eq:T1_epsilon}),  
the $SU(N)_L \times SU(N)_R \times U(1)_A$ chiral symmetry 
is effectively restored at 
\bea
\label{eq:Tc1_N}
T_{c1}^N = \max(T_c^N, T_1^N).
\eea 

At this point, we recall that, in QCD with $N_f > 2$ massless quarks, 
meson $z$-correlators for the flavor singlet and nonsinglet states 
with the same quantum numbers (i.e., scalar, pseudoscalar, vector, or axial-vector) 
become equal at temperatures above $T_c$ \cite{Evans:1996wf,Lee:1996zy,Birse:1996dx,Cohen:1996sb}. 
This equality implies that disconnected diagrams are suppressed 
in meson $z$-correlators for QCD with $N_f > 2$ massless quarks at $T > T_c$. 
However, to what extent this suppression persists in QCD with $N_f=2+1+1+1$ 
massive quarks remains unknown. 
We aim to address this question through noise estimation of all-to-all quark propagators, 
an analysis that is currently underway.

\subsection{Results of $\kappa_{VA}$ and $\kappa_{TX}$}

\begin{figure}[!h]
  \centering
  \caption{The chiral symmetry breaking parameters $(\kappa_{VA}, \kappa_{TX})$ in 
           the $(\bar u d, \bar u s, \bar s s)$ sectors.}
\begin{tabular}{@{}c@{}c@{}}
  \includegraphics[width=7.5cm,clip=true]{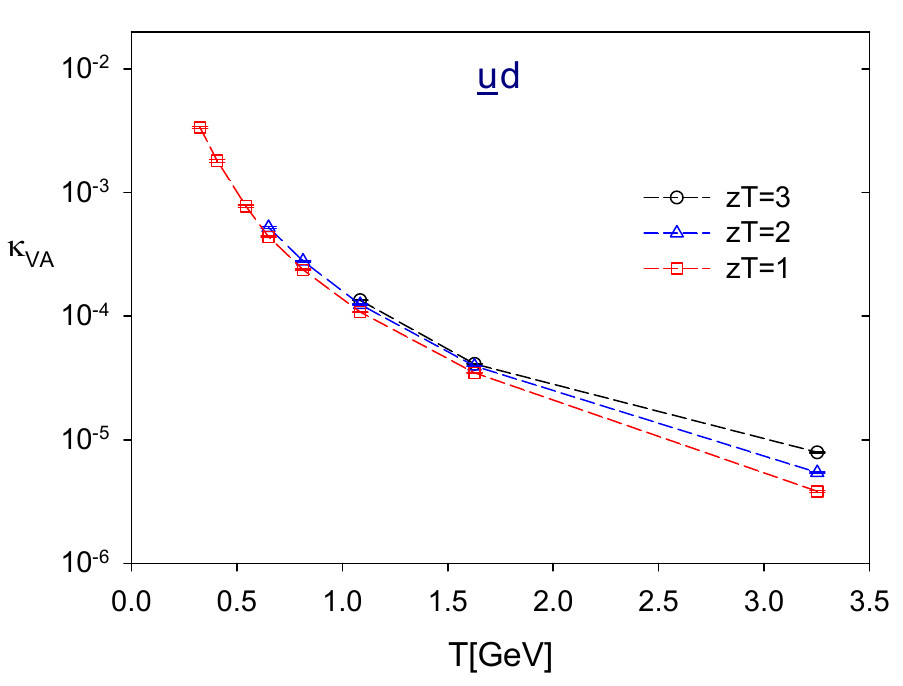}
&
  \includegraphics[width=7.5cm,clip=true]{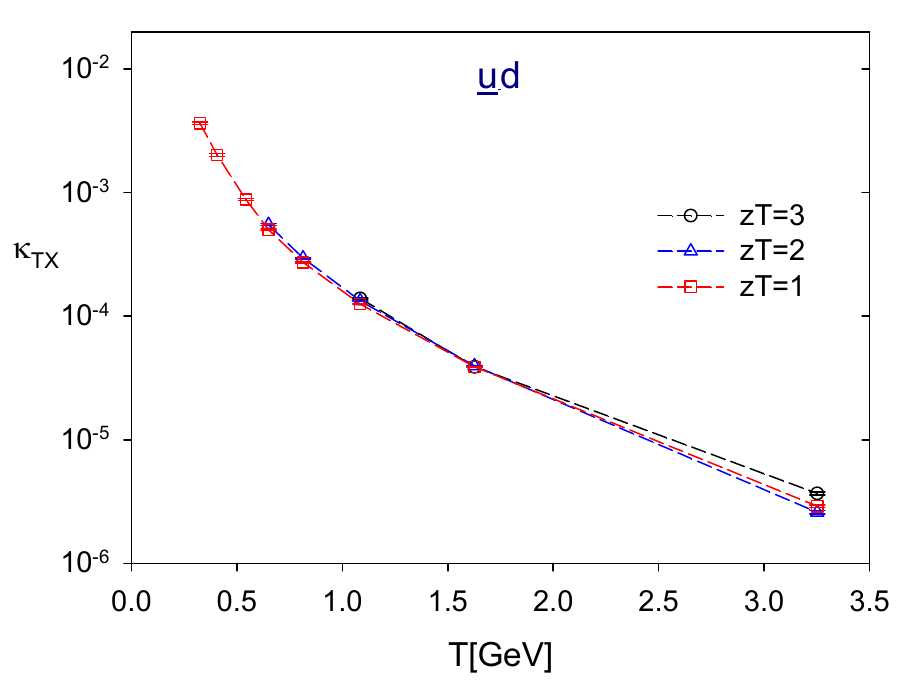} 
\vspace{-10pt} 
\\ 
  \includegraphics[width=7.5cm,clip=true]{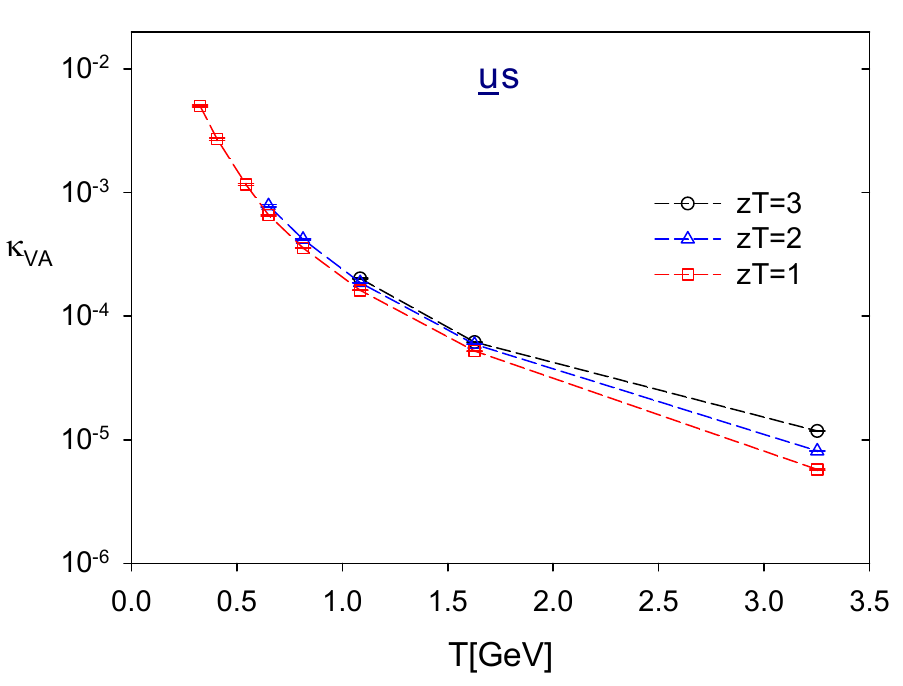}
&
  \includegraphics[width=7.5cm,clip=true]{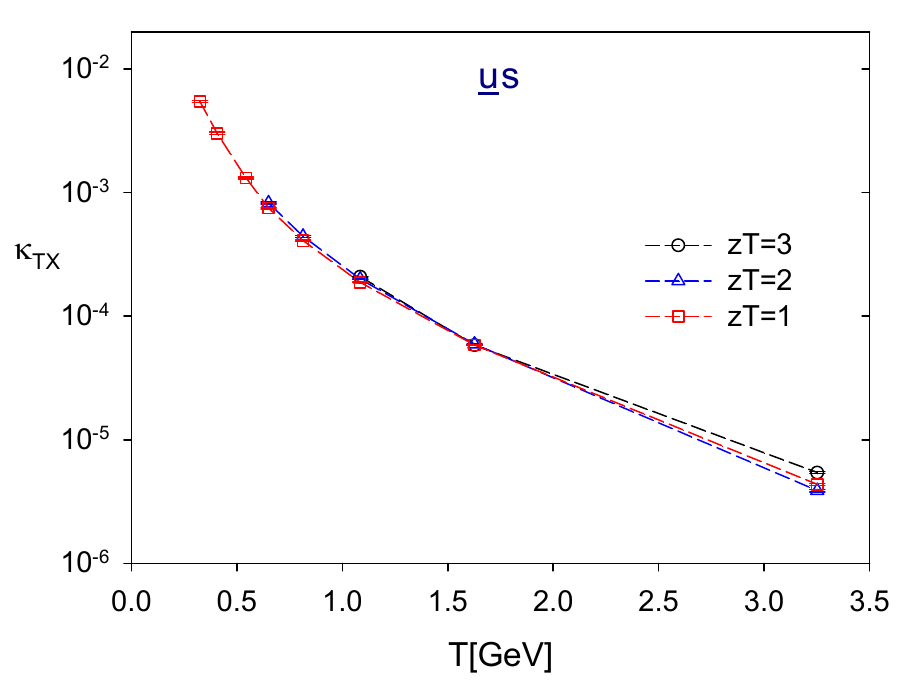} 
\vspace{-10pt} 
\\ 
  \includegraphics[width=7.5cm,clip=true]{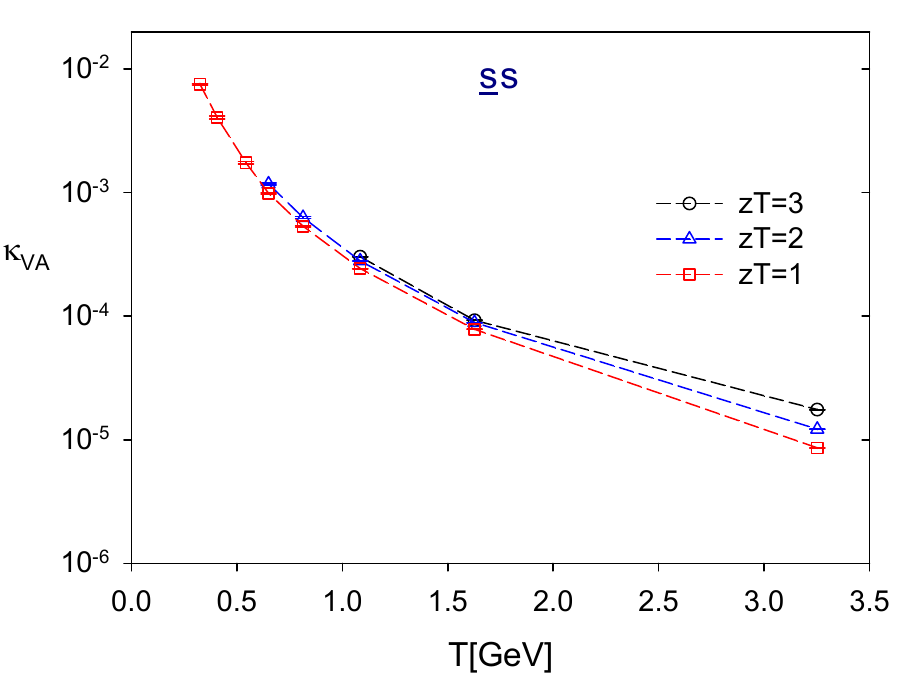}
&
  \includegraphics[width=7.5cm,clip=true]{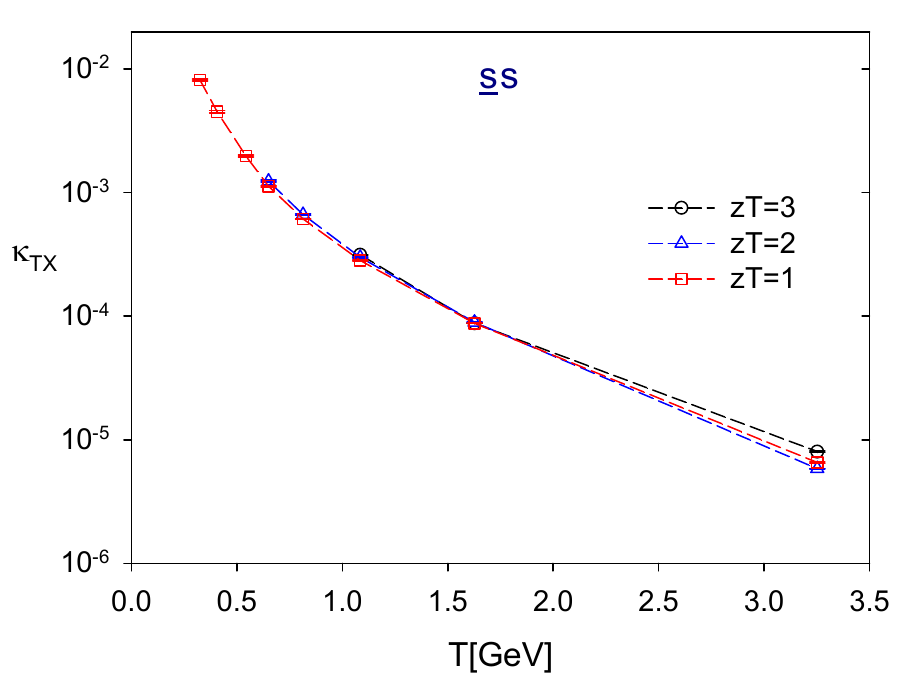} 
\end{tabular}
  \label{fig:kVA_kTX_ud_us_ss}
\end{figure}

\begin{figure}[!h]
  \centering
  \caption{The chiral symmetry breaking parameters $(\kappa_{VA}, \kappa_{TX})$ in the 
           $(\bar u c, \bar s c, \bar u b, \bar s b)$ sectors.}
\begin{tabular}{@{}c@{}c@{}}
  \includegraphics[width=7.5cm,clip=true]{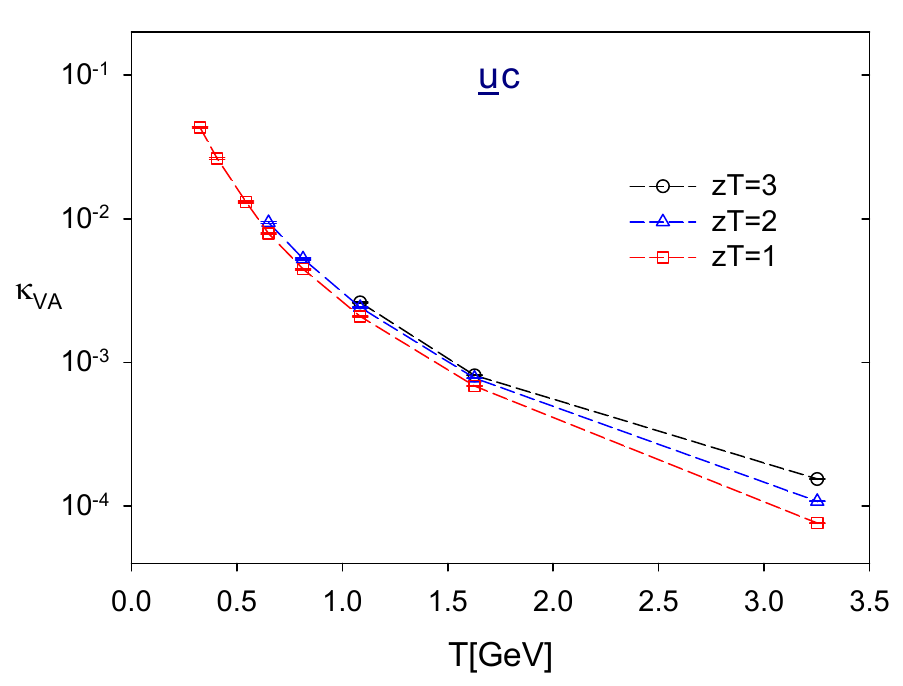}
&
  \includegraphics[width=7.5cm,clip=true]{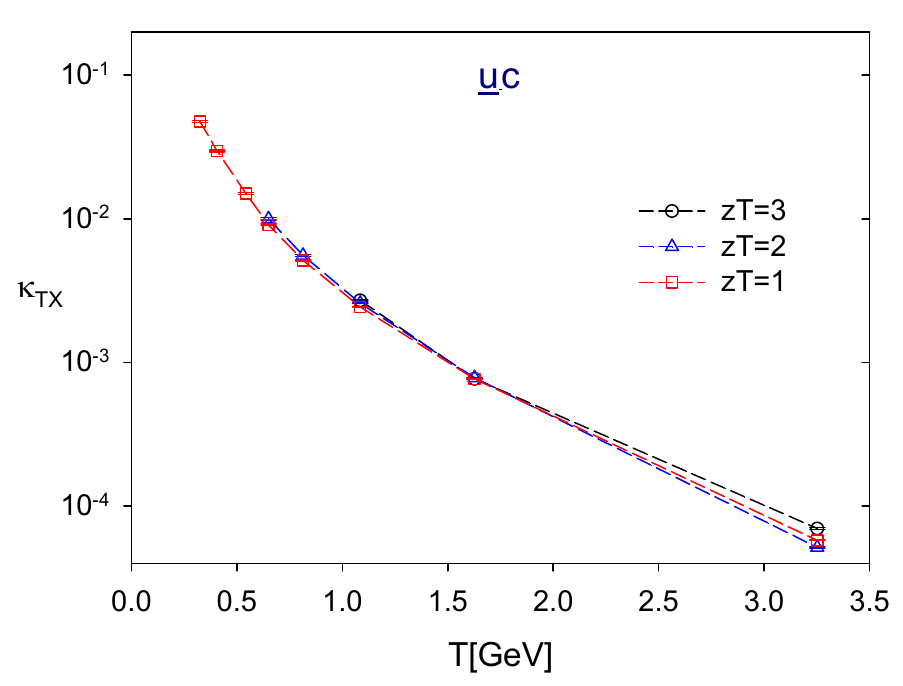} 
\\
\vspace{-10pt} 
  \includegraphics[width=7.5cm,clip=true]{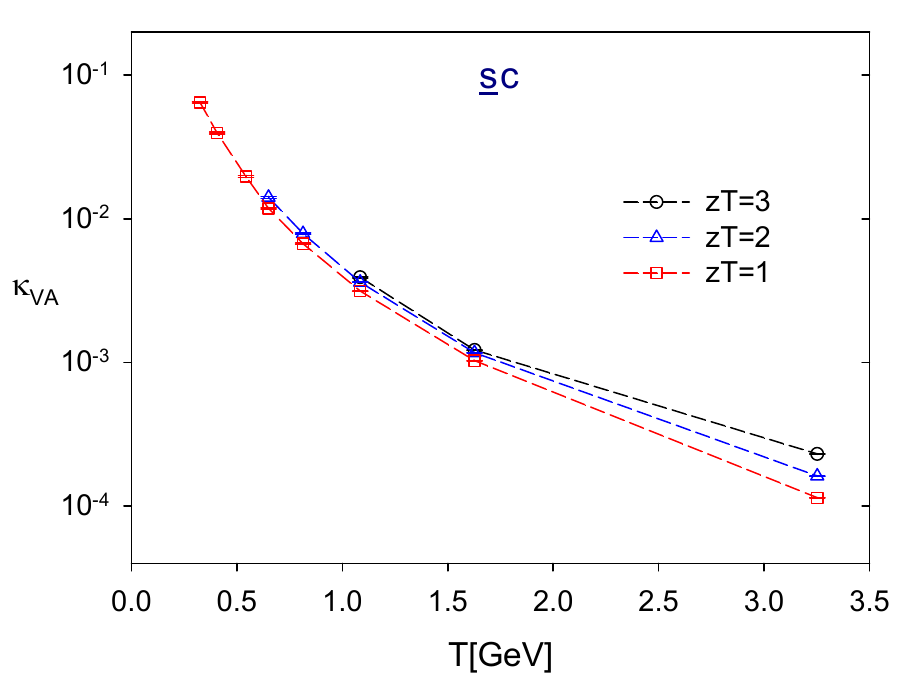}
&
  \includegraphics[width=7.5cm,clip=true]{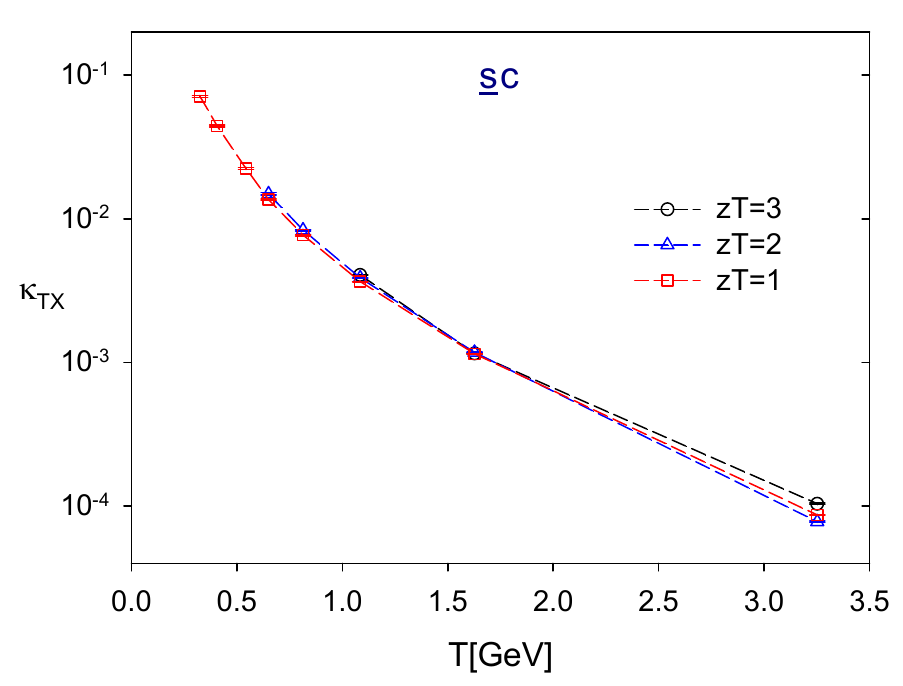} 
\vspace{-10pt} 
\\ 
  \includegraphics[width=7.5cm,clip=true]{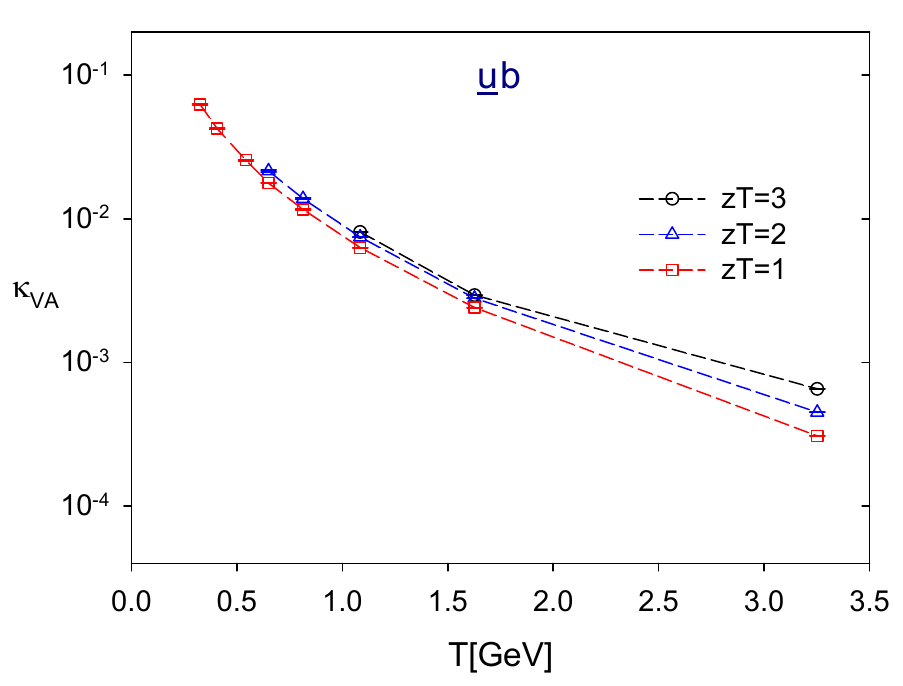}
&
  \includegraphics[width=7.5cm,clip=true]{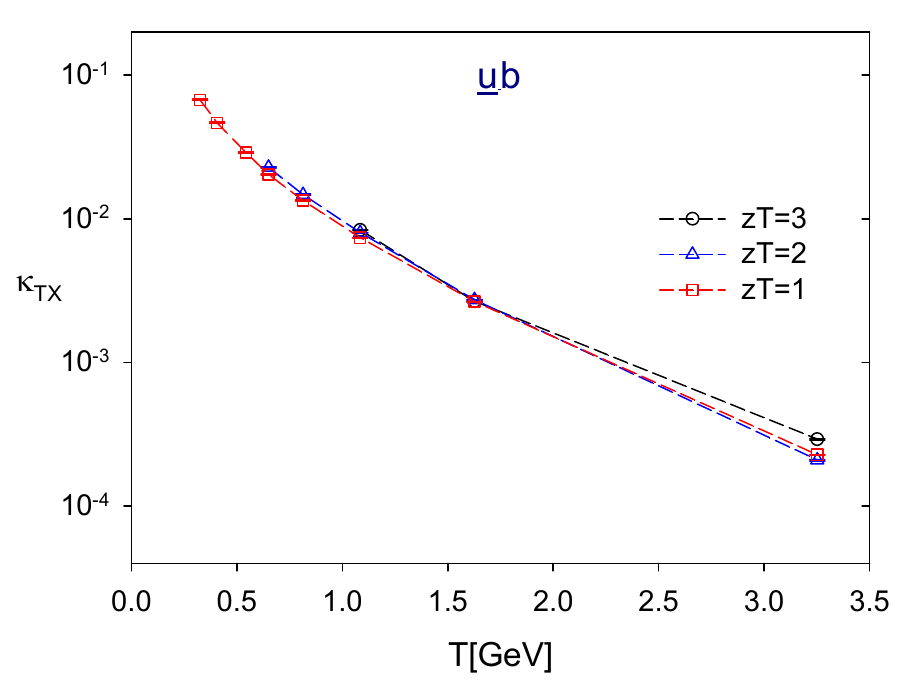} 
\vspace{-10pt} 
\\ 
  \includegraphics[width=7.5cm,clip=true]{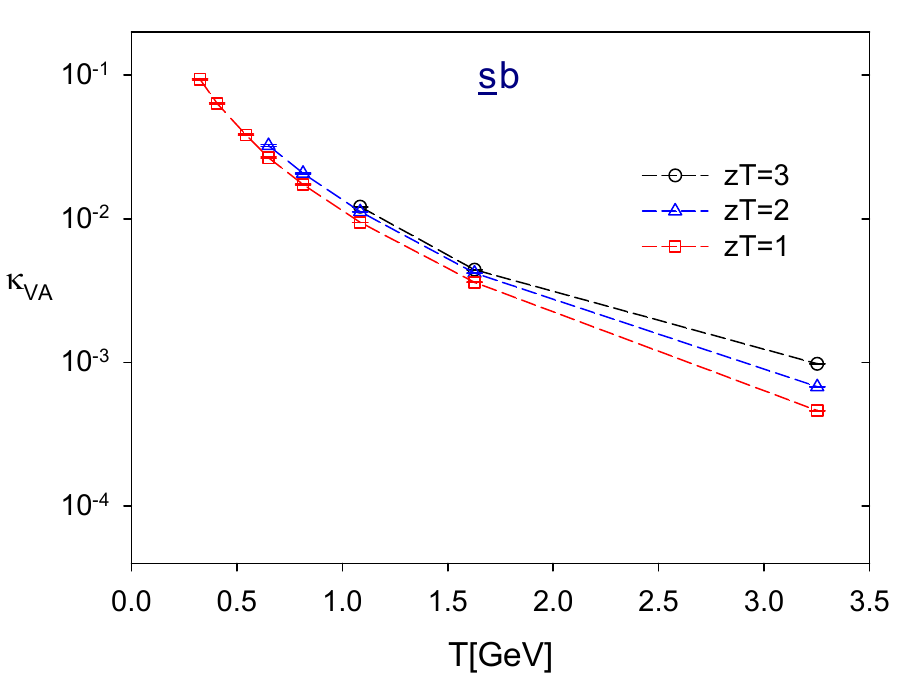}
&
  \includegraphics[width=7.5cm,clip=true]{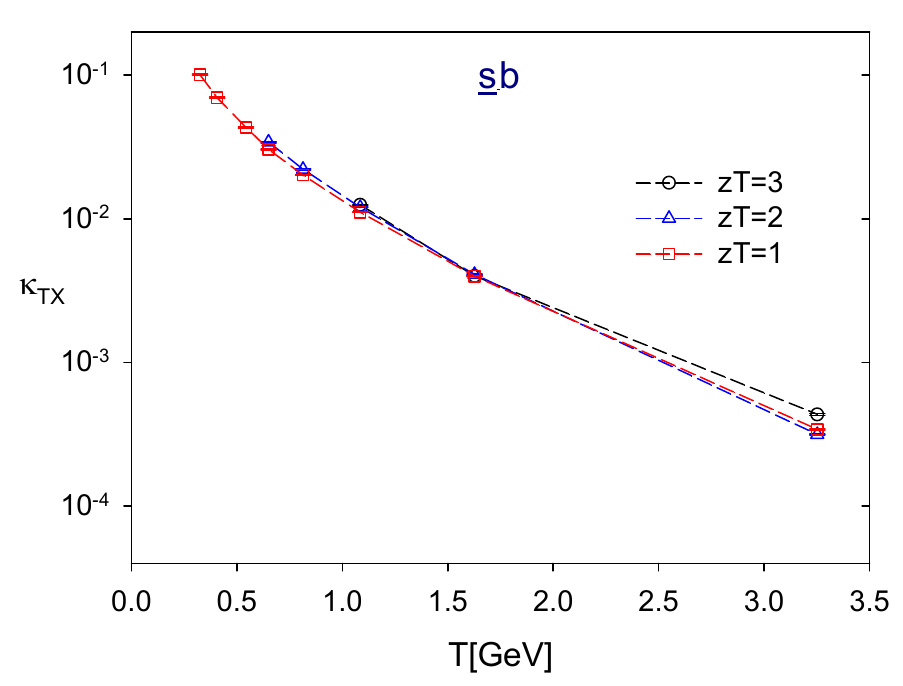} 
\end{tabular}
  \label{fig:kVA_TX_uc_sc_ub_sb}
\end{figure}

\begin{figure}[!h]
  \centering
  \caption{The chiral symmetry breaking parameters $(\kappa_{VA}, \kappa_{TX})$ in the 
           $(\bar c c, \bar c b, \bar b b)$ sectors.}
\begin{tabular}{@{}c@{}c@{}}
  \includegraphics[width=7.5cm,clip=true]{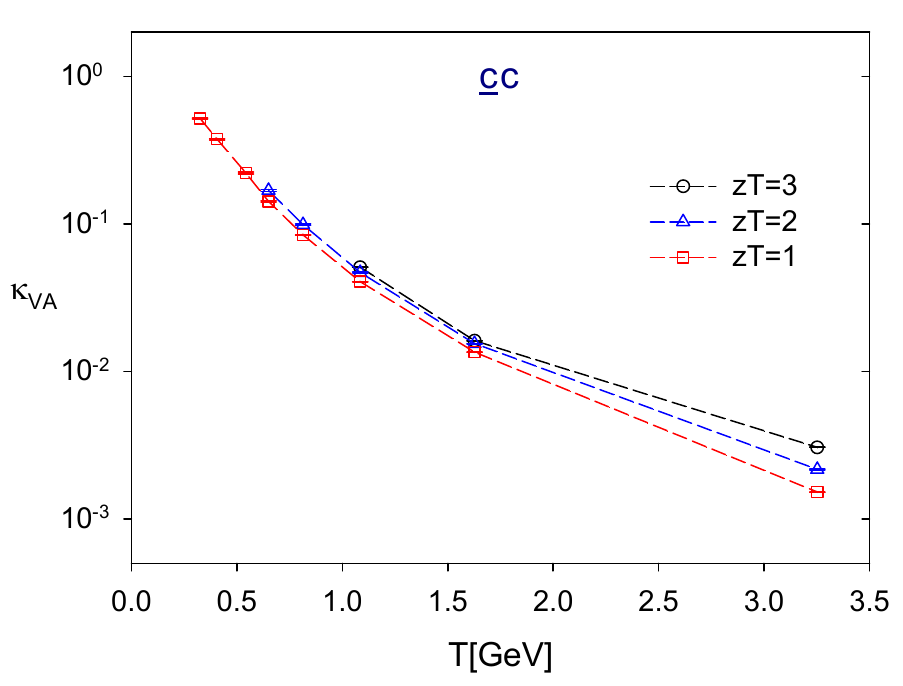}
&
  \includegraphics[width=7.5cm,clip=true]{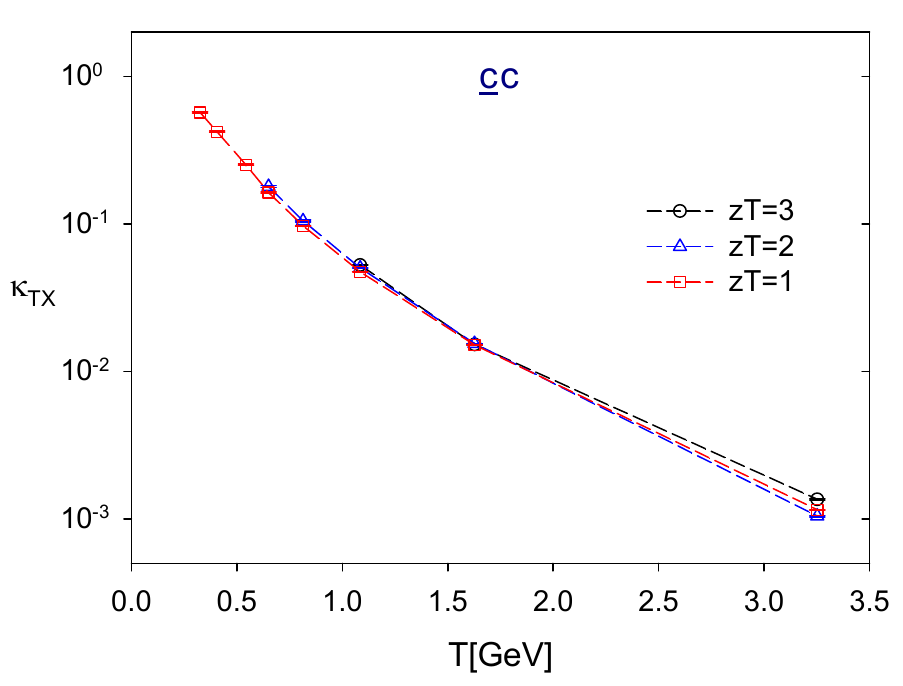} 
\vspace{-10pt} 
\\ 
  \includegraphics[width=7.5cm,clip=true]{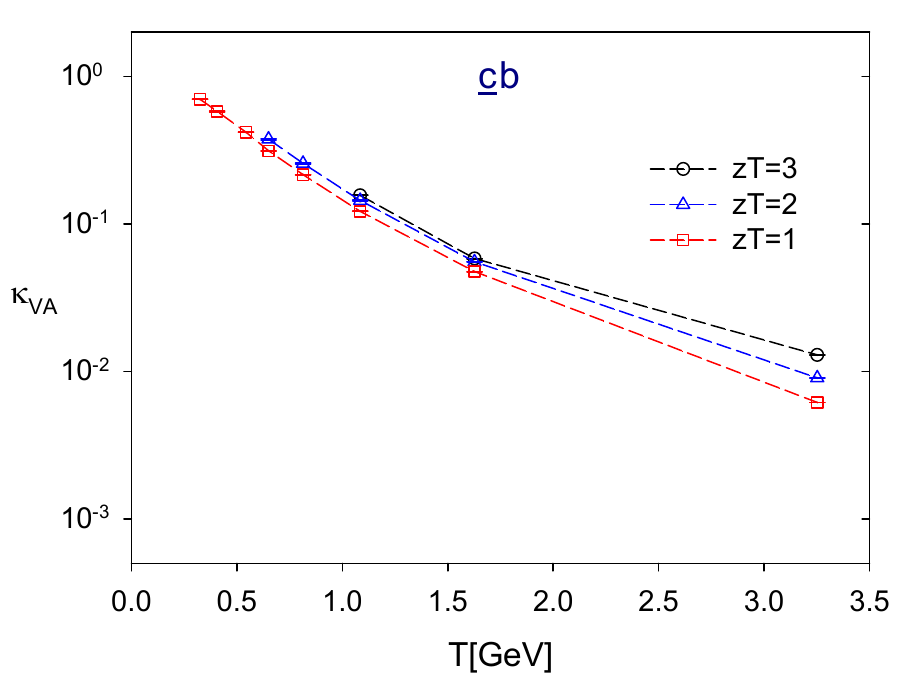}
&
  \includegraphics[width=7.5cm,clip=true]{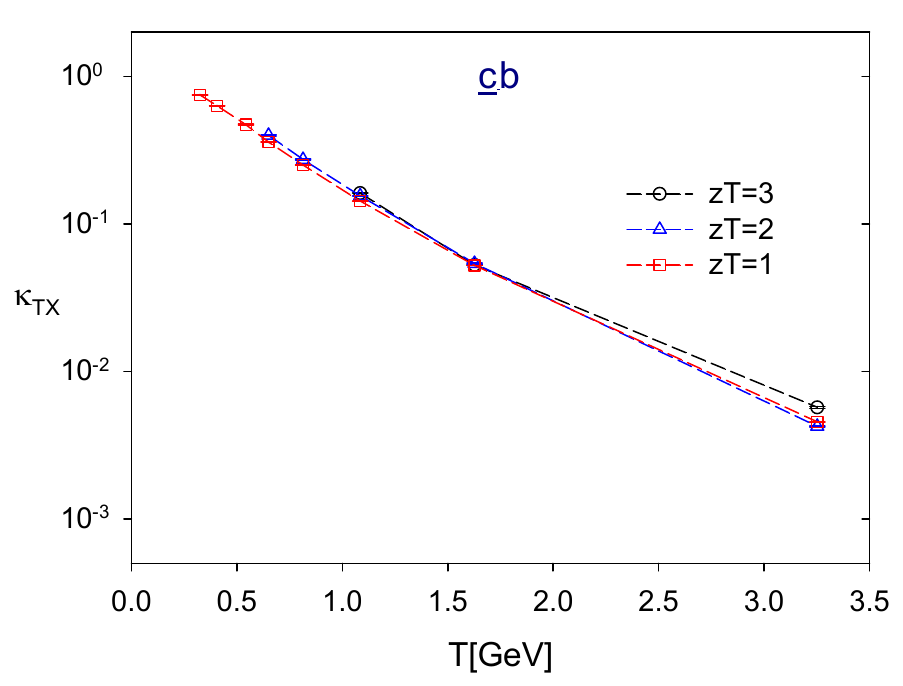} 
\vspace{-10pt} 
\\ 
  \includegraphics[width=7.5cm,clip=true]{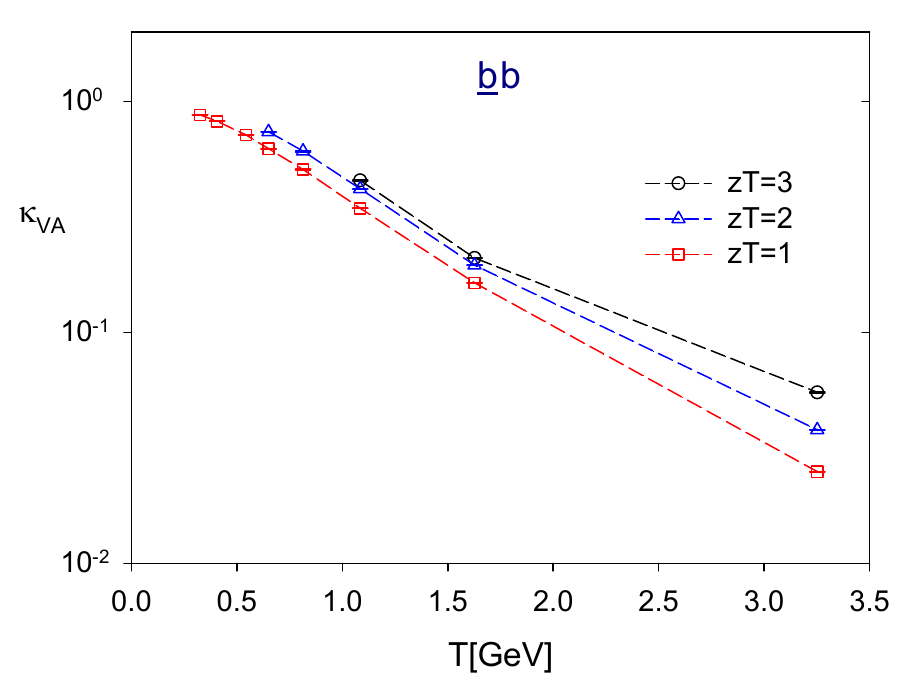}
&
  \includegraphics[width=7.5cm,clip=true]{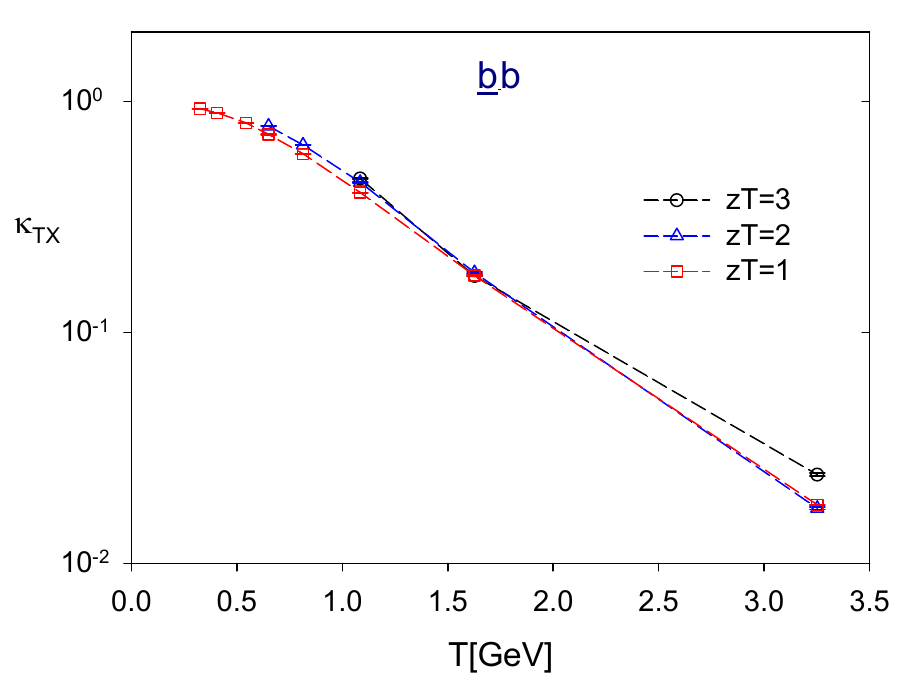} 
\end{tabular}
\label{fig:kVA_kTX_cc_cb_bb}
\end{figure}

Now we proceed to investigate the hierarchical restoration of chiral symmetry     
in $N_f=2+1+1+1$ lattice QCD with physical $s$, $c$ and $b$ quarks
but unphysically heavy $u/d$ quarks (with $M_\pi \sim 700$ MeV). 

First, we compute two sets of quark propagators with periodic and antiperiodic 
boundary conditions in the $ z $-direction, while keeping the boundary conditions 
in the $ (x,y,t) $-directions the same—periodic in $ (x,y) $ and antiperiodic in $ t $. 
Each set of quark propagators is independently used to construct the $ z $-correlators 
according to Equation~(\ref{eq:C_Gamma}), and their average is taken to obtain the final 
$ z $-correlators. This procedure effectively cancels the contributions 
of unphysical meson states at large distances \cite{Chiu:2023hnm}.  

Using these refined $ z $-correlators, we compute the chiral symmetry breaking 
parameters $ \kappa_{VA}(zT) $ and $ \kappa_{TX}(zT) $, 
plotting them as functions of $ T $ for $zT=(1,2,3)$, 
as shown in Figures~\ref{fig:kVA_kTX_ud_us_ss}--\ref{fig:kVA_kTX_cc_cb_bb}. 
The numerical values of $ \kappa_{VA} $ and $ \kappa_{TX} $ 
are provided in Tables~\ref{tab:Kappa_ud}--\ref{tab:Kappa_bb} of \ref{app:C} for each flavor sector: 
$(\bar{u} d,\bar{u} s,\bar{s} s,\bar{u} c,\bar{s} c,\bar{u} b,\bar{s} b,\bar{c} c,\bar{c} b,\bar{b} b)$.  
The statistical errors of $ \kappa_{VA} $ and $ \kappa_{TX} $ 
are estimated using the jackknife method with a bin size of $\sim 10-15$ configurations 
of which the error saturates.

For the $ z $-correlators, the possible values of $ zT $ at 
$ T = 1/(N_t a) $ are given by $ \{n_z/N_t \mid n_z = 1, 2, \dots, N_z/2 \} $. 
Thus, for $ N_z = 40 $ and $ N_t = (20, 16, 12, 10, 8, 6, 4, 2) $, 
the number of available temperature points is $ (8, 5, 3) $ for 
$ zT = (1,2,3) $, respectively, as illustrated in 
Figures~\ref{fig:kVA_kTX_ud_us_ss}--\ref{fig:kVA_kTX_cc_cb_bb} 
and Tables~\ref{tab:Kappa_ud}--\ref{tab:Kappa_bb} of \ref{app:C}.


First, for each flavor content,    
$\kappa_{VA} (zT)$ and $\kappa_{TX}(zT)$ at fixed $zT$ is a monotonic decreasing function of $T$.
At each $T$, and for fixed $zT$, the chiral symmetry breakings due to the quark masses of the 
meson operator can be seen clearly from $\kappa_{VA}$ and $\kappa_{TX}$, in the order of 
\bea
\label{eq:k_VA_TX}
\kappa_{\alpha}^{\bar u d} < \kappa_{\alpha}^{\bar u s} < \kappa_{\alpha}^{\bar s s} 
< \kappa_{\alpha}^{\bar u c} < \kappa_{\alpha}^{\bar s c}
< \kappa_{\alpha}^{\bar u b} < \kappa_{\alpha}^{\bar s b}
< \kappa_{\alpha}^{\bar c c} < \kappa_{\alpha}^{\bar c b} < \kappa_{\alpha}^{\bar bb},  
\hspace{4mm} \kappa_{\alpha} = \kappa_{VA}, \kappa_{TX}.  
\eea  
It follows that for any $\epsilon_{VA}$ in (\ref{eq:Tc_epsilon}) 
and any $\epsilon_{TX}$ in (\ref{eq:T1_epsilon}),    
the flavor dependence of $T_c$ and $T_1$ is in the order of  
\bea
\label{eq:Tc_order}
&& T_c^{\bar u d} < T_c^{\bar u s} < T_c^{\bar s s} < 
   T_c^{\bar u c} < T_c^{\bar s c} < T_c^{\bar u b} < T_c^{\bar s b} <  
   T_c^{\bar c c} < T_c^{\bar c b} < T_c^{\bar b b}, \\ 
&& T_1^{\bar u d} < T_1^{\bar u s} < T_1^{\bar s s} < 
   T_1^{\bar u c} < T_1^{\bar s c} < T_1^{\bar u b} < T_1^{\bar s b} <  
   T_1^{\bar c c} < T_1^{\bar c b} < T_1^{\bar b b}.  
\label{eq:T1_order}
\eea 
This immediately gives 
\bea
\label{eq:Tc1_order}
T_{c1}^{\bar u d} < T_{c1}^{\bar u s} < T_{c1}^{\bar s s} < 
T_{c1}^{\bar u c} < T_{c1}^{\bar s c} < T_{c1}^{\bar u b} < T_{c1}^{\bar s b} < 
T_{c1}^{\bar c c} < T_{c1}^{\bar c b} < T_{c1}^{\bar b b},   
\eea 
and the hierarchic restoration of chiral symmetry in $N_f=2+1+1+1$ QCD,  
i.e., from the restoration of $SU(2)_L \times SU(2)_R \times U(1)_A$ chiral symmetry 
of $(u, d)$ quarks at $T > T_{c1}^{\bar u d}$ to the    
the restoration of $SU(3)_L \times SU(3)_R \times U(1)_A$ chiral symmetry 
of $(u, d, s)$ quarks at $ T > T_{c1}^{\bar s s}$, then to  
the restoration of $SU(4)_L \times SU(4)_R \times U(1)_A$ chiral symmetry 
of $(u, d, s, c)$ quarks at $ T > T_{c1}^{\bar c c}$, 
and finally to $SU(5)_L \times SU(5)_R \times U(1)_A$ chiral symmetry 
of $(u, d, s, c, b)$ quarks at $ T > T_{c1}^{\bar b b}$. 

{\it
Thus, with the result of (\ref{eq:Tc1_order}), our primary objective in this exploratory study 
(see the discussion in Section~\ref{intro})—to observe the emergence of 
$ SU(5)_L \times SU(5)_R \times U(1)_A $ symmetry in QCD with $(u,d,s,c,b)$—has been fulfilled.
It is important to emphasize that our goal is to provide a qualitative understanding of the 
hierarchical restoration of chiral symmetry in $N_f=2+1+1+1$ lattice QCD, 
rather than to precisely determine the temperatures associated with this restoration. 
}

In the following, we aim to estimate approximate values of $T_c$ and $T_1$ 
for each flavor sector by solving Equations~(\ref{eq:Tc_epsilon}) and (\ref{eq:T1_epsilon}) 
through interpolation or extrapolation of the data points for $\kappa_{VA}$ and $\kappa_{TX}$. 

For example, at $ zT = 1 $, if we impose $ \epsilon_{VA} = \epsilon_{TX} \sim 0.025 $ 
as the criterion for chiral symmetry restoration, 
then the values of $ \kappa_{VA} $ and $ \kappa_{TX} $ for the $ \bar{b} b $ sector, 
as well as other flavor contents, are all below 0.025 at $ T \sim 3252 $ MeV, 
as shown in Tables~\ref{tab:Kappa_ud}--\ref{tab:Kappa_bb}. 
Consequently, the $ SU(5)_L \times SU(5)_R \times U(1)_A $ chiral symmetry 
of $ (u,d,s,c,b) $ quarks is restored at $ T_{c1}^{\bar{b} b} \sim 3252(10) $ MeV, 
in accordance with Equation~(\ref{eq:Tc1_N}).  

The next step is to determine $ T_{c1}^{\bar{c} c} $, 
at which the $ SU(4)_L \times SU(4)_R \times U(1)_A $ chiral symmetry of $ (u,d,s,c) $ 
quarks is restored. From Tables~\ref{tab:Kappa_uc}, \ref{tab:Kappa_sc}, 
and \ref{tab:Kappa_cc}, the values of $ \kappa_{VA} $ and $ \kappa_{TX} $ 
for the $ (\bar{u} c, \bar{s} c, \bar{c} c) $ sectors decrease to approximately 0.025 
at three different temperatures: $ 406 < T_{c1}^{\bar{u} c} < 542 $ MeV, 
$ 406 < T_{c1}^{\bar{s} c} < 542 $ MeV, and $ 1084 < T_{c1}^{\bar{c} c} < 1626 $ MeV, following 
the hierarchy $ T_{c1}^{\bar{u} c} < T_{c1}^{\bar{s} c} < T_{c1}^{\bar{c} c} $ in (\ref{eq:Tc1_order}).
Consequently, the $ SU(4)_L \times SU(4)_R \times U(1)_A $ chiral symmetry of 
$ (u,d,s,c) $ quarks is restored at $ T_{c1}^{\bar{c} c} \sim 1385(50) $ MeV, 
estimated via piecewise linear interpolation of $ \kappa_{VA} $ and $ \kappa_{TX} $ 
between 1084 MeV and 1626 MeV. The uncertainty in $T_{c}$ ($T_1$) is estimated by 
comparing results from two different schemes: 
piecewise linear interpolation of $\kappa_{VA}$ ($\kappa_{TX}$) 
and piecewise linear interpolation of $\log(\kappa_{VA})$ ($\log(\kappa_{TX})$).  

Similarly, from Tables~\ref{tab:Kappa_ud}--\ref{tab:Kappa_ss} for 
the $ (\bar{u} d, \bar{u} s, \bar{s} s) $ sectors, 
the values of $ \kappa_{VA} $ and $ \kappa_{TX} $ fall below 0.025 
at $ T \sim 325 $ MeV, the lowest temperature among the eight gauge ensembles listed 
in Table~\ref{tab:8_ensembles}. 
This indicates that the $ SU(3)_L \times SU(3)_R \times U(1)_A $ chiral symmetry 
of $ (u,d,s) $ quarks has been restored at $ T < 325 $ MeV.  
By applying piecewise linear extrapolation of $ \log(\kappa_{VA}) $ 
and $ \log(\kappa_{TX}) $, we estimate $ T_{c1}^{\bar{s} s} \sim 167(8) $ MeV. 
Here the logarithmic scale is preferred due to the observed linear behavior of 
$\log(\kappa_{VA})$ and $\log(\kappa_{TX})$ versus $T$ 
for the three lowest temperature data points at $ T = (325, 406, 542) $ MeV. 
Therefore, the $ SU(3)_L \times SU(3)_R \times U(1)_A $ chiral symmetry of
$ (u,d,s) $ quarks is restored at $ T_{c1}^{\bar{s} s} \sim 167(8) $ MeV, 
implying that the $ SU(2)_L \times SU(2)_R \times U(1)_A $ chiral symmetry 
of $ (u,d) $ quarks should be restored at $ T_{c1}^{\bar{u} d} < 167 $ MeV.  
However, we do not attempt to estimate $ T_{c1}^{\bar{u} d} $ via extrapolation, 
given the unphysically heavy $ u/d $ quarks used in this study.

To summarize the hierarchical restoration of chiral symmetry for 
$ \epsilon_{VA} = \epsilon_{TX} \sim 0.025 $ and $ zT = 1 $:  
\begin{itemize}
\item 
First, the $ SU(2)_L \times SU(2)_R \times U(1)_A $ chiral symmetry 
of $ (u, d) $ quarks is expected to be restored at $ T_{c1}^{\bar{u} d} < 167 $ MeV, 
but its precise determination is beyond the scope of this study.   
\item 
As the temperature increases, the $ SU(3)_L \times SU(3)_R \times U(1)_A $ 
chiral symmetry of $ (u, d, s) $ quarks is restored at 
$ T_{c1}^{\bar{s} s} \sim 167(8) $ MeV.  
\item 
With further temperature increase, the $ SU(4)_L \times SU(4)_R \times U(1)_A $ 
chiral symmetry of $ (u, d, s, c) $ quarks is restored 
at $ T_{c1}^{\bar{c} c} \sim 1385(50) $ MeV.   
\item 
Finally, the $ SU(5)_L \times SU(5)_R \times U(1)_A $ chiral symmetry 
of $ (u, d, s, c, b) $ quarks is restored at $ T_{c1}^{\bar{b} b} \sim 3252(10) $ MeV.
\end{itemize}

It should be emphasized that our results for $ T_{c1}^{\bar q Q} $ 
are subject to systematic uncertainties arising from unphysically heavy $ u/d $ quarks, 
as well as discretization and finite volume effects. These uncertainties cannot be quantified
in the present study, as the available gauge ensembles include 
only a single unphysical $ u/d $ quark mass, one spatial volume, 
and a single lattice spacing. {\it Our goal is not to provide a precise determination 
of $ T_c $ (or $ T_1 $) for each flavor content in $ N_f=2+1+1+1 $ lattice QCD, 
but rather to offer a qualitative picture of the hierarchical restoration of chiral symmetry
in $ N_f=2+1+1+1 $ lattice QCD, as demonstrated above.} 
This work represents a first step toward more precise determinations of $ T_{c1} $
with controlled systematics in future lattice studies, 
which will require simulations at the physical point and sufficiently large spatial volumes 
(i.e., $ > 180^3 \times N_t $). 

Next, we demonstrate how $ T_c$ and $T_1 $ depend on $ \epsilon_{VA} $ 
and $ \epsilon_{TX} $ in (\ref{eq:Tc_epsilon}) and (\ref{eq:T1_epsilon}). 
Since $ \kappa_{VA}^{\bar q_1 q_2} $ ($ \kappa_{TX}^{\bar q_1 q_2} $) 
at fixed $ zT $ is a monotonically decreasing function of $ T $, 
it follows that $ T_c$ ($T_1$) increases as $ \epsilon_{VA} $ 
($ \epsilon_{TX} $) decreases (i.e., the precision of chiral symmetry improves).  

For example, consider the case at $ zT=1 $ when both $ \epsilon_{VA} $ 
and $ \epsilon_{TX} $ are decreased from 0.025 to 0.015. 
According to Tables~\ref{tab:Kappa_ud}--\ref{tab:Kappa_ss}, 
at $ T = 325 $ MeV (the lowest temperature of the gauge ensembles), 
the values of $ \kappa_{VA} $ and $ \kappa_{TX} $ for the 
$ (\bar{u} d, \bar{u} s, \bar{s} s) $ sectors are all 
significantly below 0.015. This suggests that 
the $ SU(3)_L \times SU(3)_R \times U(1)_A $ chiral symmetry of $ (u, d, s) $ 
quarks must have been restored at $ T < 325 $ MeV. 
Using piecewise linear extrapolation of $ \log(\kappa_{VA}) $ 
and $ \log(\kappa_{TX}) $, we estimate $ T_{c1}^{\bar{s} s} \sim 240(10) $ MeV. 
The choice of using logarithmic values instead of linear ones follows the same 
reasoning as in the case where $ \epsilon_{VA} = \epsilon_{TX} \sim 0.025 $ at $ zT=1 $. 
Thus, the restoration of $ SU(3)_L \times SU(3)_R \times U(1)_A $ chiral symmetry 
occurs at $ T_{c1}^{\bar{s} s} \sim 240(10) $ MeV.  

At higher temperatures, Tables~\ref{tab:Kappa_uc}, \ref{tab:Kappa_sc}, 
and \ref{tab:Kappa_cc} show that the values of $ \kappa_{VA} $ 
and $ \kappa_{TX} $ for the $ (\bar{u} c, \bar{s} c, \bar{c} c) $ sectors 
decrease to approximately 0.015 at temperatures $ T_{c1}^{\bar{u} c} \sim 542 $ MeV, 
{542~MeV $< T_{c1}^{\bar{s} c} < $ 650~MeV}, and $ T_{c1}^{\bar{c} c} \sim 1626 $ MeV. 
Consequently, the $ SU(4)_L \times SU(4)_R \times U(1)_A $ chiral symmetry 
of $ (u,d,s,c) $ quarks is restored at $ T_{c1}^{\bar{c} c} \sim 1626(20) $ MeV. 

Turning to the $ \bar{b} b $ sector, Table~\ref{tab:Kappa_bb} indicates that 
for $ zT=1 $, the values of $ \kappa_{VA} $ and $ \kappa_{TX} $ 
remain above 0.015 even at $ T = 3252 $ MeV, the highest temperature among 
the eight gauge ensembles studied. 
This implies that the $ SU(5)_L \times SU(5)_R \times U(1)_A $ chiral symmetry 
of $ (u,d,s,c,b) $ quarks is restored only at $ T_{c1}^{\bar{b} b} > 3252 $ MeV. 
Using piecewise linear extrapolation of $ \kappa_{VA} $ and $ \kappa_{TX} $, 
we estimate $ T_{c1}^{\bar{b} b} \sim 3370(50) $ MeV.  

This analysis demonstrates the hierarchical restoration of chiral symmetry 
in $ N_f=2+1+1+1 $ QCD for $ \epsilon_{VA} = \epsilon_{TX} \sim 0.015 $ 
and $ zT=1 $, progressing from the restoration of $ SU(3)_L \times SU(3)_R \times U(1)_A $ 
for $ (u,d,s) $ quarks at $ T_{c1}^{\bar{s} s} \sim 240(10) $ MeV, 
to $ SU(4)_L \times SU(4)_R \times U(1)_A $ for $ (u,d,s,c) $ quarks 
at $ T_{c1}^{\bar{c} c} \sim 1626(20) $ MeV, 
and finally to $ SU(5)_L \times SU(5)_R \times U(1)_A $ for $ (u,d,s,c,b) $ quarks 
at $ T_{c1}^{\bar{b} b} \sim 3370(50) $ MeV.  

Clearly, regardless of how small $ \epsilon_{VA} $ and $ \epsilon_{TX} $ become, 
the hierarchical restoration of chiral symmetry in QCD with $ (u,d,s,c,b) $ quarks 
will occur at progressively higher temperatures.

\begin{figure}[th!]
  \centering
  \caption{Comparison of the chiral symmetry breaking parameters, $\kappa_{VA}$ and $\kappa_{TX}$,  
   in the $(\bar{u}d, \bar{u}s, \bar{u}c)$ sectors for lattice QCD with $N_f = 2+1+1+1$ (this work) 
   and $N_f = 2+1+1$ at the physical point \cite{Chiu:2024jyz}.}
\begin{tabular}{@{}c@{}c@{}}
  \includegraphics[width=7.5cm,clip=true]{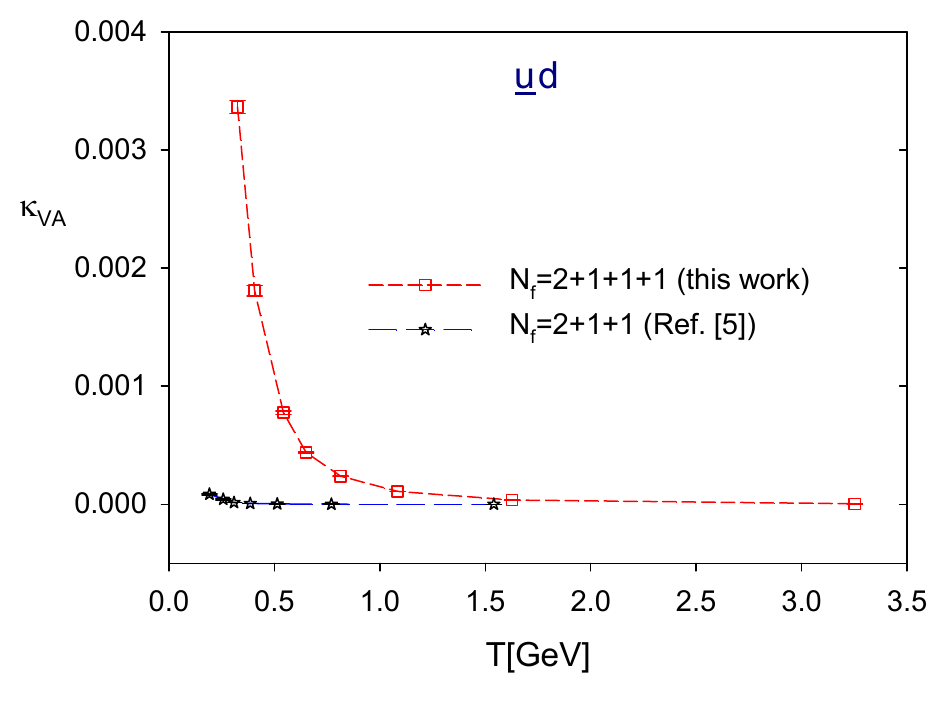}
&
  \includegraphics[width=7.5cm,clip=true]{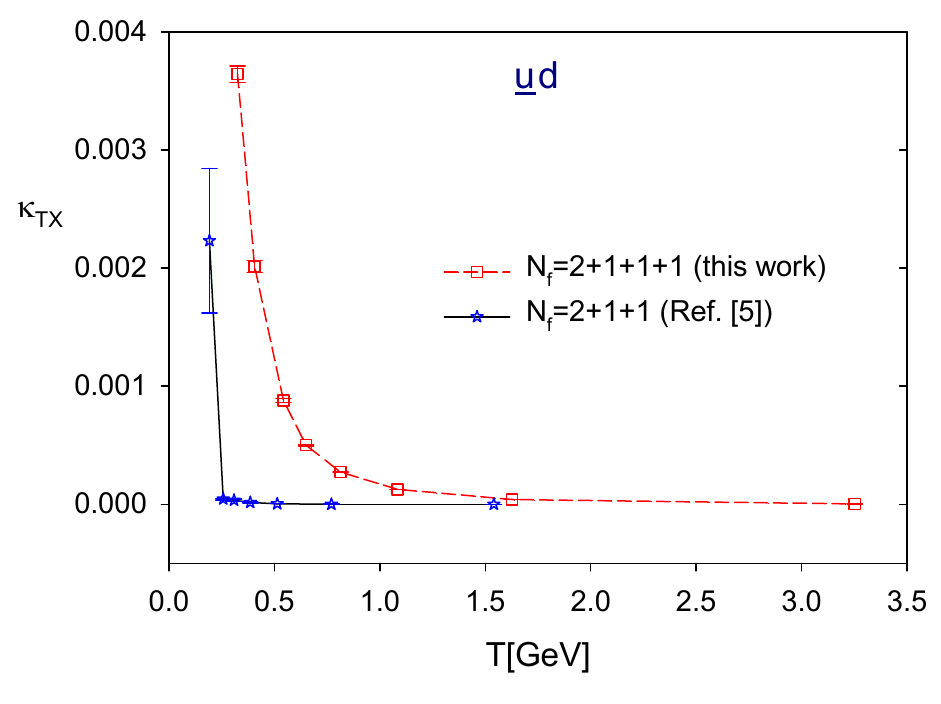}
\vspace{-10pt}
\\
  \includegraphics[width=7.5cm,clip=true]{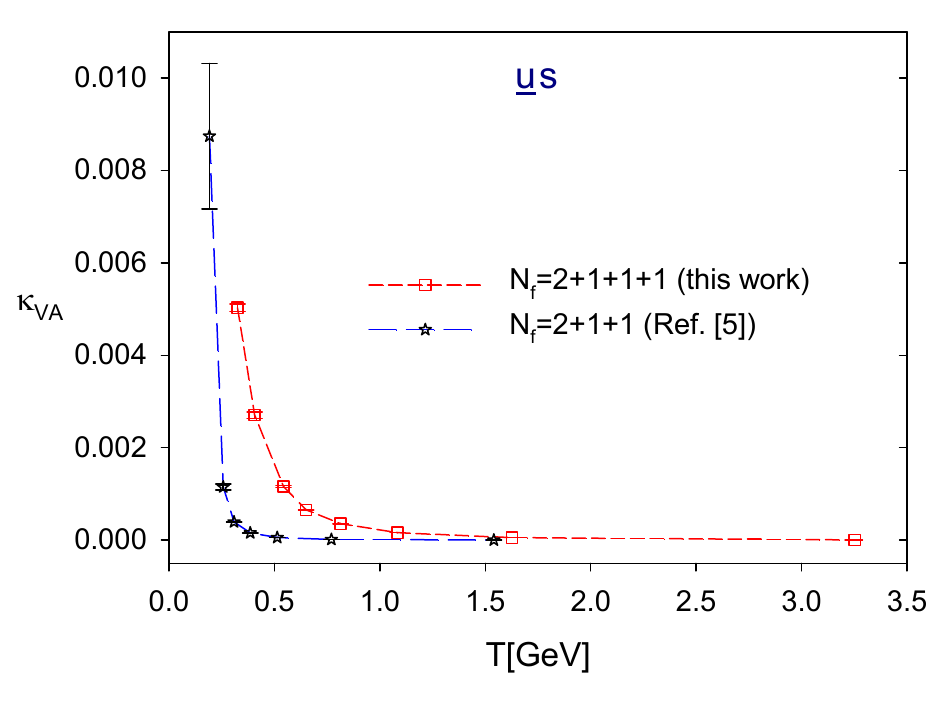}
&
  \includegraphics[width=7.5cm,clip=true]{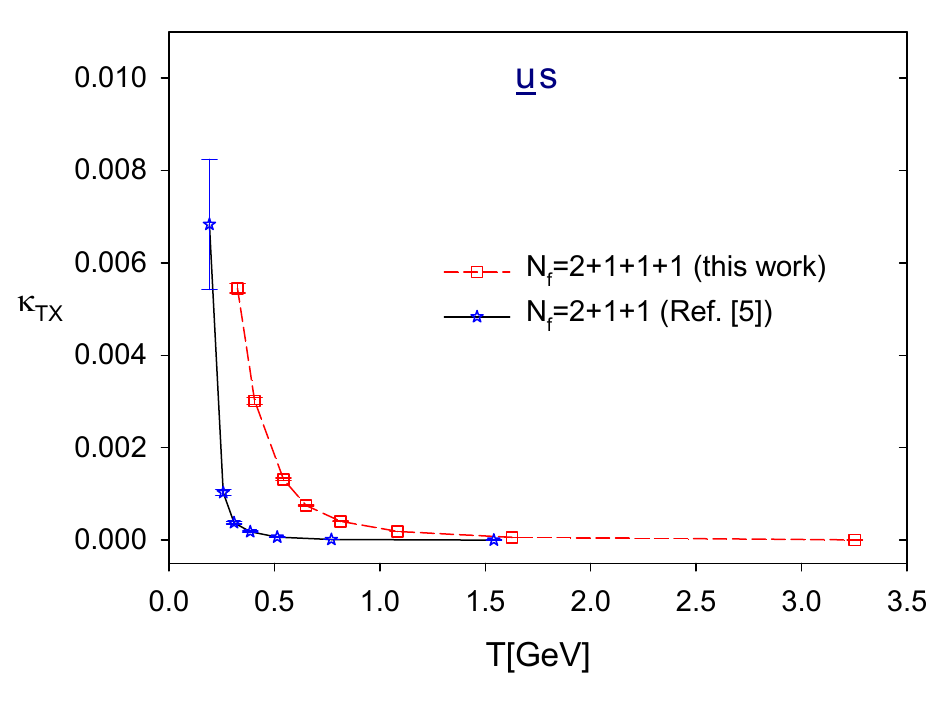}
\vspace{-10pt}
\\
  \includegraphics[width=7.5cm,clip=true]{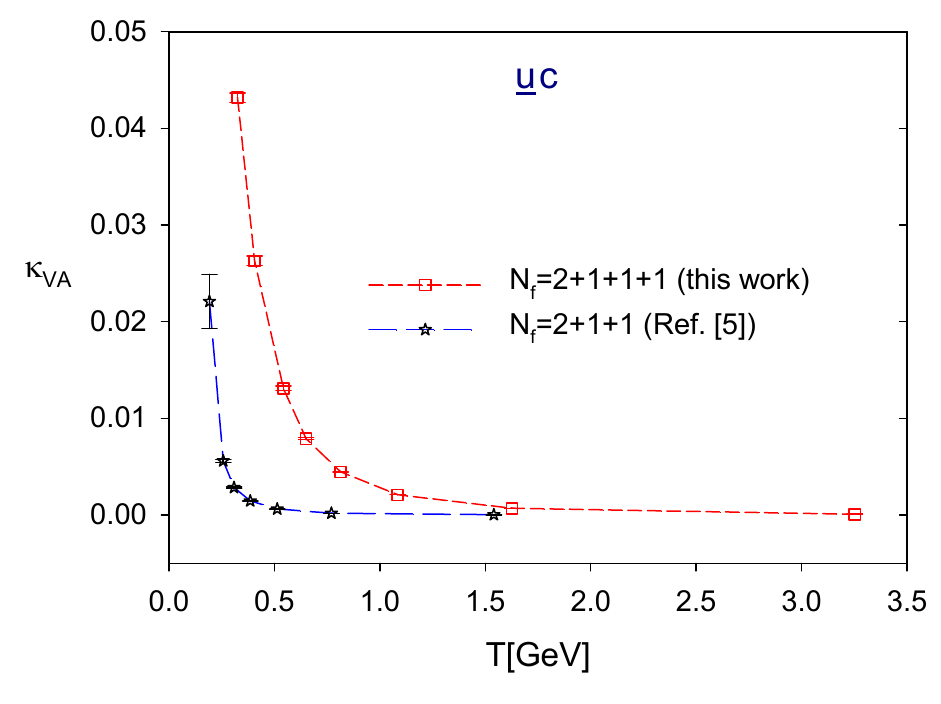}
&
  \includegraphics[width=7.5cm,clip=true]{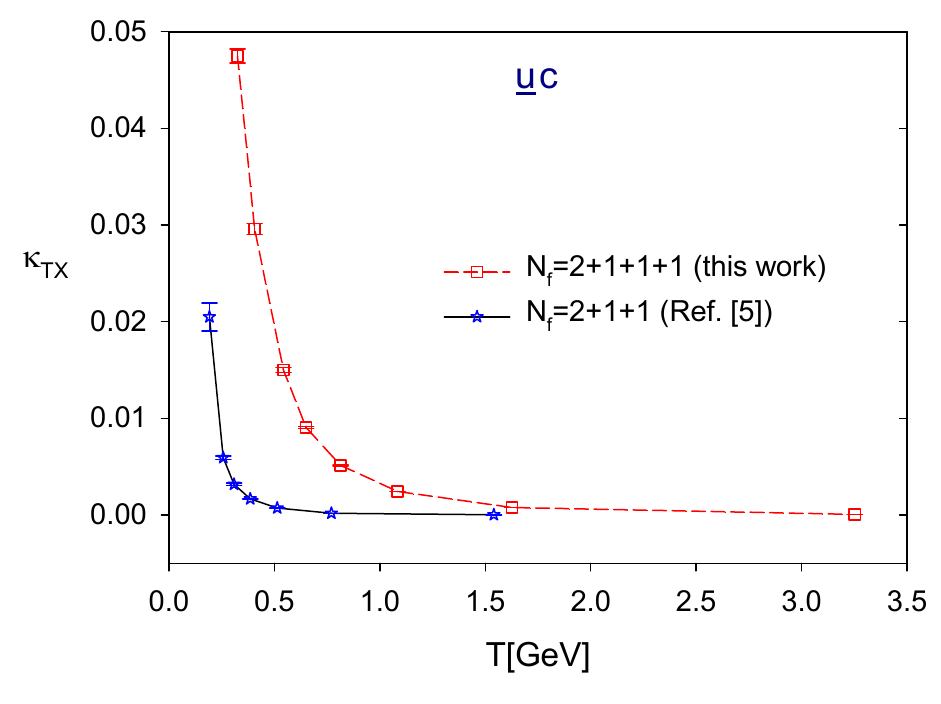}
\end{tabular}
\label{fig:kVA_kTX_ud_us_uc_compare}
\end{figure}

\begin{figure}[th!]
  \centering
  \caption{Comparison of the chiral symmetry breaking parameters, $\kappa_{VA}$ and $\kappa_{TX}$, in the
           $(\bar s s, \bar s c, \bar c c)$ sectors for lattice QCD with $N_f=2+1+1+1$ (this work)
           and $N_f=2+1+1$ at the physical point \cite{Chiu:2024jyz}.}
  \includegraphics[width=15cm,clip=true]{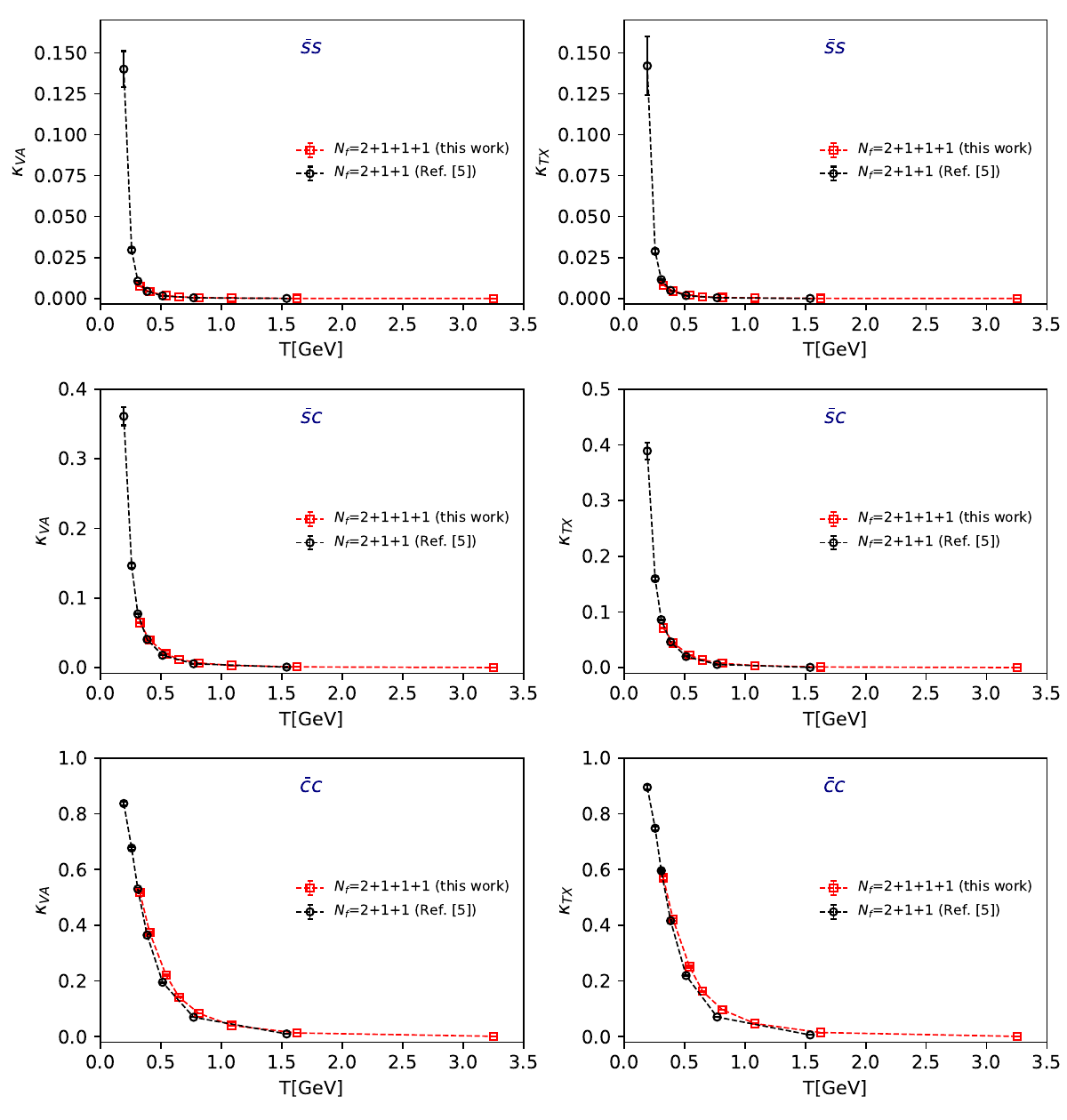}
\label{fig:kVA_kTX_ss_sc_cc_compare}
\end{figure}


\subsection{Comparison with $N_f=2+1+1$ lattice QCD at the physical point} 

Now we compare the chiral symmetry breaking parameters, 
$ \kappa_{VA} $ and $ \kappa_{TX} $, in $ N_f=2+1+1+1 $ lattice QCD 
with those in $ N_f=2+1+1 $ lattice QCD at the physical point \cite{Chiu:2024jyz}. 
The numerical values for $ N_f=2+1+1+1 $ QCD are presented 
in Tables~\ref{tab:Kappa_ud}--\ref{tab:Kappa_bb}, 
while those for $ N_f=2+1+1 $ QCD can be found in 
Tables~\ref{tab:Kappa_ud_Nf2p1p1}--\ref{tab:Kappa_cc_Nf2p1p1}.  
Figures~\ref{fig:kVA_kTX_ud_us_uc_compare} and \ref{fig:kVA_kTX_ss_sc_cc_compare} 
show the values of $ \kappa_{VA} $ and $ \kappa_{TX} $ at $ zT=1 $ 
for both lattice setups. 

For the $(\bar{u} d, \bar{u} s, \bar{u} c)$ sectors, significant discrepancies are observed, 
which can be attributed to the unphysically heavy $u/d$ quarks contributing to both 
the valence quark propagators and the vacuum fluctuations in the sea. 

For the $\bar{s} s$ and $\bar{s} c$ sectors, the values of $\kappa_{VA}$ and $\kappa_{TX}$ 
in $N_f=2+1+1+1$ lattice QCD are in good agreement with those in $N_f=2+1+1$ lattice QCD 
at the physical point, despite the presence of unphysically heavy $u/d$ quarks in the sea. 

For the $\bar{c} c$ sector, the values of $\kappa_{VA}$ and $\kappa_{TX}$ in $N_f=2+1+1+1$ lattice QCD 
are in reasonable agreement with those in $N_f=2+1+1$ lattice QCD at the physical point. 
However, discrepancies in the temperature range $T \sim 400$–$1200$~MeV are more pronounced 
compared to those in the $\bar{s}s$ and $\bar{s}c$ sectors. 
At a fixed temperature, $\kappa_{VA}$ and $\kappa_{TX}$ in $N_f=2+1+1+1$ lattice QCD is consistently 
larger than those in $N_f=2+1+1$ lattice QCD. 
Due to the limited number of data points in both lattice setups, 
precisely quantifying these discrepancies remains challenging. Since both setups are subject to 
discretization and finite volume uncertainties, these effects are likely the primary sources 
of the observed differences. A more thorough understanding would require taking the continuum 
and infinite volume limits for both lattice setups, which is beyond the scope of this paper.  

Next, we compare $T_c$ ($T_1$) between the two lattice setups for a given 
$\epsilon_{VA}$ ($\epsilon_{TX}$). 
This generally requires interpolation or extrapolation $\kappa_{VA}$ ($\kappa_{TX}$) 
to solve Equation~(\ref{eq:Tc_epsilon}) or (\ref{eq:T1_epsilon}), 
which may introduce significant uncertainties due to the limited number of data points in both setups.  

This issue becomes particularly severe when $\epsilon_{VA}$ or $\epsilon_{TX}$ falls below 0.01, 
as the approximate solutions for $T_c$ and $T_1$ extend into or beyond the range of the  
two highest-temperature data points: $(770, 1540)$~MeV for $N_f=2+1+1$ QCD 
and $(1626, 3252)$~MeV for $N_f=2+1+1+1$ QCD. 
Consequently, interpolation or extrapolation of $\kappa_{VA}$ and $\kappa_{TX}$ may introduce 
large uncertainties, leading to discrepancies of approximately 100–300 MeV in the estimated values 
of $T_c$ and $T_1$.  
Furthermore, these two highest-temperature data points correspond to the smallest values 
of $N_t = 4$ and 2, which can introduce significant discretization errors and distort the $z$-correlators, 
thereby affecting $\kappa_{VA}$ and $\kappa_{TX}$. In other words, interpolation or extrapolation 
using only the two highest-temperature data points in each lattice setup is prone to large 
systematic errors. As a result, the discrepancies in $T_c$ and $T_1$ between the two lattice setups 
may increase as $\epsilon_{VA}$ and $\epsilon_{TX}$ decrease—that is, 
as higher precision in chiral symmetry is pursued.

In view of the above discussion, we set $\epsilon_{VA}$ and $\epsilon_{TX}$ to 0.1 and 0.05 
(both smaller than 0.01) and estimate approximate values of $T_c$ and $T_1$ 
in the ($\bar s s, \bar s c, \bar c c)$ sectors for two lattice setups.  

\begin{table}[htbp]
\caption{Comparison of $T_{c}^{\bar{s}s}$, $T_c^{\bar s c}$, and $T_{c}^{\bar{c}c}$ between
         $N_f=2+1+1$ and $N_f=2+1+1+1$ lattice QCD,
         for $zT = 1$ and $\epsilon_{VA}=0.1$ and 0.05, respectively.}
\vspace{2mm}
\centering
\begin{tabular}{|c|cc|cc|}
\hline
   &\multicolumn{2}{c|}{$N_f=2+1+1$ \cite{Chiu:2024jyz}} &\multicolumn{2}{c|}{$N_f=2+1+1+1$ (This Work)} \\
\hline
$\epsilon_{VA}$    & 0.1 & 0.05 & 0.1 & 0.05 \\
\hline
$T_{c}^{\bar s s}$[MeV] & 210(5) & 240(10) & \LB      & \LB        \\
$T_{c}^{\bar s c}$[MeV] & 285(10)& 360(10) & 250(40)  & 370(10)    \\
$T_{c}^{\bar c c}$[MeV] & 710(30)& 965(50) & 760(20)  & 1010(20)   \\
\hline
\end{tabular}
\label{tab:compare_Tc}
\end{table}

\begin{table}[h!]
\caption{Comparison of $T_{1}^{\bar{s}s}$, $T_1^{\bar s c}$, and $T_{1}^{\bar{c}c}$ between
         $N_f=2+1+1$ and $N_f=2+1+1+1$ lattice QCD,
         for $zT = 1$ and $\epsilon_{TX}=0.1$ and 0.05, respectively.}
\vspace{2mm}
\centering
\begin{tabular}{|c|cc|cc|}
\hline
    &\multicolumn{2}{c|}{$N_f=2+1+1$ \cite{Chiu:2024jyz}} &\multicolumn{2}{c|}{$N_f=2+1+1+1$ (This Work)} \\
\hline
$\epsilon_{TX}$ & 0.1 & 0.05 & 0.1 & 0.05 \\
\hline
$T_{1}^{\bar s s}$[MeV] & 210(5)  & 240(10)  & \LB      & \LB      \\
$T_{1}^{\bar s c}$[MeV] & 295(10) & 370(10)  & 270(30)  & 380(10)  \\
$T_{1}^{\bar c c}$[MeV] & 720(30) & 1020(30) & 800(50)  & 1060(20) \\
\hline
\end{tabular}
\label{tab:compare_T1}
\end{table}

In Tables \ref{tab:compare_Tc} and \ref{tab:compare_T1}, we compare $T_c$ and $T_1$ 
for the $(\bar{s}s, \bar{s}c, \bar{c}c)$ sectors in two lattice setups at $zT = 1$, 
with $\epsilon_{VA} = \epsilon_{TX} = 0.1$ and 0.05. The values of $T_c$ and $T_1$ are obtained 
by solving Equations~(\ref{eq:Tc_epsilon}) and (\ref{eq:T1_epsilon}) through interpolation or extrapolation 
of $\kappa_{VA}$ and $\kappa_{TX}$. 
The uncertainty in each $T_c$ ($T_1$) is estimated by 
comparing results from two different schemes: 
piecewise linear interpolation/extrapolation of $\kappa_{VA}$ ($\kappa_{TX}$) 
and piecewise linear interpolation/extrapolation of $\log(\kappa_{VA})$ ($\log(\kappa_{TX})$).  

First, consider the $\bar{s}s$ sector.
For $\epsilon_{VA}=\epsilon_{TX} = 0.1$ and 0.05, $T_c$ and $T_1$ of $N_f=2+1+1+1$ lattice QCD
cannot be determined using any of the two extrapolation schemes mentioned above,
as the values fall well below 325~MeV (the lowest temperature of the gauge ensembles in this study).
Consequently, a comparison in this case is not possible.

Next, consider the $\bar{s}c$ and $\bar c c$ sectors. For $\epsilon_{VA}=\epsilon_{TX}=0.1$ and 0.05,
the values of $T_c$ ($T_1$) from $N_f=2+1+1+1$ and $N_f=2+1+1$ lattice QCD are in good agreement,
within the uncertainties due to interpolation.

Overall, the reasonable agreement of $\kappa_{VA}$ and $\kappa_{TX}$ as well as $T_c$ and $T_1$ 
between $N_f=2+1+1+1$ lattice QCD and $N_f=2+1+1$ lattice QCD at the physical point \cite{Chiu:2024jyz} 
for the $(\bar{s}s, \bar{s}c, \bar{c}c)$ sectors highlights {\it the consistency between these two 
lattice setups for the physical $s$ and $c$ quarks.}

\section{$SU(2)_{CS}$ chiral-spin symmetry}
\label{SU2_CS}

In this section, we explore the emergence of approximate $ SU(2)_{CS} $ chiral-spin symmetry 
in $ N_f = 2+1+1+1 $ lattice QCD. Our results are subject to systematic uncertainties 
arising from unphysically heavy $ u/d $ quarks, as well as discretization and finite volume effects. 
These uncertainties cannot be quantified within this study, as the gauge ensembles include 
only a single unphysical $ u/d $ quark mass, one spatial volume, and a single lattice spacing. 
Rather than precisely determining the temperatures at which approximate $ SU(2)_{CS} $ chiral-spin symmetry 
emerges in each flavor sector, our aim is to provide a qualitative picture of its behavior 
in $ N_f = 2+1+1+1 $ lattice QCD.  

First, we recall the $SU_{CS}$ symmetry breaking and fading parameters $(\kappa_{CS}, \kappa)$ as 
defined in ref.~\cite{Chiu:2024jyz}, following the same notations and conventions.         

In general, to examine the emergence of $SU(2)_{CS}$ symmetry, it needs to measure 
the splittings in the $SU_{CS}(2)$ multiplet $(A_1, T_4, X_4)$. 
Since the splitting of $T_4$ and $X_4$ has been measured by the 
$U(1)_A$ symmetry breaking parameter $\kappa_{TX}$ (\ref{eq:k_TX_z}) with $k=4$, 
it remains to measure the splitting of $A_1$ and $X_4$ with  
\bea
\label{eq:k_AX_z}
\kappa_{AX}(zT) = \frac{\left| C_{A_1}(zT) - C_{X_4}(zT) \right|}{C_{A_1}(zT) + C_{X_4}(zT)}, 
\hspace{4mm} z > 0,   
\eea
then taking the maximum of $\kappa_{AT}$ and $\kappa_{TX}$ 
as the $SU(2)_{CS}$ symmetry breaking parameter,  
\bea
\label{eq:k_CS_z}
\kappa_{CS} = \max(\kappa_{AX}, \kappa_{TX}). 
\eea
Note that $\kappa_{AX}$ in (\ref{eq:k_AX_z}) is exactly the same as 
$\kappa_{AT}$ in ref.~\cite{Chiu:2024jyz}. Here we just change the subscript from $AT$
to $AX$ for consistency, since it refers to the splitting of the axial vector $A_1$ 
and the axial-tensor vector $X_4$.
    
As the temperature $T$ is increased, the separation between the multiplets of 
$SU(2)_{CS}$ and $U(1)_A$ is decreased.
Therefore, at sufficiently high temperatures $T > T_c^{\bar q Q}$, 
the $U(1)_A$ multiplet $M_0 = (P, S)$ and the 
$SU(2)_{CS} \times SU(2)_L \times SU(2)_R$ multiplet $M_2=(V_1, A_1, T_4, X_4)$ merge together, 
then the $SU(2)_{CS}$ symmetry becomes washed out, 
and only the $SU(2)_L \times SU(2)_R \times U(1)_A $ chiral symmetry remains.  
Note that the $SU(2)_{CS} \times SU(2)_L \times SU(2)_R$ multiplet  
$ M_4 = (V_4, A_4, T_1, X_1) $ never merges with 
$M_0$ and $M_2$ even in the limit $T \to \infty$, as discussed in ref.~\cite{Chiu:2023hnm}. 
Thus $M_4$ is irrelevant to the fading of the $SU(2)_{CS} $ symmetry.  
Following ref.~\cite{Chiu:2024jyz}, we use $\kappa(zT)$
to measure the fading of the $SU(2)_{CS}$ symmetry. 
\bea
\kappa(zT) = \left|\frac{C_{A_1}(zT) - C_{X_4}(zT)}{C_{M_0}(zT) - C_{M_2}(zT)} \right|, \hspace{4mm} z > 0, 
\label{eq:kappa_z}
\eea
where
\BAN
&& C_{M_0}(zT) \equiv \frac{1}{2} \left[ C_P(zT) + C_S(zT) \right],  \\
&& C_{M_2}(zT) \equiv \frac{1}{4} \left[ C_{V_1}(zT) + C_{A_1}(zT) + C_{T_4}(zT) + C_{X_4}(zT) \right].   
\EAN

Thus, to determine to what extent the $SU(2)_{CS}$ symmetry
is manifested in the $z$-correlators, it is necessary to examine whether 
both $\kappa(zT)$ and $\kappa_{CS}(zT)$ are sufficiently small. 
For a fixed $zT$, the following condition
\bea
\label{eq:SU2_CS_crit_z}
\left(~\kappa_{CS}(zT) < \epsilon_{cs} ~\right)~\land ~\left(~\kappa(zT) < \epsilon_{fcs} ~\right)
\eea
serves as a criterion for the $SU(2)_{CS}$ symmetry 
in the $z$-correlators, where $ \epsilon_{cs} $ is for the $SU(2)_{CS}$ symmetry breaking, 
while $ \epsilon_{fcs} $ for the $SU(2)_{CS}$ symmetry fading. 
For fixed $zT$, (\ref{eq:SU2_CS_crit_z}) gives a window of $T$ for the 
$SU(2)_{CS}$ symmetry. Obviously, the size of this window depends on  
$\epsilon_{cs}$ and $\epsilon_{fcs}$. That is, larger $\epsilon_{cs}$ or $\epsilon_{fcs}$
gives a wider window of $T$, and conversely, smaller $\epsilon_{cs}$ or $\epsilon_{fcs}$ 
gives a narrower window of $T$.

\begin{figure}[h!]
  \centering
  \caption{The chiral-spin symmetry breaking and fading parameters of the
           $(\bar u d, \bar u s, \bar s s)$ sectors.}
\begin{tabular}{@{}c@{}c@{}}
  \includegraphics[width=7.5cm,clip=true]{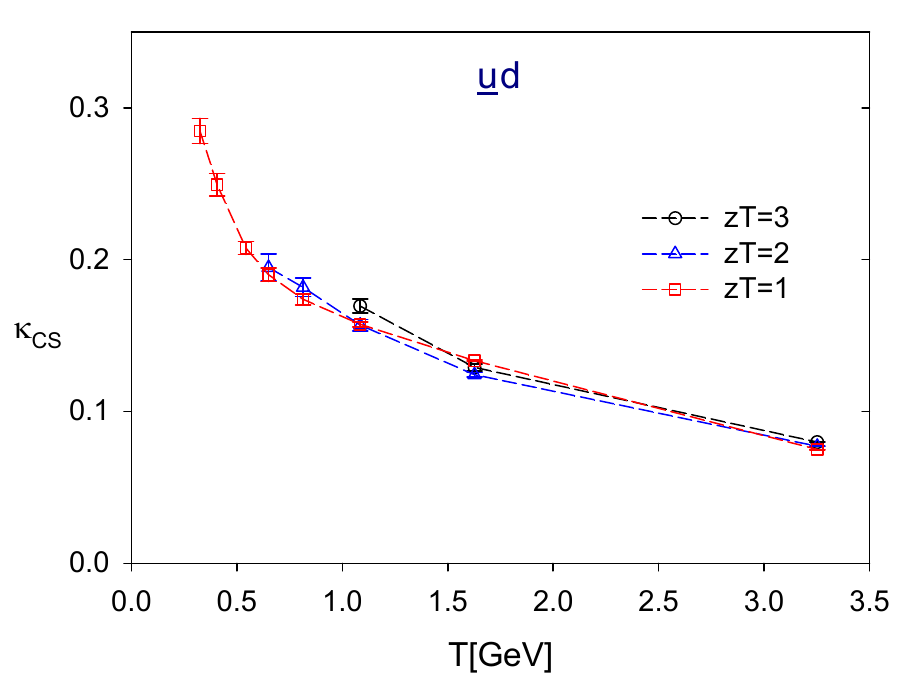}
&
  \includegraphics[width=7.5cm,clip=true]{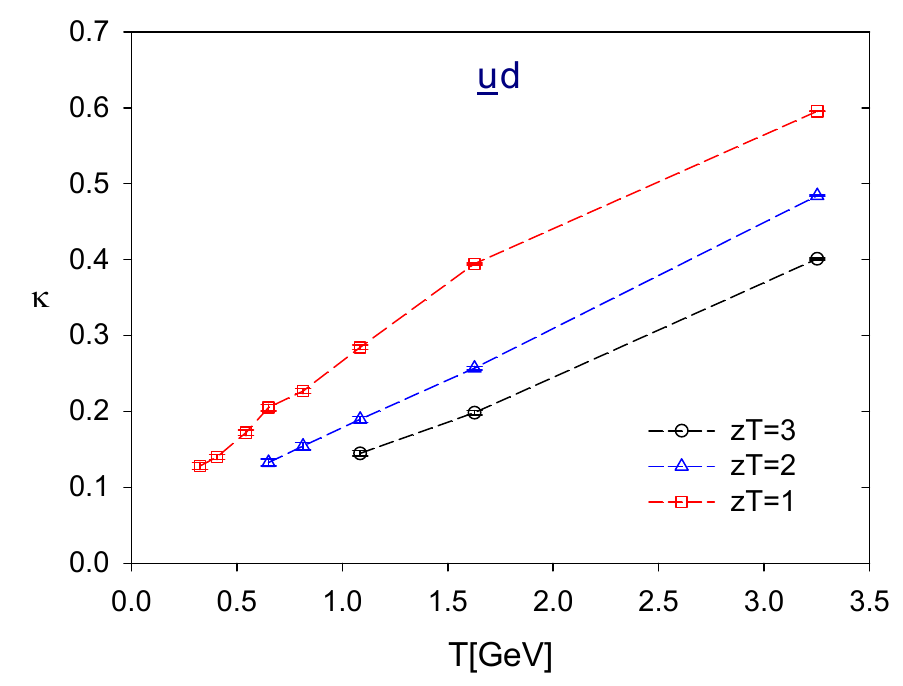}
\vspace{-10pt}
\\
  \includegraphics[width=7.5cm,clip=true]{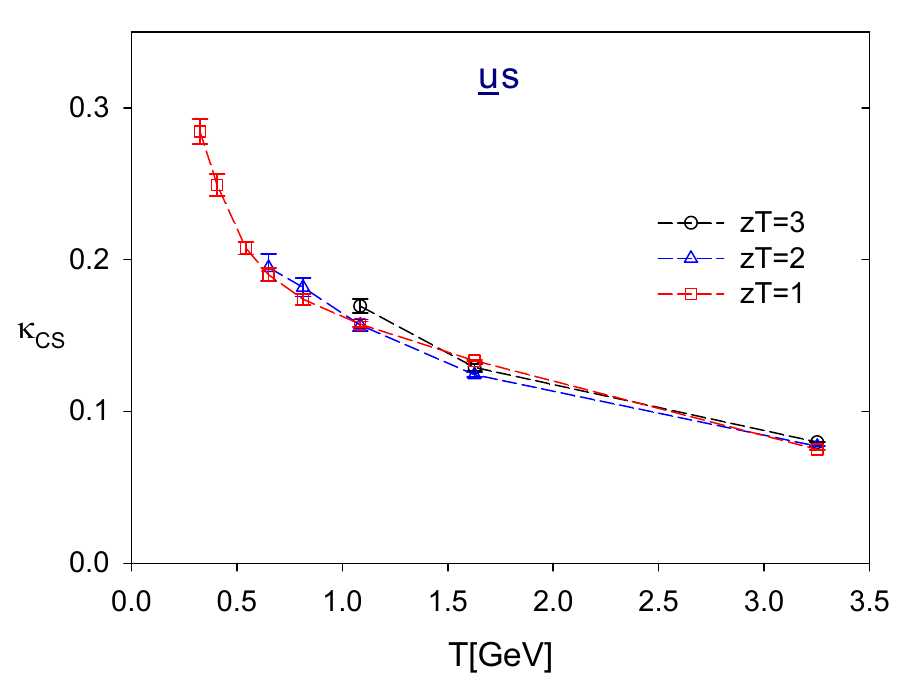}
&
  \includegraphics[width=7.5cm,clip=true]{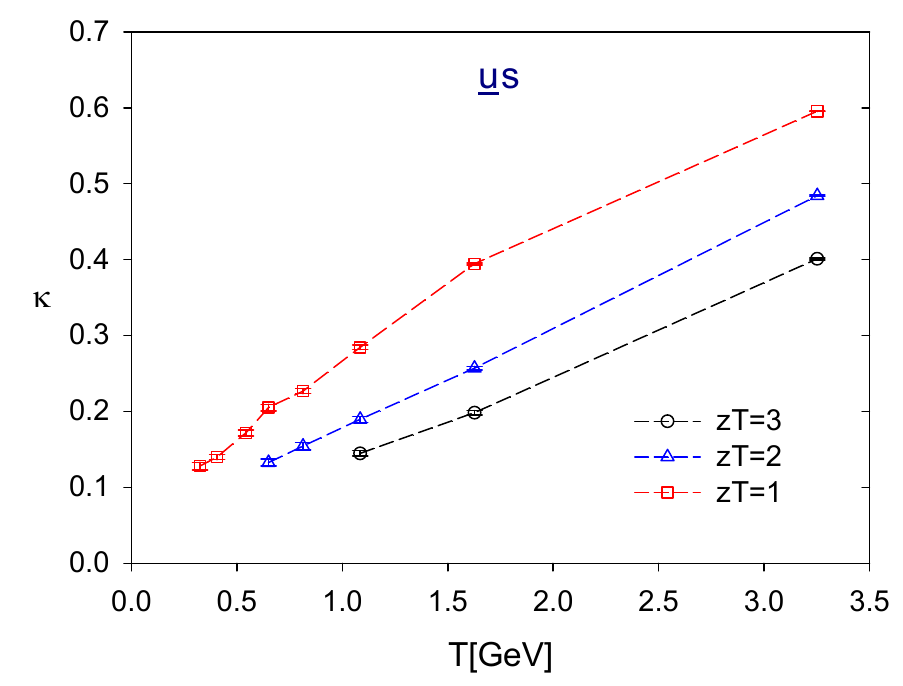}
\vspace{-10pt}
\\
  \includegraphics[width=7.5cm,clip=true]{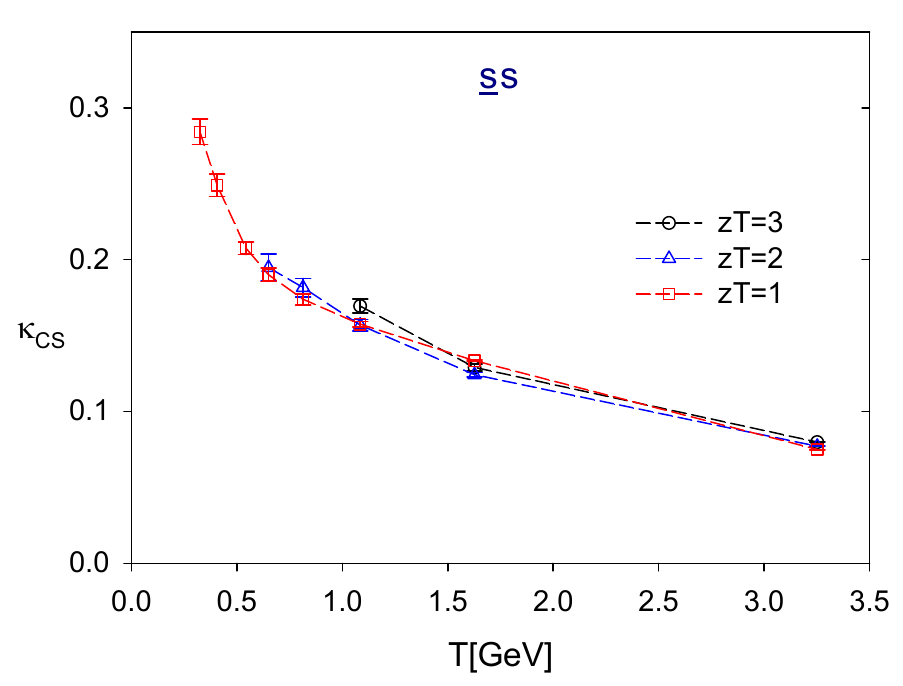}
&
  \includegraphics[width=7.5cm,clip=true]{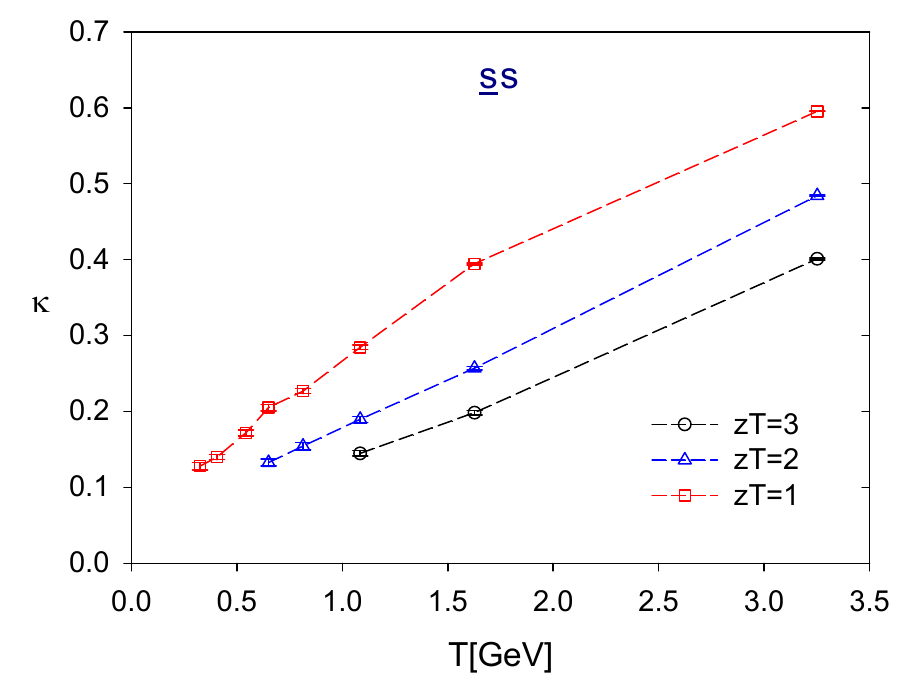}
\vspace{-10pt}
\end{tabular}
  \label{fig:kCS_K_ud_us_ss}
\end{figure}

\begin{figure}[h!]
  \centering
  \caption{The chiral-spin symmetry breaking and fading parameters of the
           $(\bar u c, \bar s c, \bar u b, \bar s b)$ sectors.}
\begin{tabular}{@{}c@{}c@{}}
  \includegraphics[width=7.5cm,clip=true]{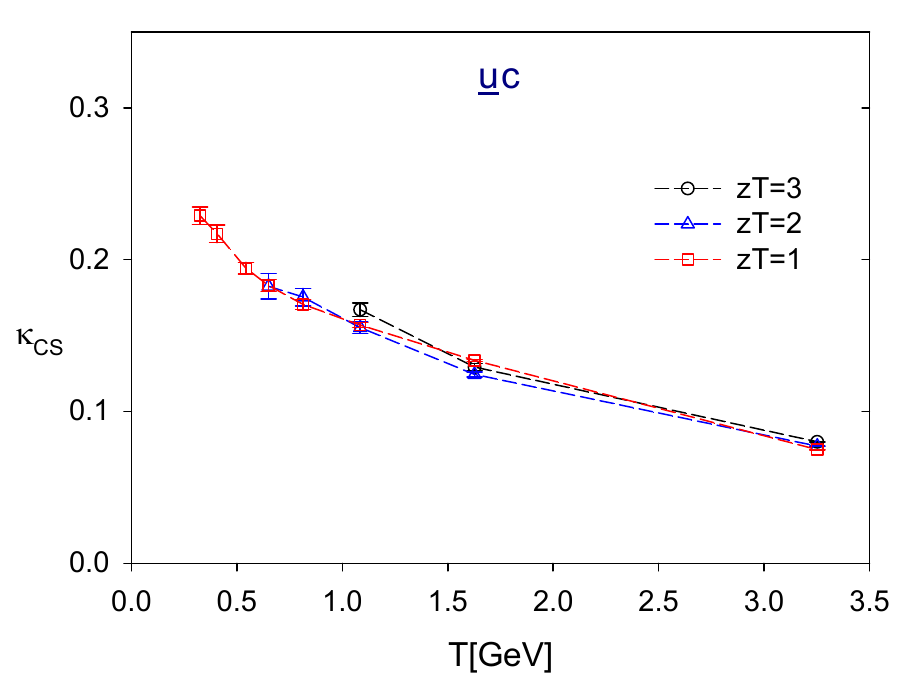}
&
  \includegraphics[width=7.5cm,clip=true]{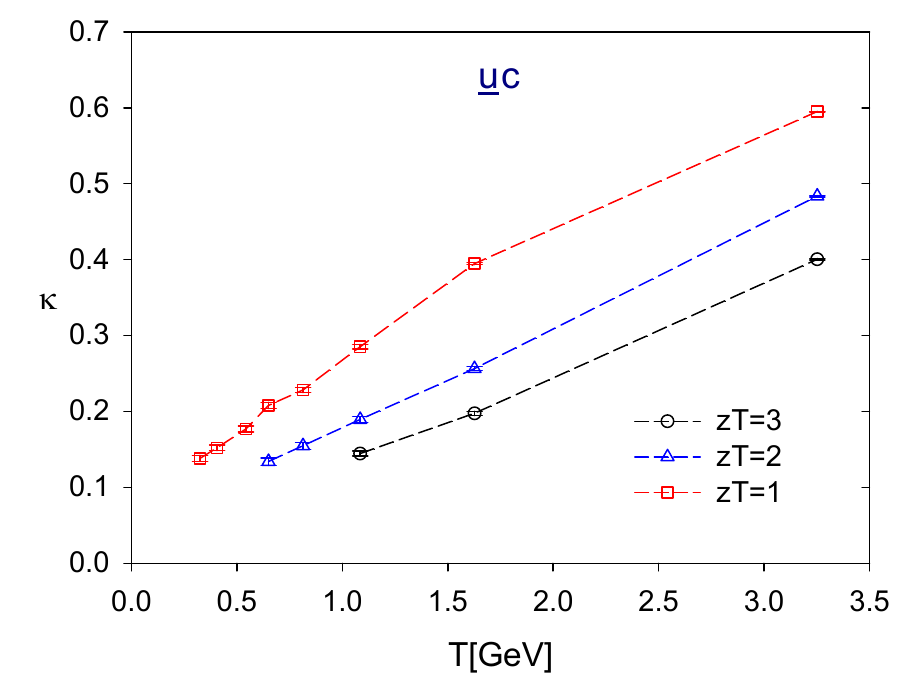} 
\vspace{-10pt} 
\\ 
  \includegraphics[width=7.5cm,clip=true]{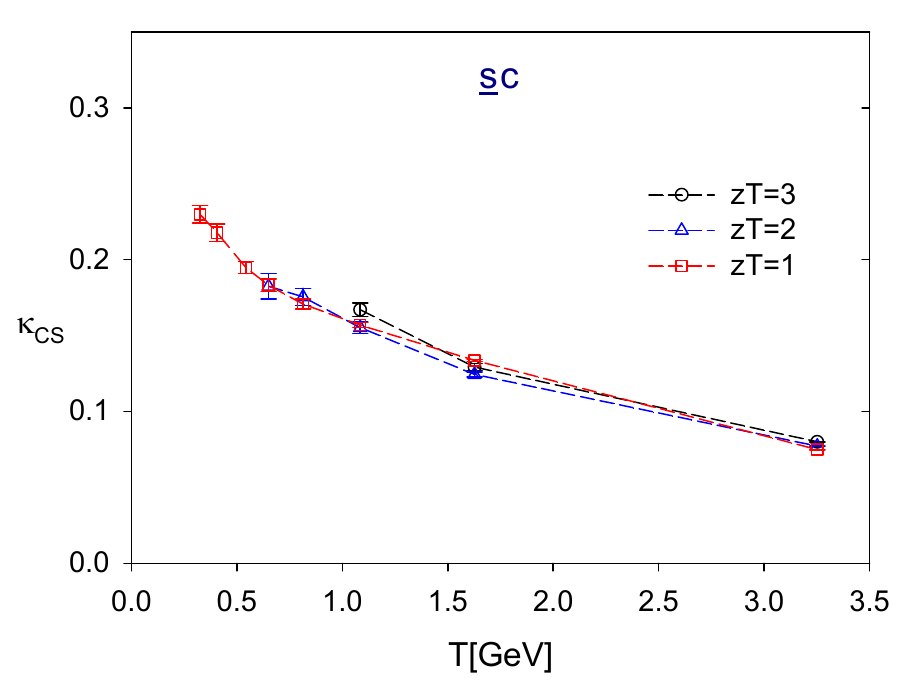}
&
  \includegraphics[width=7.5cm,clip=true]{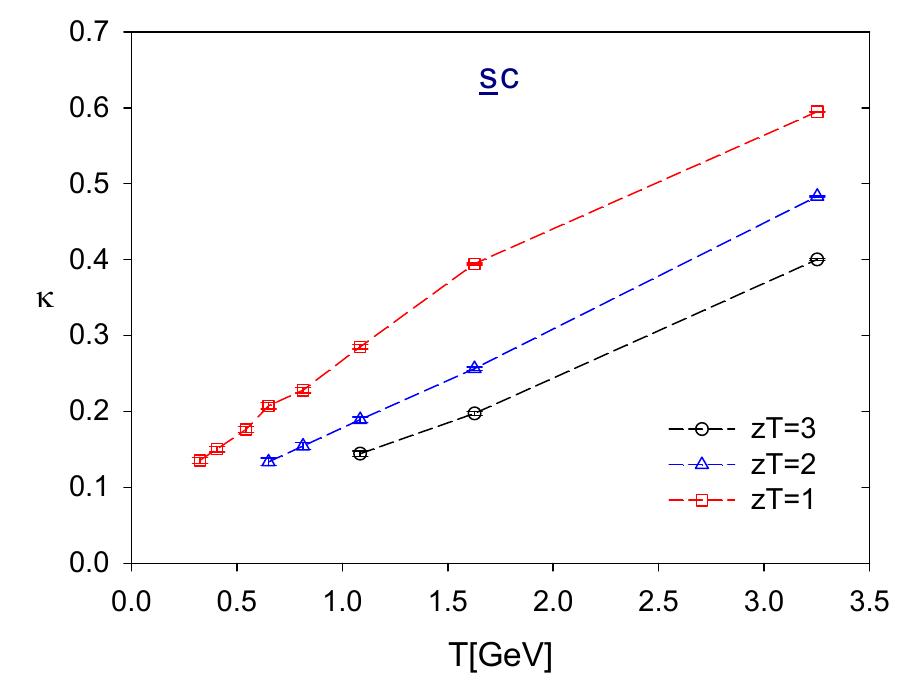} 
\vspace{-10pt} 
\\ 
  \includegraphics[width=7.5cm,clip=true]{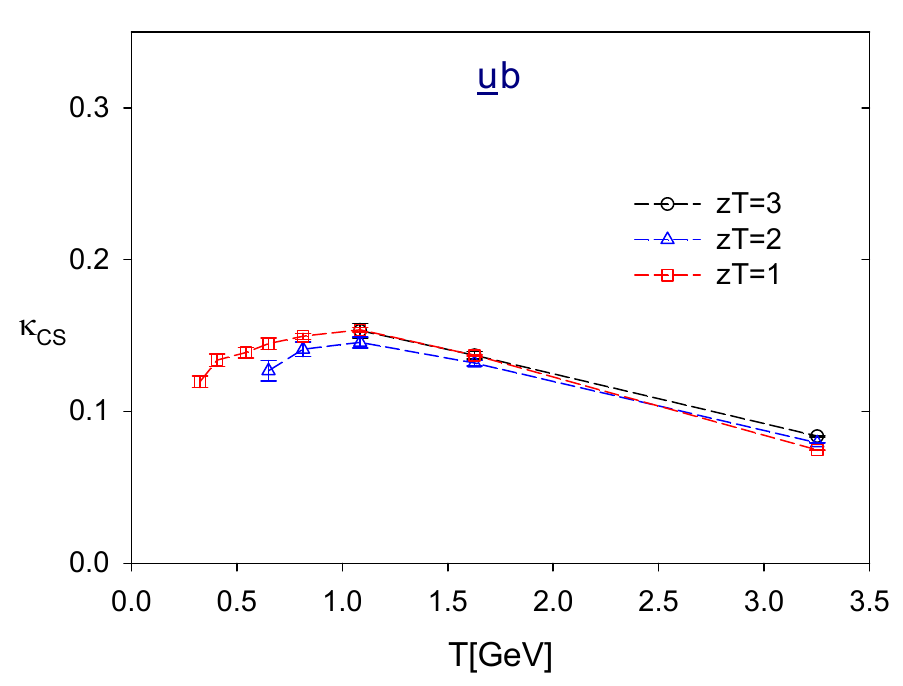}
&
  \includegraphics[width=7.5cm,clip=true]{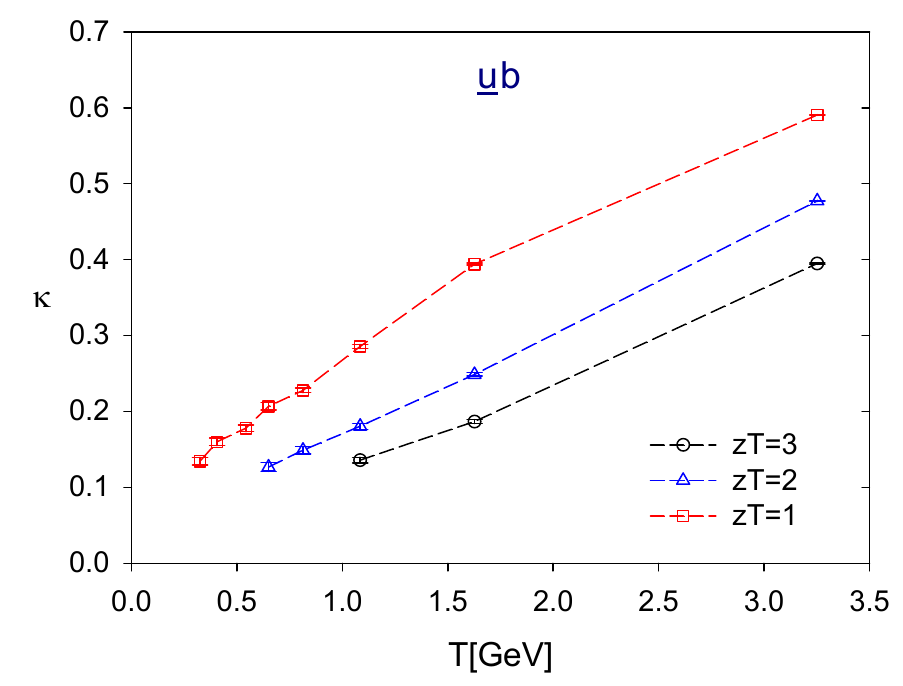} 
\\ 
  \includegraphics[width=7.5cm,clip=true]{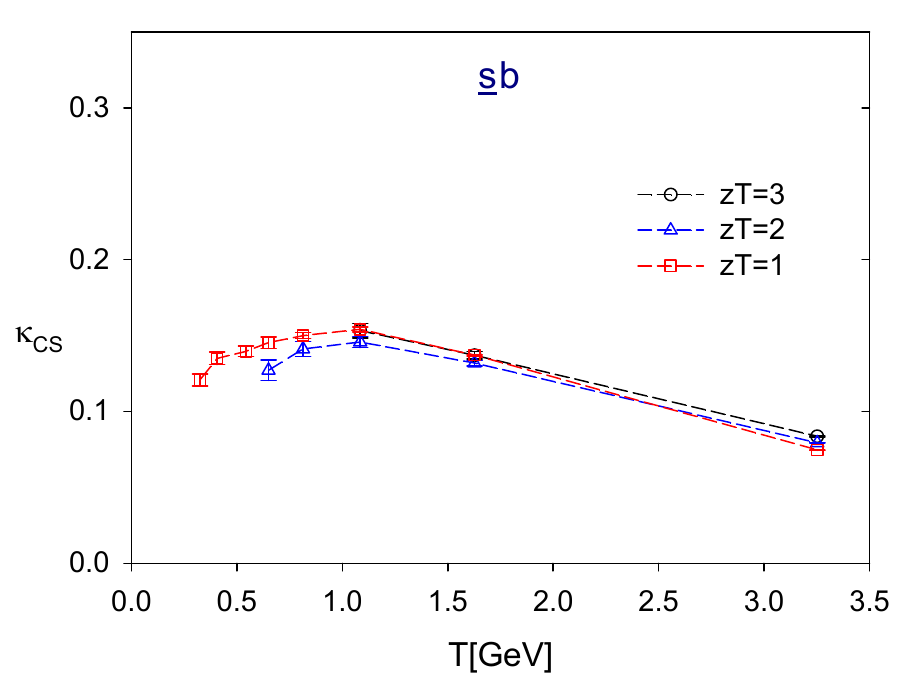}
&
  \includegraphics[width=7.5cm,clip=true]{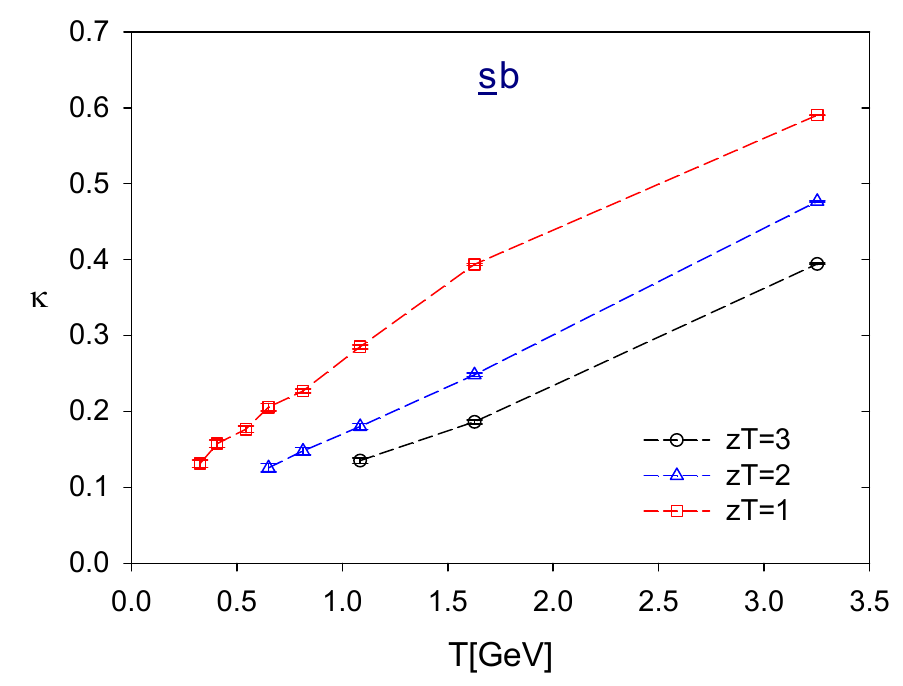} 
\end{tabular}
  \label{fig:kCS_K_uc_sc_ub_sb}
\end{figure}

\begin{figure}[h!]
  \centering
  \caption{The chiral-spin symmetry breaking and fading parameters of the
           $(\bar s c, \bar s b, \bar c b)$ sectors.}
\begin{tabular}{@{}c@{}c@{}}
  \includegraphics[width=7.5cm,clip=true]{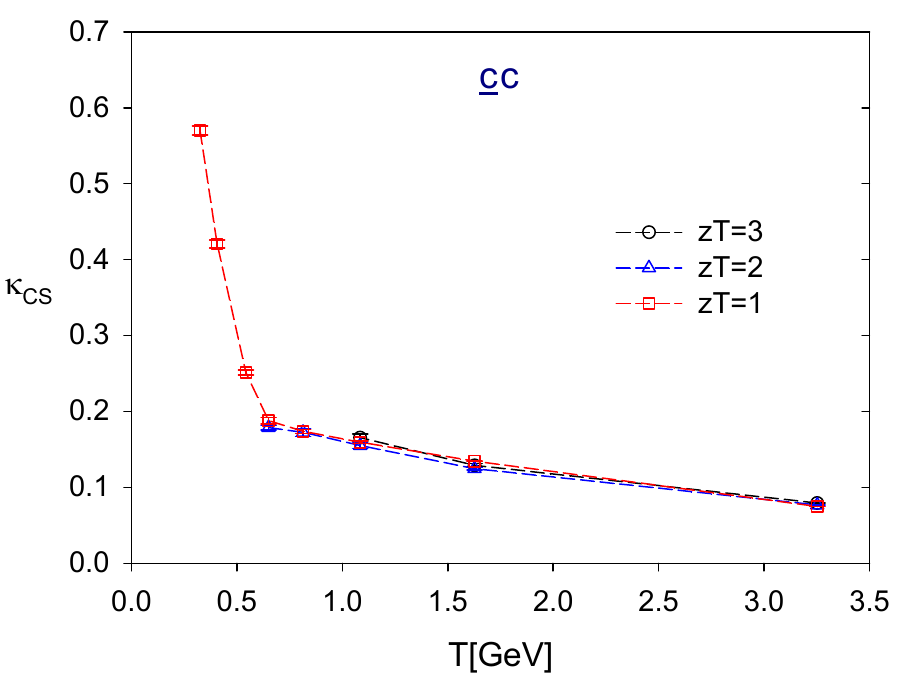}
&
  \includegraphics[width=7.5cm,clip=true]{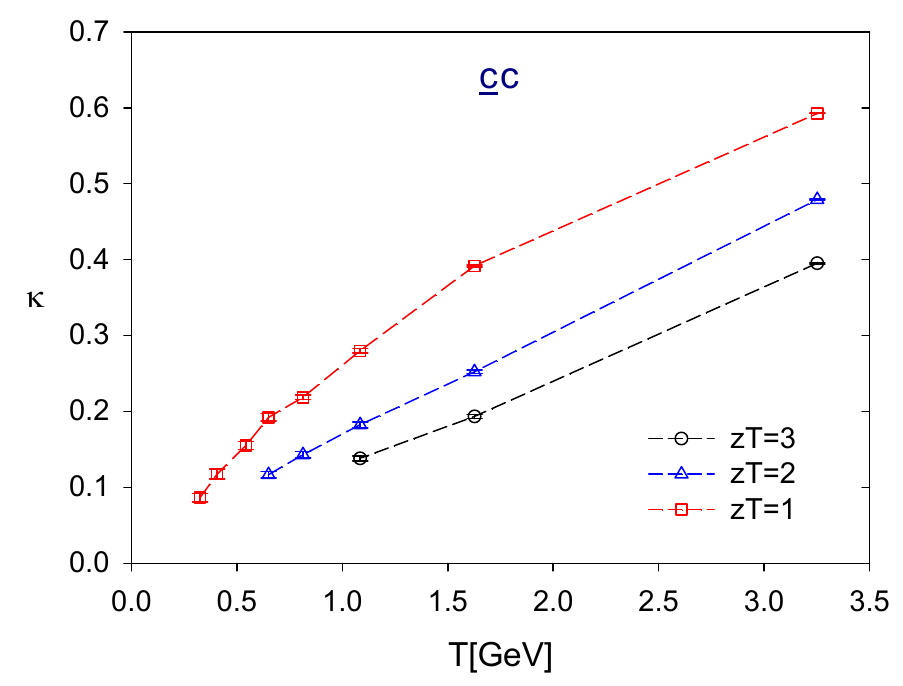} 
\vspace{-10pt} 
\\ 
  \includegraphics[width=7.5cm,clip=true]{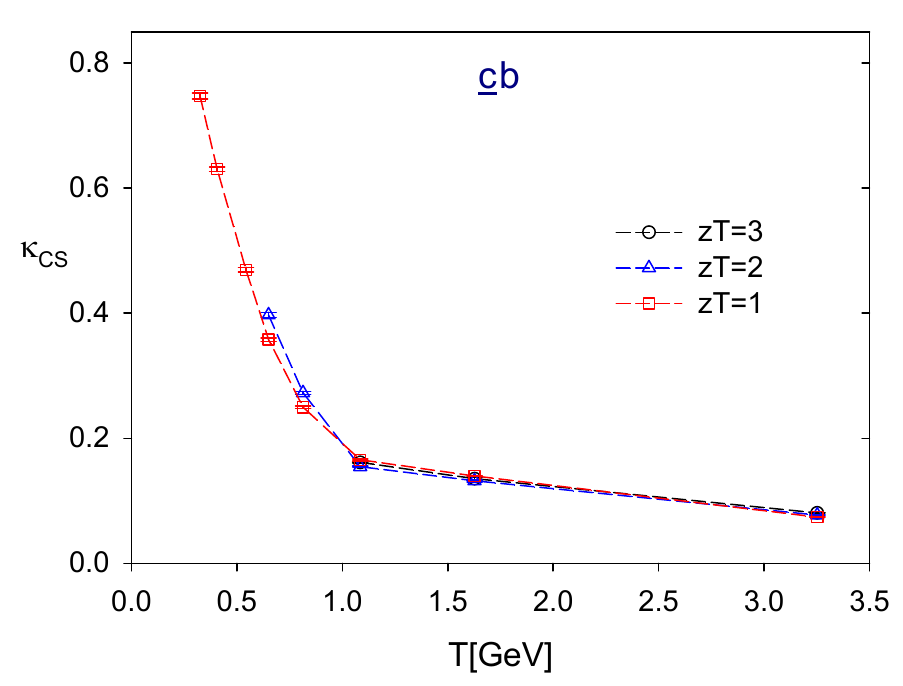}
&
  \includegraphics[width=7.5cm,clip=true]{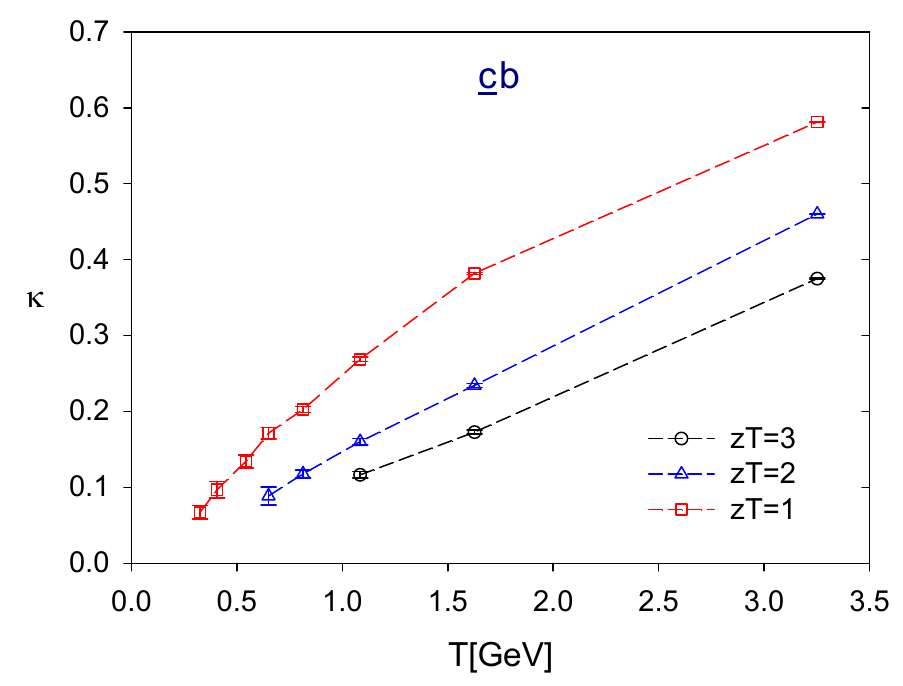} 
\vspace{-10pt} 
\\ 
  \includegraphics[width=7.5cm,clip=true]{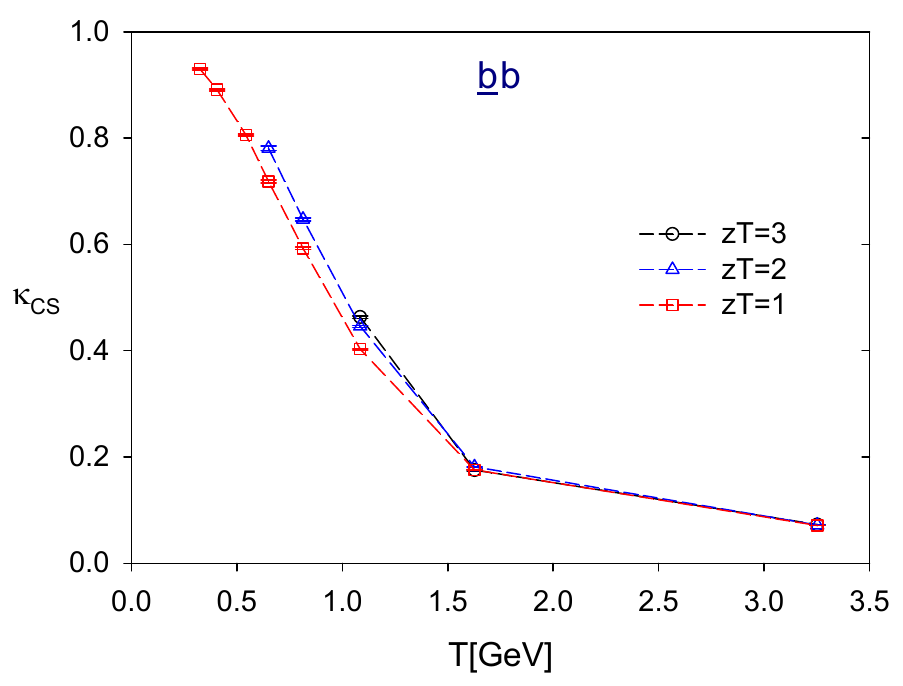}
&
  \includegraphics[width=7.5cm,clip=true]{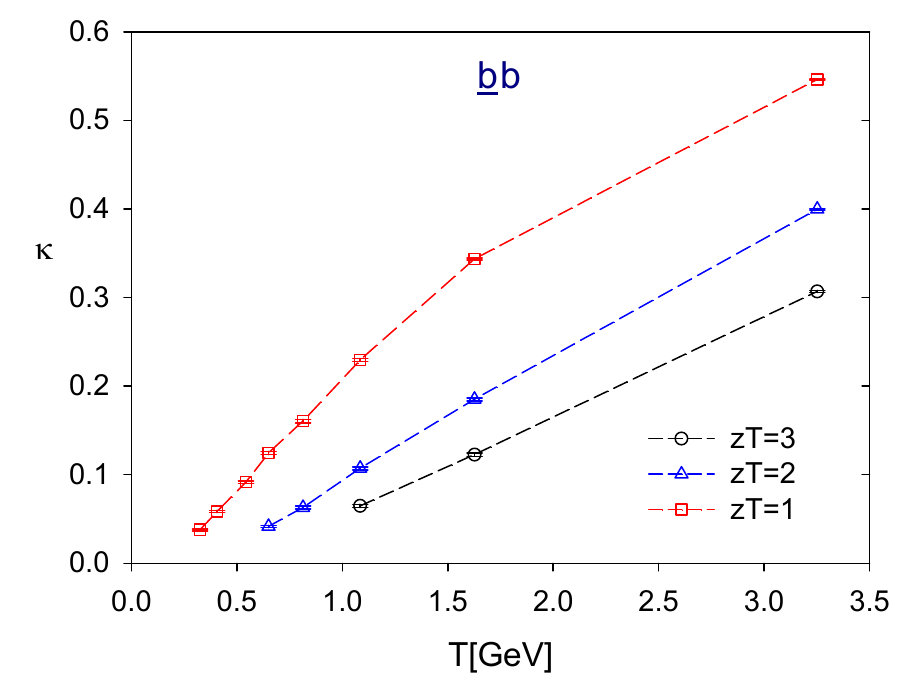} 
\vspace{-10pt} 
\end{tabular}
  \label{fig:kCS_K_cc_cb_bb}
\end{figure}

\subsection{Results of $\kappa_{CS}$ and $\kappa$}

We now proceed to study the $SU(2)_{CS}$ symmetry in $N_f=2+1+1+1$ lattice QCD, 
incorporating physical $s$, $c$, and $b$ quarks 
but unphysically heavy $u/d$ quarks ($M_{\pi} \sim 700$ MeV).  

To this end, we first compute the $z$-correlators following the procedure outlined in Section~\ref{HRCS}. 
We then evaluate the $SU(2)_{CS}$ symmetry-breaking parameter ($\kappa_{CS}$) 
and the symmetry-fading parameter ($\kappa$) and plot them as functions of temperature $T$ 
for $zT=(1,2,3)$, as shown in Figures~\ref{fig:kCS_K_ud_us_ss}--\ref{fig:kCS_K_cc_cb_bb}. 
The numerical values of $\kappa_{CS}$ and $\kappa$ for each flavor sector  
$(\bar u d, \bar u s, \bar s s, \bar u c, \bar s c, \bar u b, \bar s b, \bar c c, \bar c b, \bar b b)$  
are tabulated in Tables~\ref{tab:Kappa_ud}--\ref{tab:Kappa_bb} in \ref{app:C}.  
The statistical error for each $\kappa_{CS}$ or $\kappa$ is estimated using the jackknife method 
with a bin size of $\sim 10$--$15$ configurations, of which the statistical error saturates. 

For the $z$-correlators, the possible values of $zT$ at $T=1/(N_t a)$ are  
\[
\left\{ \frac{n_z}{N_t} \mid n_z = 1, 2, \dots, \frac{N_z}{2} \right\}.
\]  
Thus, for $N_z=40$ and $N_t = (20, 16, 12, 10, 8, 6, 4, 2)$, the number of available temperature points 
is $(8,5,3)$ for $zT=(1,2,3)$, respectively, as illustrated in 
Figures~\ref{fig:kCS_K_ud_us_ss}--\ref{fig:kCS_K_cc_cb_bb} and Tables~\ref{tab:Kappa_ud}--\ref{tab:Kappa_bb}. 

For the $(\bar u d, \bar u s, \bar s s, \bar u c, \bar s c, \bar u b, \bar s b)$ sectors, 
we find that  
\[
\kappa_{AX}(zT) > \kappa_{TX}(zT) \quad \text{for all } zT \text{ at the same } T.
\]  
Thus, (\ref{eq:k_CS_z}) gives $\kappa_{CS} = \kappa_{AX}$. 
However, for the $(\bar c c, \bar c b, \bar b b)$ sectors, we observe that  
\[
\kappa_{AX}(zT) < \kappa_{TX}(zT)
\]  
at low temperatures, while $\kappa_{AX}(zT) > \kappa_{TX}(zT)$ at high temperatures.  
Thus (\ref{eq:k_CS_z}) gives $\kappa_{CS} = \kappa_{TX}$ at low temperatures, but 
$\kappa_{CS} = \kappa_{AX}$ at high temperatures.
This results in an abrupt transition at some intermediate temperature. 
This transition is evident in the left panels of Figure~\ref{fig:kCS_K_cc_cb_bb}.  

In general, for any flavor content and at fixed $zT$,  
$\kappa$ is a monotonically increasing function of $T$, while  
$\kappa_{CS}$ is a monotonically decreasing function of $T$,  
except for the $\bar u b$ and $\bar s b$ sectors, as seen in Figure~\ref{fig:kCS_K_uc_sc_ub_sb}.  
Thus, for any given $\epsilon_{cs}$ and $\epsilon_{fcs}$, we can determine the temperature window 
satisfying the criterion (\ref{eq:SU2_CS_crit_z}) for each flavor content. 
Moreover, as $\epsilon_{cs}$ or $\epsilon_{fcs}$ decreases, the window of $T$ narrows 
and eventually vanishes.  

We estimate the approximate $ T $-window for each flavor sector by solving (\ref{eq:SU2_CS_crit_z}) 
through interpolation or extrapolation of the available data points for $ \kappa_{CS} $ and $ \kappa $, 
as tabulated in Tables~\ref{tab:Kappa_ud}--\ref{tab:Kappa_bb} in \ref{app:C}. 
For $ zT = 1, 2, $ and $ 3 $, we estimate the temperature windows for all ten flavor sectors, 
as presented in Tables~\ref{tab:TCS_all_zT10000}--\ref{tab:TCS_all_zT30000}, 
across all values of $ \epsilon_{cs} $ and $ \epsilon_{fcs} $, sampled from (0.1, 0.15, 0.20).  
Each $ T $-window is expressed in MeV, with uncertainties of approximately 10–20 MeV on both ends, 
combining statistical and interpolation/extrapolation uncertainties in quadrature. 
If the lower bound of a $ T $-window cannot be reliably determined by extrapolation below 325 MeV, 
it is denoted as “$<325$ MeV”. Likewise, temperatures that cannot be reliably extrapolated 
below 650 MeV are represented as “$<650$ MeV”.

%
\begin{table}[htbp]
\caption{The approximate ranges of $T$ satisfying the criterion (\ref{eq:SU2_CS_crit_z}) at $zT=1$ 
         for ten flavor contents. 
         The table lists all nonzero windows of $T$ for all values   
         of $\epsilon_{cs}$ and $\epsilon_{fcs}$ sampling from $(0.1, 0.15, 0.20)$.
         Each $T$ window is in units of MeV, with uncertainties $\sim 10-20$~MeV on both ends of the window.
}
\setlength{\tabcolsep}{4pt}
\vspace{2mm}
\centering
\scalebox{0.76}{  
\begin{tabular}{|cc|cccccccccc|}
\hline
    $\epsilon_{cs}$
  & $\epsilon_{fcs}$
  & $\bar u d$
  & $\bar u s$ 
  & $\bar s s$ 
  & $\bar u c$ 
  & $\bar s c$ 
  & $\bar u b$
  & $\bar s b$ 
  & $\bar c c$ 
  & $\bar c b$ 
  & $\bar b b$ \\
\hline
\hline
0.20&0.20&590-635 &590-635 &590-635 &510-620 &510-625 &\LB-625 &\LB-630 &\BB-700 & NULL   & NULL \\
0.20&0.15& NULL   & NULL   & NULL   & NULL   & NULL   &\LB-375 &\LB-385 & NULL   & NULL   & NULL \\
0.15&0.20& NULL   & NULL   & NULL   & NULL   & NULL   &\LB-625 &\LB-630 & NULL   & NULL   & NULL \\
0.15&0.15& NULL   & NULL   & NULL   & NULL   & NULL   &\LB-375 &\LB-385 & NULL   & NULL   & NULL \\
0.10&0.20& NULL   & NULL   & NULL   & NULL   & NULL   &\LB-625 &\LB-630 & NULL   & NULL   & NULL \\
0.10&0.15& NULL   & NULL   & NULL   & NULL   & NULL   &\LB-375 &\LB-385 & NULL   & NULL   & NULL \\
0.10&0.10& NULL   & NULL   & NULL   & NULL   & NULL   & NULL   & NULL   & NULL   & NULL   & NULL \\
\hline
\end{tabular}
}
\label{tab:TCS_all_zT10000}
\end{table}

%
\begin{table}[htbp]
\caption{Same as Table \ref{tab:TCS_all_zT10000} except for $zT=2$.}
\setlength{\tabcolsep}{4pt}
\vspace{2mm}
\centering
\scalebox{0.76}{  
\begin{tabular}{|cc|cccccccccc|}
\hline
    $\epsilon_{cs}$
  & $\epsilon_{fcs}$
  & $\bar u d$
  & $\bar u s$ 
  & $\bar s s$ 
  & $\bar u c$ 
  & $\bar s c$ 
  & $\bar u b$
  & $\bar s b$ 
  & $\bar c c$ 
  & $\bar c b$ 
  & $\bar b b$ \\
\hline
\hline
%
%
0.20&0.20&\BB-1170 &\BB-1170 &\BB-1170 &\BB-1170 &\BB-1170 &\LB-1240&\LB-1245&\BB-1220 &980-1375&1590-1740\\
0.20&0.15&\BB-780  &\BB-780  &\BB-780  &\BB-780  &\BB-780  &\LB-825 &\LB-830 &\BB-860  &980-1020& NULL \\
0.20&0.10& NULL    & NULL    & NULL    &\BB-385  &\BB-390  &\LB-450 &\LB-460 & NULL    & NULL   & NULL \\
0.15&0.20& NULL    & NULL    & NULL    & NULL    & NULL    &\LB-1240&\LB-1245&1175-1220&1190-1375& NULL \\
0.15&0.15& NULL    & NULL    & NULL    & NULL    & NULL    &\LB-825 &\LB-830 & NULL    & NULL   & NULL \\
0.15&0.10& NULL    & NULL    & NULL    & NULL    & NULL    &\LB-450 &\LB-460 & NULL    & NULL   & NULL \\
0.10&0.20& NULL    & NULL    & NULL    & NULL    & NULL    &\LB-340 &\LB-335 & NULL    & NULL   & NULL \\
0.10&0.15& NULL    & NULL    & NULL    & NULL    & NULL    &\LB-340 &\LB-335 & NULL    & NULL   & NULL \\
0.10&0.10& NULL    & NULL    & NULL    & NULL    & NULL    &\LB-340 &\LB-335 & NULL    & NULL   & NULL \\
\hline
\end{tabular}
}
\label{tab:TCS_all_zT20000}
\end{table}

%
%
\begin{table}[htbp]
\caption{Same as Table \ref{tab:TCS_all_zT10000} except for $zT=3$.}
\setlength{\tabcolsep}{4pt}
\vspace{2mm}
\centering
\scalebox{0.76}{  
\begin{tabular}{|cc|cccccccccc|}
\hline
    $\epsilon_{cs}$
  & $\epsilon_{fcs}$
  & $\bar u d$
  & $\bar u s$ 
  & $\bar s s$ 
  & $\bar u c$ 
  & $\bar s c$ 
  & $\bar u b$
  & $\bar s b$ 
  & $\bar c c$ 
  & $\bar c b$ 
  & $\bar b b$ \\
\hline
\hline
%
%
0.20&0.20&675-1640 &675-1640&675-1640&610-1650 &610-1650&\LB-1730 &\LB-1735 &\BB-1680&980-1850&1580-2310\\
0.20&0.15&675-1140 &675-1140&675-1140&610-1140 &610-1140&\LB-1235 &\LB-1240 &\BB-1200&980-1410&1580-1870\\
0.15&0.20&1345-1640&1345-1640&1345-1640&1330-1650&1330-1650&1190-1730&1190-1735&1315-1680&1315-1850&2030-2310\\
0.20&0.10& NULL   & NULL   & NULL   &610-630 &610-630 &\LB-705  &\LB-710  &\BB-710 & NULL    & NULL \\
0.15&0.15& NULL   & NULL   & NULL   & NULL   & NULL   &1190-1235&1190-1240& NULL   &1315-1410& NULL \\
0.15&0.10& NULL   & NULL   & NULL   & NULL   & NULL   &NULL     &NULL     & NULL   &NULL     & NULL \\
0.10&0.10& NULL   & NULL   & NULL   & NULL   & NULL   &NULL     &NULL     & NULL   &NULL     & NULL \\
\hline
\end{tabular}
}
\label{tab:TCS_all_zT30000}
\end{table}

Tables~\ref{tab:TCS_all_zT10000}--\ref{tab:TCS_all_zT30000} show that as $ (\epsilon_{cs}, \epsilon_{fcs}) $ 
decrease from (0.20, 0.20) to (0.15, 0.15), and further to (0.10, 0.10), the $ T $-windows 
for all flavor sectors progressively shrink and eventually vanish, 
except for the $ \bar{u}b $ and $ \bar{s}b $ sectors, which retain nonzero $ T $-windows. 
This indicates that the $ T $-windows of the emergent $ SU(2)_{CS} $ symmetry are primarily dominated 
by the $ \bar{u}b $ and $ \bar{s}b $ sectors, 
{\it composed of the heaviest $ b $ quark and the light quarks of the system}.  

Notably, in lattice QCD with $ (u,d,s,c) $ quarks, the $ T $-windows of the 
emergent $ SU(2)_{CS} $ symmetry are predominantly governed by the  
$ \bar{u}c $ and $ \bar{s}c $ sectors, 
{\it composed of the heaviest $ c $ quark and the light quarks of the system}, 
as reported in ref.~\cite{Chiu:2024jyz}. Comparing these two lattice setups suggests 
an important universal feature of any QCD system: {\it the $ T $-windows of the emergent 
chiral-spin symmetry are primarily dominated by the sectors involving the heaviest quark 
and the light quarks of the system}.  

The results in Tables~\ref{tab:TCS_all_zT10000}--\ref{tab:TCS_all_zT30000} also indicate that 
the most favorable channels for detecting the emergence of $ SU(2)_{CS} $ symmetry 
in QCD with $ (u,d,s,c,b) $ quarks are in vector mesons with flavor contents    
$ \bar{u}b $ ($\bar{d}b$) and $ \bar{s}b $. 
This finding may have {\it phenomenological implications} for observing $ SU(2)_{CS} $ symmetry 
in relativistic heavy-ion collisions at experiments such as the LHC and RHIC.  

Moreover, this suggests that hadron-like objects, particularly vector mesons with flavor contents 
$ \bar{s}b $ and $ \bar{u}b $ ($ \bar{d}b $), are more likely to be 
{\it predominantly bound by chromoelectric interactions into color singlets 
at temperatures within their respective $ T $-windows of the emergent $ SU(2)_{CS} $ symmetry}. 
This is notable because neither the chromomagnetic part of the quark-gluon interaction 
nor the noninteracting theory with free quarks possesses any $ SU(2)_{CS} $ symmetry.

%

It is important to note that since the $u/d$ quarks are unphysical 
and the gauge ensembles are limited to a single lattice spacing and spatial volume, 
we cannot determine the $T$ windows of any flavor sector in the physical limit. 
However, we expect that, in the physical limit, the vector mesons in the 
$\bar{s}b$ sector will remain one of the most favorable channels for 
detecting the emergent $SU(2)_{CS}$ chiral-spin symmetry, 
and these hadron-like objects will predominantly be bound by chromoelectric interactions 
into color singlets. 


\begin{figure}[!h]
  \centering
  \caption{Comparison of the ${SU(2)}_{CS}$ chiral-spin symmetry breaking and fading parameters
           $(\kappa_{CS}, \kappa)$ at $zT=1$, for the $(\bar u d, \bar u s, \bar u c)$ sectors
           of lattice QCD with $N_f=2+1+1+1$ (this work)
           and $N_f=2+1+1$ at the physical point \cite{Chiu:2024jyz}.}
\begin{tabular}{@{}c@{}c@{}}
  \includegraphics[width=7.5cm,clip=true]{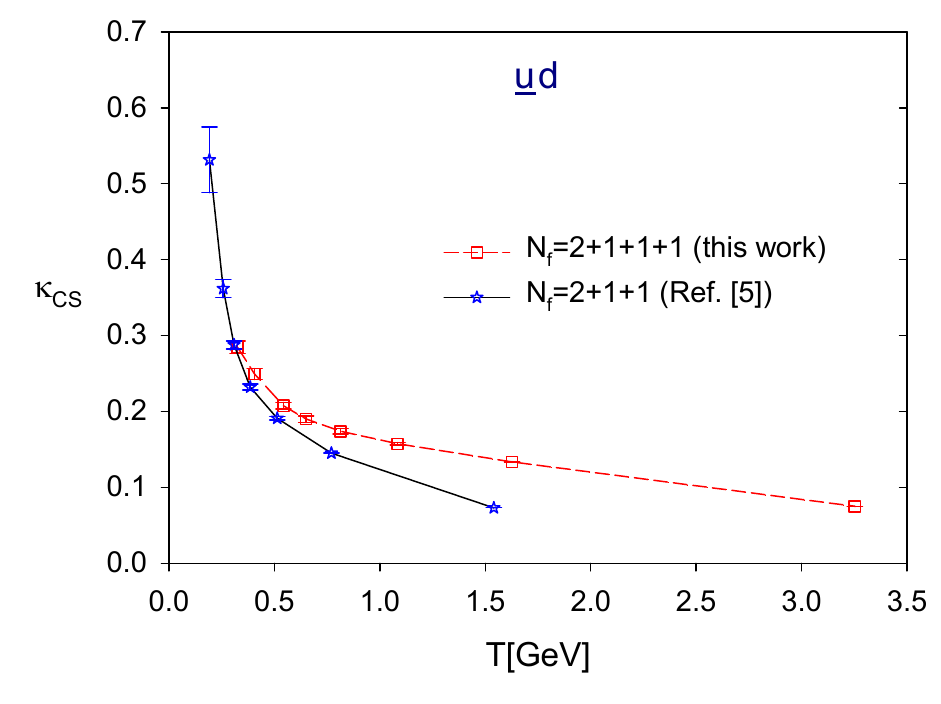}
&
  \includegraphics[width=7.5cm,clip=true]{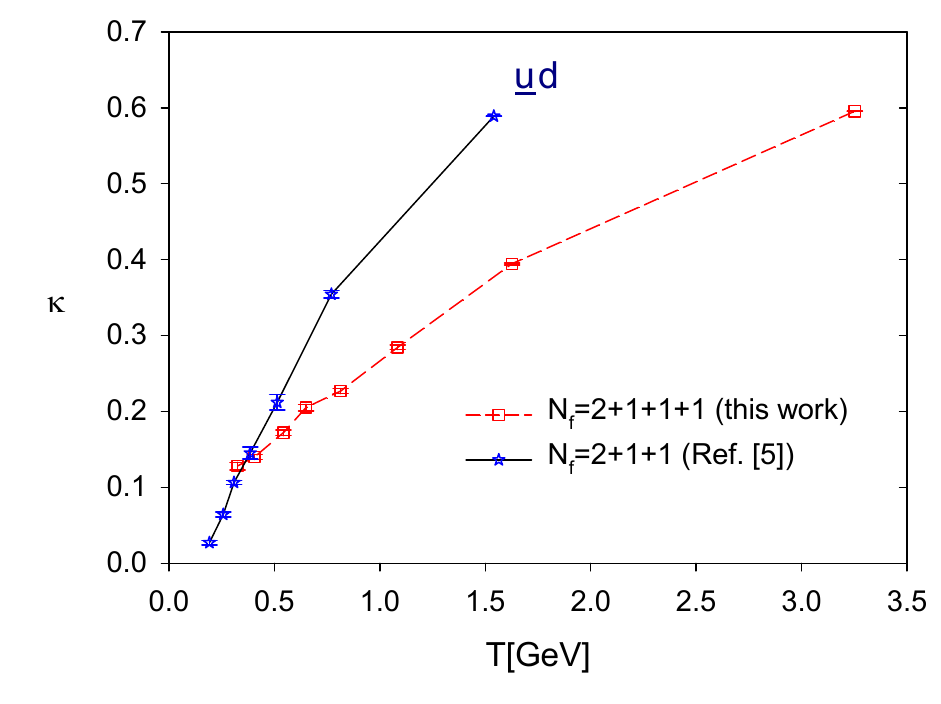}
\vspace{-10pt}
\\
  \includegraphics[width=7.5cm,clip=true]{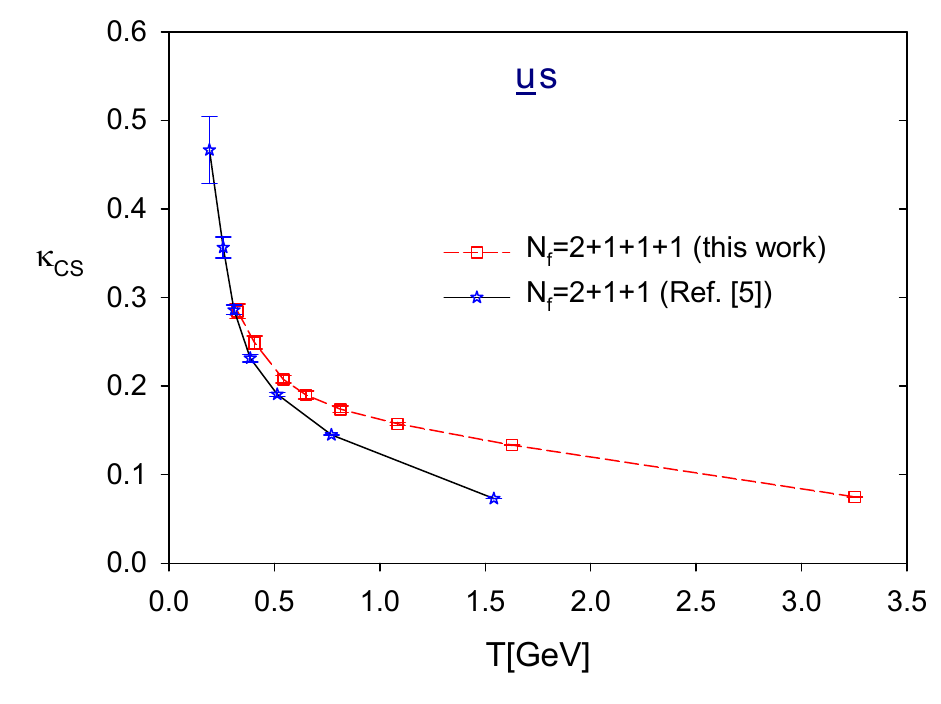}
&
  \includegraphics[width=7.5cm,clip=true]{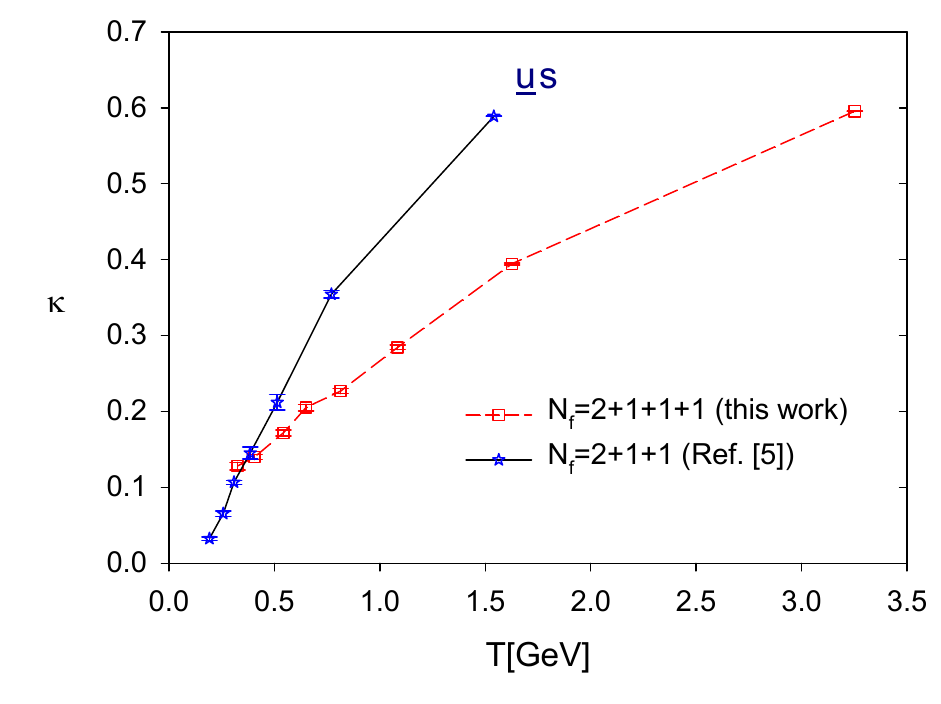}
\vspace{-10pt}
\\
  \includegraphics[width=7.5cm,clip=true]{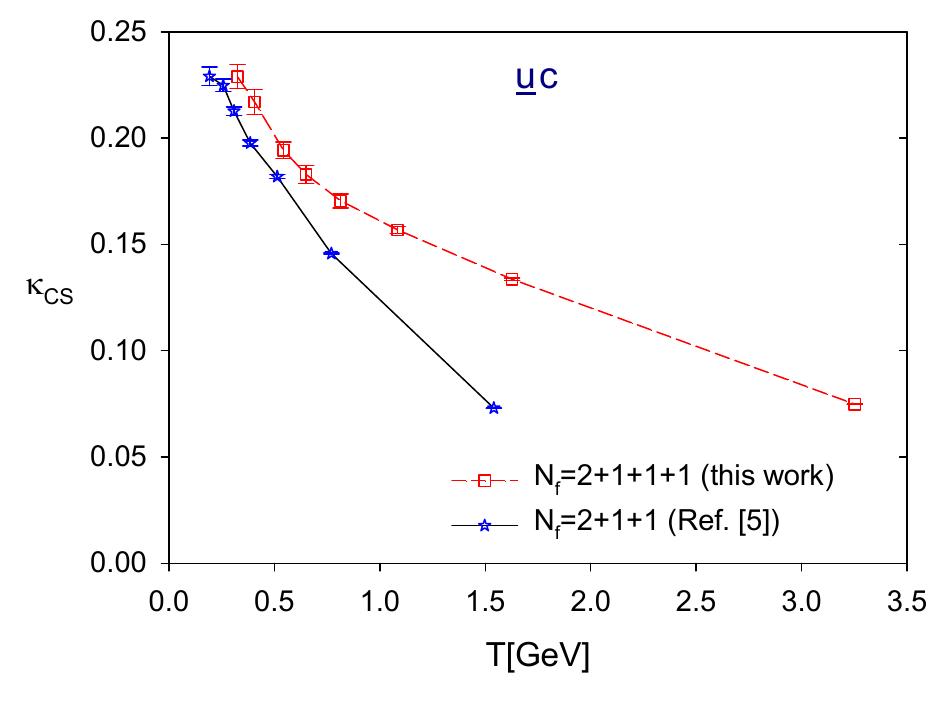}
&
  \includegraphics[width=7.5cm,clip=true]{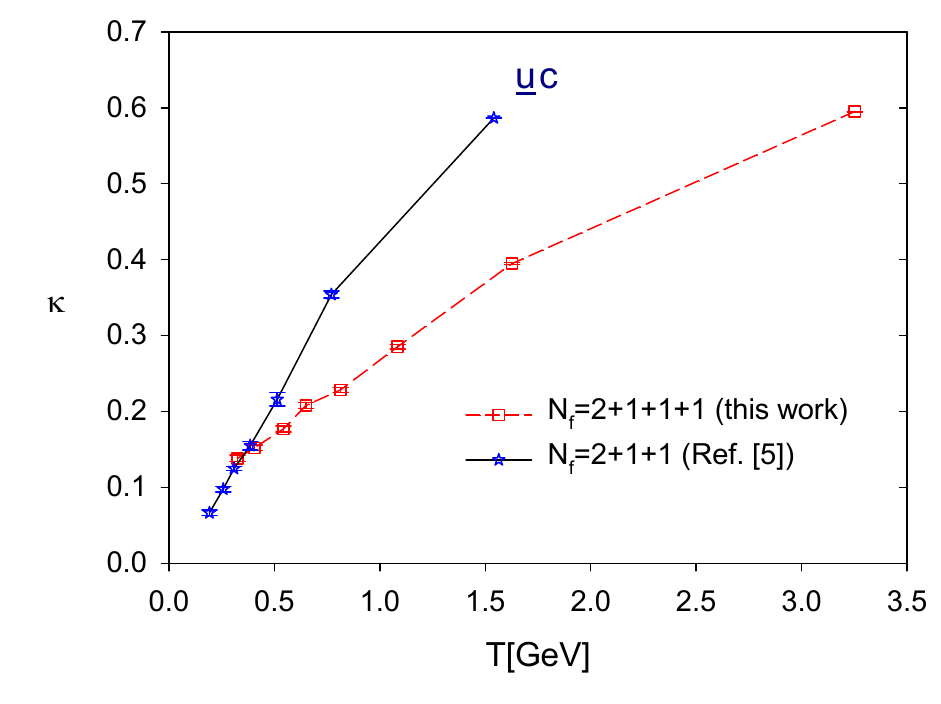}
\end{tabular}
\label{fig:kCS_k_ud_us_uc_compare_zT1}
\end{figure}

\begin{figure}[!h]
  \centering
  \caption{Comparison of the ${SU(2)}_{CS}$ chiral-spin symmetry breaking and fading parameters
           $(\kappa_{CS}, \kappa)$ at $zT=1$, for the $(\bar s s, \bar s c, \bar c c)$ sectors
           of lattice QCD with $N_f=2+1+1+1$ (this work)
           and $N_f=2+1+1$ at the physical point \cite{Chiu:2024jyz}.}
\begin{tabular}{@{}c@{}c@{}}
  \includegraphics[width=7.5cm,clip=true]{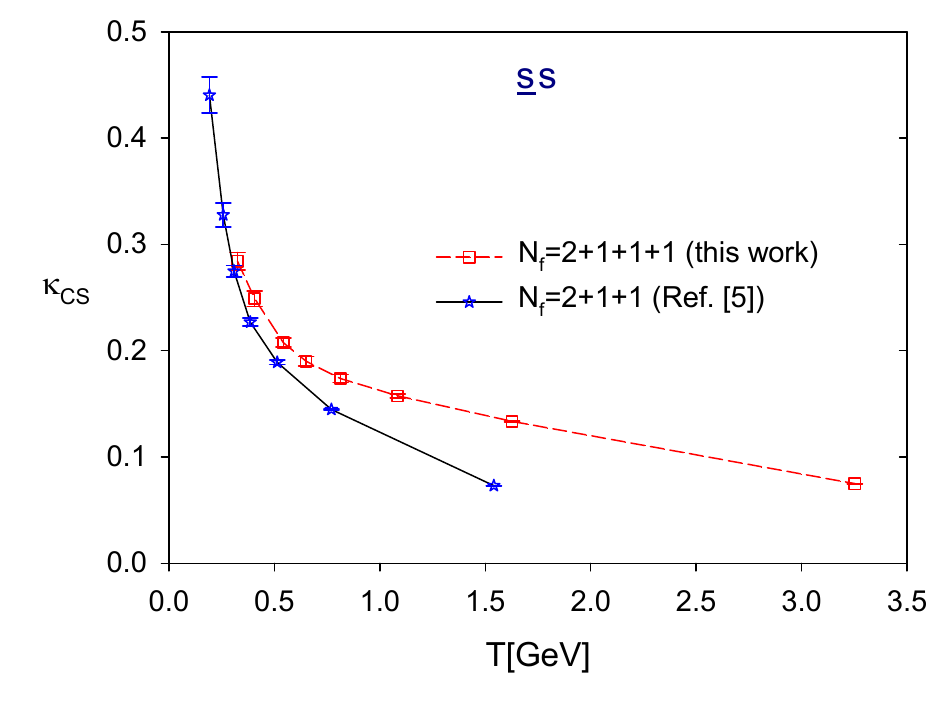}
&
  \includegraphics[width=7.5cm,clip=true]{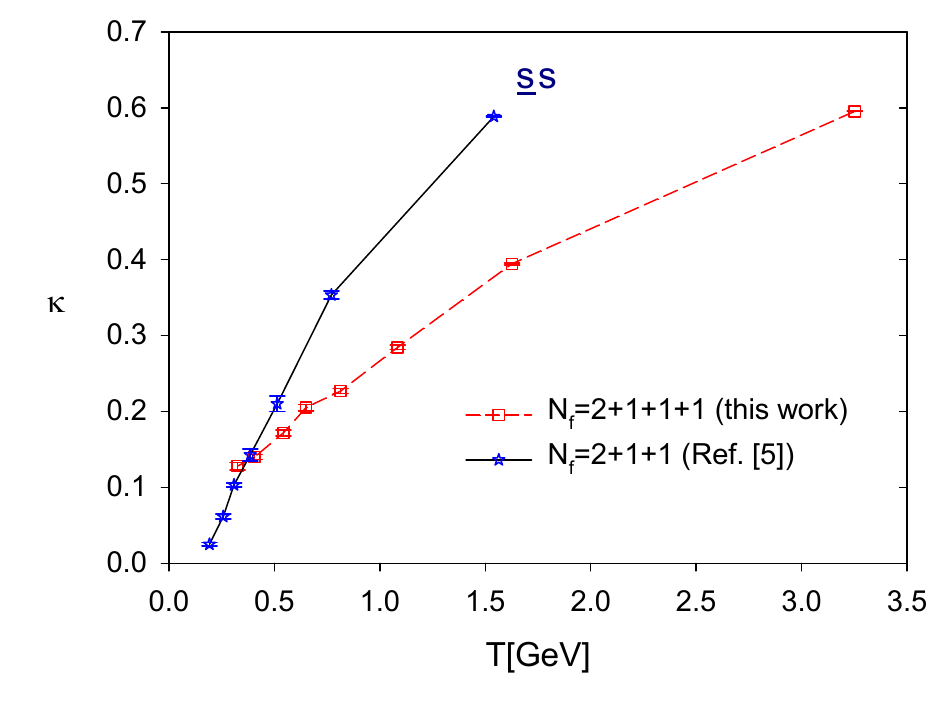}
\vspace{-10pt}
\\
  \includegraphics[width=7.5cm,clip=true]{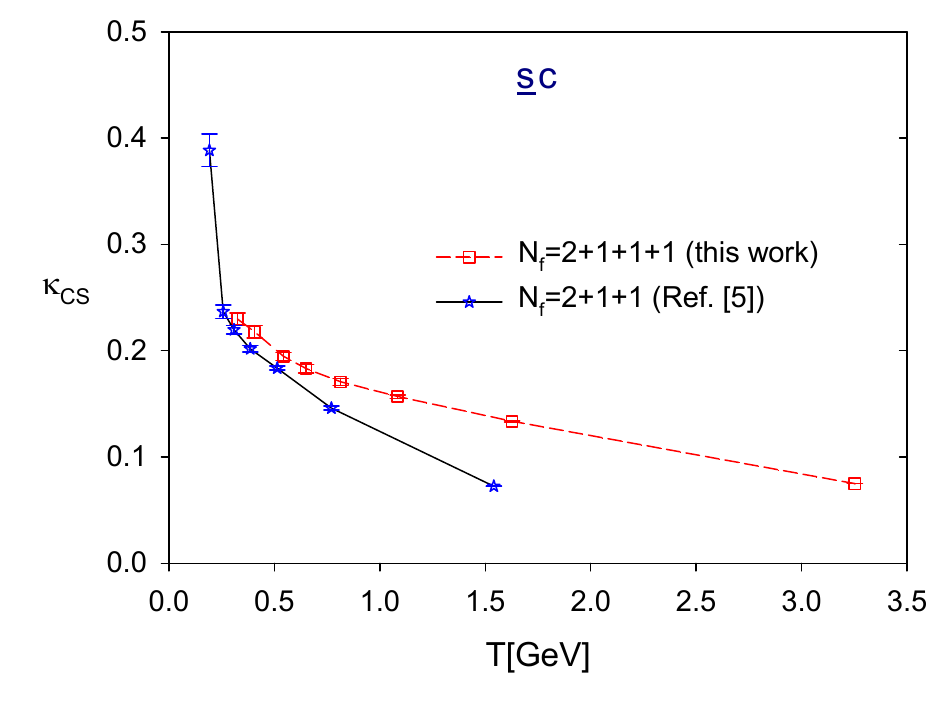}
&
  \includegraphics[width=7.5cm,clip=true]{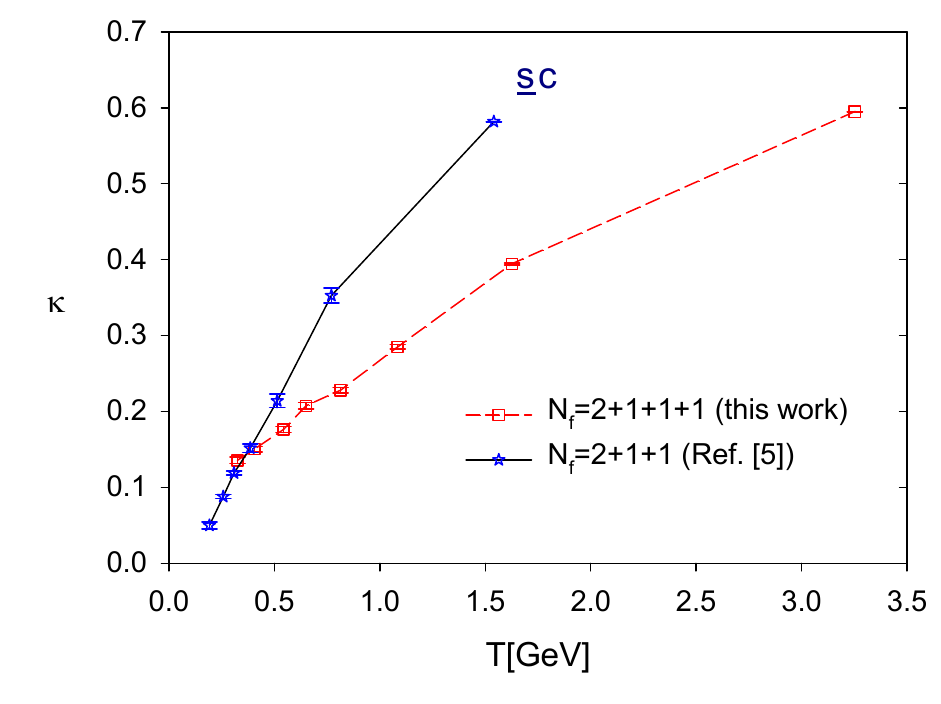}
\vspace{-10pt}
\\
  \includegraphics[width=7.5cm,clip=true]{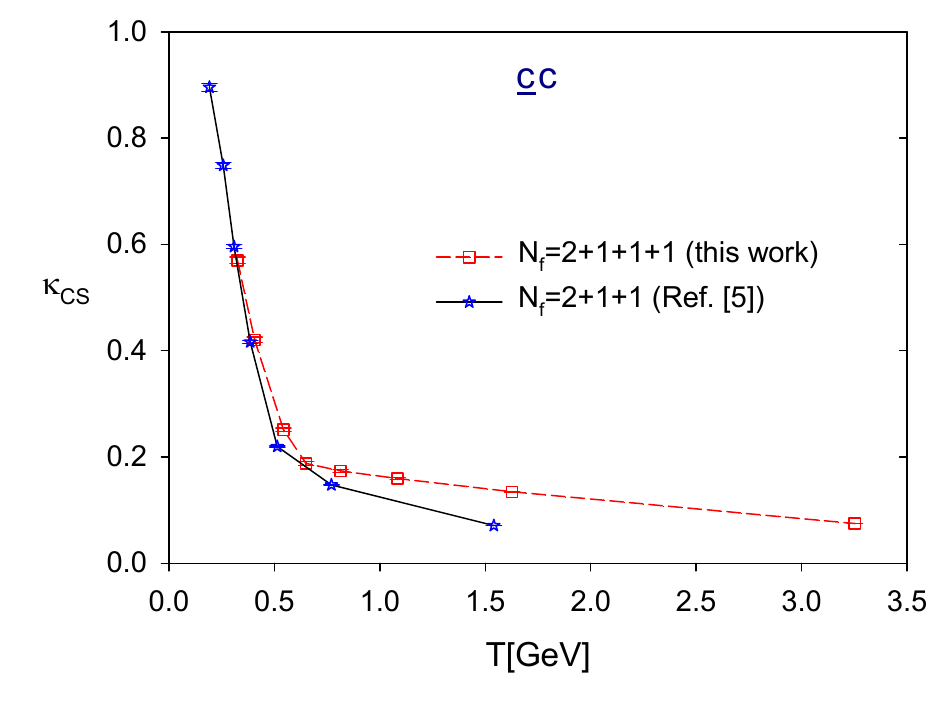}
&
  \includegraphics[width=7.5cm,clip=true]{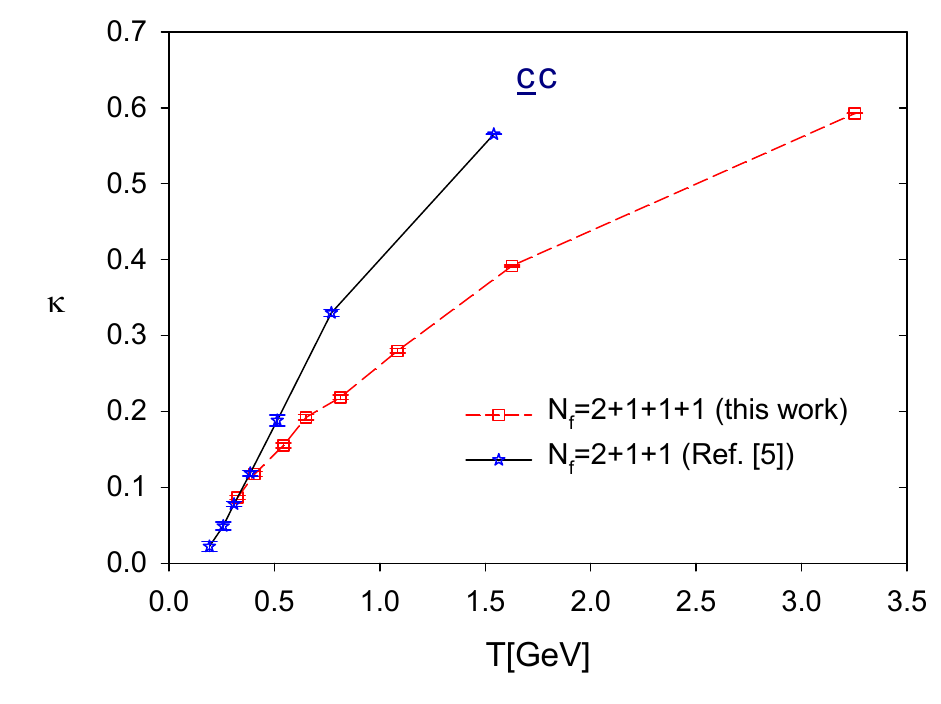}
\end{tabular}
\label{fig:kCS_k_ss_sc_cc_compare_zT1}
\end{figure}

\subsection{Comparison with $N_f=2+1+1$ lattice QCD at the physical point}

In the following, 
we compare the $SU(2)_{CS}$ symmetry breaking and fading parameters ($\kappa_{CS}$ and $\kappa$), 
as well as the temperature windows for the emergent $SU(2)_{CS}$ chiral-spin symmetry, 
between $N_f=2+1+1+1$ QCD (this work) and $N_f=2+1+1$ QCD at the physical point \cite{Chiu:2024jyz}.

The numerical values of $\kappa_{CS}$ and $\kappa$ for $N_f=2+1+1+1$ QCD are provided 
in Tables~\ref{tab:Kappa_ud}--\ref{tab:Kappa_bb} of \ref{app:C}, while those for $N_f=2+1+1$ QCD can be found 
in Tables~\ref{tab:Kappa_ud_Nf2p1p1}--\ref{tab:Kappa_cc_Nf2p1p1} of \ref{app:D}.

As an example, we compare the $SU(2)_{CS}$ symmetry breaking and fading parameters 
($\kappa_{CS}$ and $\kappa$) at $zT=1$ for both lattice setups, 
as shown in Figures~\ref{fig:kCS_k_ud_us_uc_compare_zT1} and \ref{fig:kCS_k_ss_sc_cc_compare_zT1}. 

First, we observe that for any flavor sector with $T > 325$ MeV, $\kappa_{CS}$ in $N_f=2+1+1+1$ QCD 
is larger than in $N_f=2+1+1$ QCD, while $\kappa$ in $N_f=2+1+1+1$ QCD is smaller than in $N_f=2+1+1$ QCD.
Since $\kappa_{CS}$ is a monotonically decreasing function of $T$, 
while $\kappa$ is a monotonically increasing function of $T$, 
it follows that for any given $\epsilon_{cs}$ and $\epsilon_{fcs}$ in (\ref{eq:SU2_CS_crit_z}), 
both the lower and upper bounds of each $T$-window for the emergent $SU(2)_{CS}$ symmetry 
in $N_f=2+1+1+1$ QCD occur at higher temperatures than those in $N_f=2+1+1$ QCD.

\begin{table}[h!]
\caption{Comparison of the approximate temperature windows for the $SU(2)_{CS}$ emergent symmetry
         between $N_f=2+1+1$ and $N_f=2+1+1+1$ lattice QCD,
         for $zT=1$ and $\epsilon_{cs}=\epsilon_{fcs} = 0.20$.
         All temperatures are given in MeV.}
\vspace{2mm}
\centering
\begin{tabular}{|c|c|c|}
\hline
Flavor Content      & $N_f=2+1+1$ \cite{Chiu:2024jyz} & $N_f=2+1+1+1$ (This Work) \\
\hline
$\bar u d$          &  485(10)-490(20) &  590(20)-635(15) \\
$\bar u s$          &  485(10)-490(20) &  590(20)-635(15) \\
$\bar u c$          &  370(10)-480(20) &  510(20)-620(15)  \\
$\bar s s$          &  475(10)-495(20) &  590(20)-635(15)  \\
$\bar s c$          &  400(20)-485(15) &  510(20)-625(15)  \\
$\bar c c$          &  NULL            &  630(10)-700(20)  \\
\hline
\end{tabular}
\label{tab:compare_TCS_zT1}
\end{table}

For instance, for $\epsilon_{CS} = \epsilon_{FCS} = 0.20$, the corresponding results 
are summarized in Table~\ref{tab:compare_TCS_zT1}. The lower and upper bounds of each $T$-window, 
along with their uncertainties, are estimated through piecewise linear interpolation 
of $\kappa_{CS}$ and $\kappa$. Clearly, for any flavor sector, the $T$-window in $N_f=2+1+1+1$ QCD 
shifts to a higher temperature range compared to that in $N_f=2+1+1$ QCD, while also expanding in size. 

To better understand this behavior, we compare the $T$-window for the emergent $SU(2)_{CS}$ symmetry 
in the $\bar{u}d$ sector between $N_f=2+1+1$ lattice QCD at the physical point \cite{Chiu:2024jyz} 
and $N_f=2$ lattice QCD near the physical point \cite{Rohrhofer:2019qwq}. 

Specifically, for $\epsilon_{cs} = \epsilon_{fcs} = 0.2$ at $zT=2$, the $T$-window in $N_f=2$ lattice QCD 
spans approximately $320$–$500$ MeV, whereas in $N_f=2+1+1$ lattice QCD, 
it shifts to $610(15)$–$730(15)$ MeV. This indicates that the presence of dynamical $s$ and $c$ quarks, 
which are significantly heavier than the light $u$ and $d$ quarks, raises both the lower and upper bounds 
of the $T$-window in the $\bar{u}d$ sector while also reducing its size.

Synthesizing these findings with our earlier discussions, we obtain a universal feature 
of the emergent chiral-spin symmetry in any QCD system: \\
\noindent {\it Increasing the number of dynamical heavy quarks shifts the $T$-windows 
for $SU(2)_{CS}$ symmetry to higher temperature ranges, and these windows are primarily dominated by 
the sectors involving the heaviest quark and the light quarks of the system.} \\
\noindent This constitutes one of the interesting findings of our study.
However, given the different lattice spacings in the $N_f=2/2+1+1/2+1+1+1$ ensembles
and the associated discretization errors---as well as the unphysical $u/d$ quark masses
in the $N_f=2+1+1+1$ lattices---it remains an open question whether this feature persists
in the physical limit; that is, with $(u,d,s,c,b)$ quarks at the physical point across all
lattice setups, in the continuum and infinite volume limits. If confirmed, this would reflect
a nontrivial realization of nonperturbative QCD dynamics
of increasing the number of heavy dynamical quarks at high temperatures.
This effect is revealed through changes of the splitting in the $SU(2)_{CS}$ multiplet $(A_1, T_4, X_4)$   
as measured by the $SU(2)_{CS}$ symmetry breaking parameter $\kappa_{AX}$ (\ref{eq:k_AX_z}),  
and the ratio of the splitting to the distance between   
the $U(1)_A$ multiplet $M_0=(P,S)$ and the $SU(2)_{CS} \times SU(2)_L \times SU(2)_R $ multiplet 
$M_2 = (V_1, A_1, T_4, X_4)$ as measured by the $SU(2)_{CS}$ symmetry fading parameter 
$\kappa$ (\ref{eq:kappa_z}).

\subsection{Comparison between $\kappa_{VA}$ and $\kappa_{CS}$}

\begin{figure}[h!]
    \centering
    \caption{
      The ratio of the chiral symmetry breaking parameter $\kappa_{VA}$ and the chiral-spin symmetry 
      breaking parameter $\kappa_{CS}$ as a function of $T$, for all flavor combinations and 
      $zT=1$,2, and 3 respectively.  
    }
    \hspace{-0.05\textwidth} 
    \begin{subfigure}[b]{0.3\textwidth}
           \includegraphics[height=0.2\textheight]{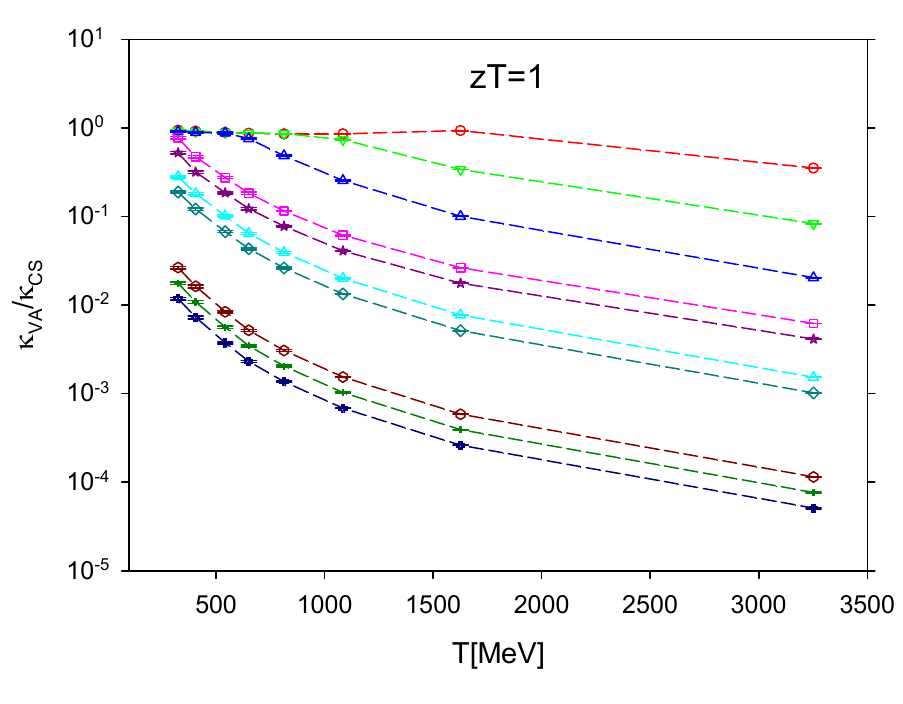}
    \end{subfigure}
    \hspace{0.05\textwidth} 
    \begin{subfigure}[b]{0.3\textwidth}
           \includegraphics[height=0.2\textheight]{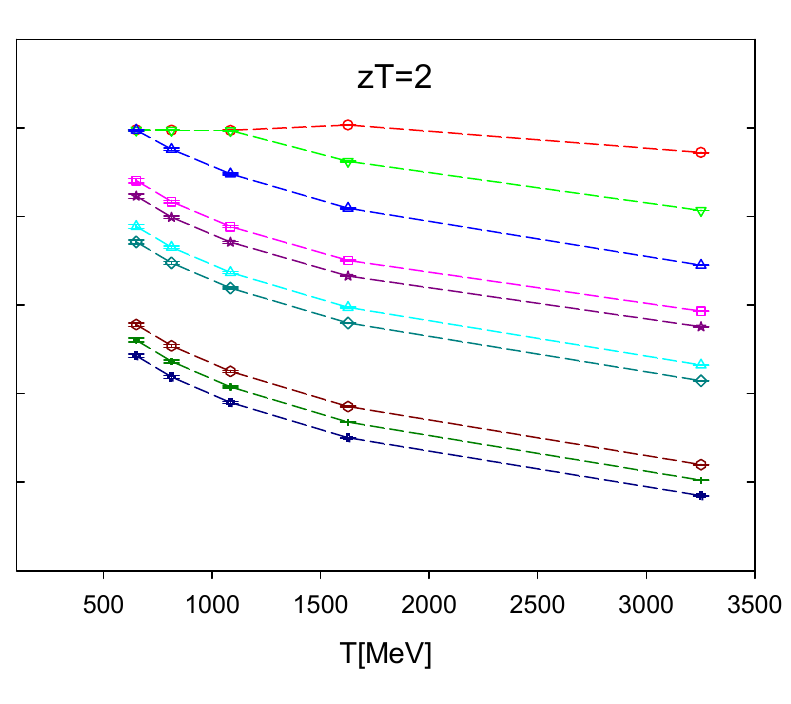}
    \end{subfigure}
    \hspace{0.00\textwidth} 
    \begin{subfigure}[b]{0.3\textwidth}
           \includegraphics[height=0.2\textheight]{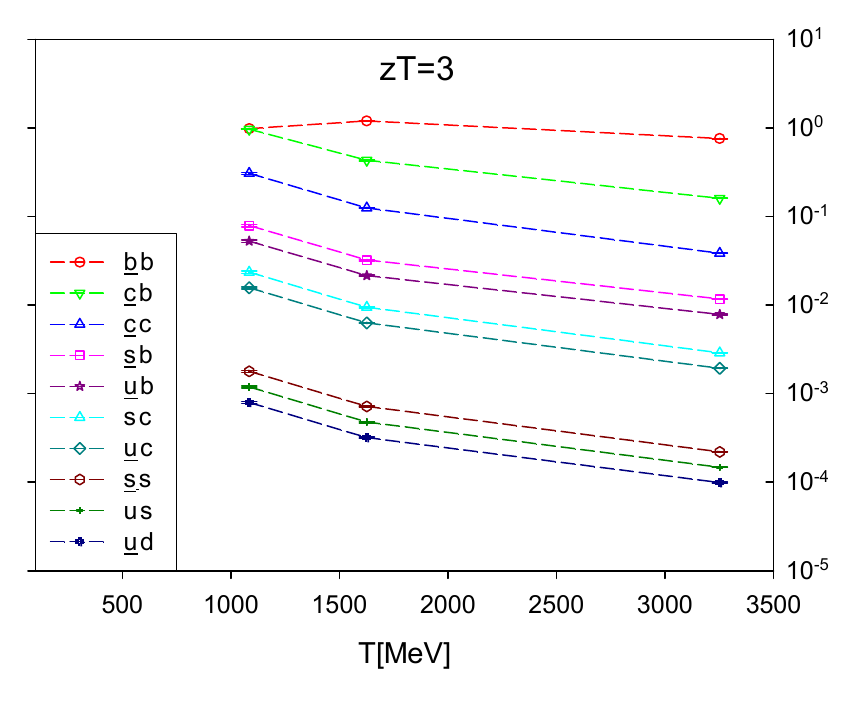}
    \end{subfigure}
\label{fig:ratio_kVA_kCS} 
\end{figure}

Finally, we compare the precision of symmetry between the $SU(2)_{CS}$ chiral-spin symmetry 
and the $SU(2)_L \times SU(2)_R$ chiral symmetry. This comparison provides critical insights 
into the interplay of chiral and chiral-spin symmetry manifestations in lattice QCD.

To this end, we compute the ratio of their symmetry-breaking parameters, $\kappa_{VA}/\kappa_{CS}$, 
for all flavor contents, as shown in Figure~\ref{fig:ratio_kVA_kCS} for $zT = 1$, 2, and 3. 
Notably, for each flavor content, the ratio $\kappa_{VA}/\kappa_{CS}$ decreases monotonically 
with $T$ and remains nearly constant across all $zT$ at a fixed $T$. This behavior strongly suggests 
the emergence of an $SU(2)_{CS} \times SU(2)_L \times SU(2)_R$ symmetry once the $SU(2)_{CS}$ symmetry 
arises within the temperature windows satisfying the criterion (\ref{eq:SU2_CS_crit_z}). 
Furthermore, this observation hints at the possible manifestation of a larger $SU(4)$ symmetry, 
which contains $SU(2)_{CS} \times SU(2)_L \times SU(2)_R$ 
as a subgroup \cite{Glozman:2014mka,Glozman:2015qva}.

To fully investigate the $SU(4)$ symmetry, it is necessary to examine the degeneracies of 
$SU(4)$ multiplets, including flavor singlets of $J=1$ mesons. 
However, the $z$-correlators for these multiplets involve disconnected diagrams, 
which are not included in this study. Instead, we will address this by analyzing the degeneracies 
in the chiral susceptibilities of the multiplets using all-to-all quark propagators estimated 
with $Z_2$ noise.

Moreover, the hierarchy of the ratio $R \equiv \kappa_{VA}/\kappa_{CS}$ follows the same order 
as (\ref{eq:Tc_order}) and (\ref{eq:T1_order}), i.e.,  
\bea
\label{eq:ratio_kVA_kCS_order}
R^{\bar u d} <
R^{\bar u s} <
R^{\bar s s} <
R^{\bar u c} <
R^{\bar s c} <
R^{\bar u b} <
R^{\bar s b} <
R^{\bar c c} <
R^{\bar c b} <
R^{\bar b b},
\eea
for all $zT$ at fixed $T$. However, this hierarchy does not necessarily imply that the emergence 
of $SU(2)_{CS}$ symmetry follows the same order, as the fading of $SU(2)_{CS}$ symmetry 
must also be considered. 
This is evident from the temperature windows for the emergence of $SU(2)_{CS}$ symmetry, 
as detailed in Tables~\ref{tab:TCS_all_zT10000}--\ref{tab:TCS_all_zT30000}.

If we compare $\kappa_{CS}$ and $\kappa_{VA}$ on an equal footing, 
Figure~\ref{fig:ratio_kVA_kCS} reveals that the $SU(2)_L \times SU(2)_R$ chiral symmetry 
(as well as $U(1)_A$, since $\kappa_{TX} \simeq \kappa_{VA}$) is significantly more precise than 
the $SU(2)_{CS}$ chiral-spin symmetry in the $(\bar u d, \bar u s, \bar s s)$ sectors 
involving light quarks, with $\kappa_{VA}/\kappa_{CS} \lesssim 10^{-2}$. 
In the heavy-light quark sectors $(\bar u c, \bar s c, \bar u b, \bar s b)$, 
the chiral symmetry remains slightly more precise than the chiral-spin symmetry, 
with $\kappa_{VA}/\kappa_{CS} \lesssim 0.5$. 
In contrast, for sectors containing only heavy quarks $(\bar c c, \bar c b, \bar b b)$, 
the precision of the chiral and chiral-spin symmetries becomes comparable, 
as indicated by $\kappa_{VA}/\kappa_{CS} \lesssim 1$. This provides a qualitative picture of 
how the relative strength of the chiral-spin symmetry versus the chiral symmetry varies 
with quark content.

\section{Concluding remarks}
\label{conclusions}

In this study, we have generated eight gauge ensembles of $N_f=2+1+1+1$ lattice QCD 
with physical $(s,c,b)$ quarks but unphysically heavy $u/d$ quarks with $M_\pi \sim 700$~MeV,
on the $40^3 \times (20, 16, 12, 10, 8, 6, 4, 2)$ lattices
with lattice spacing $a \sim 0.03$~fm,
for temperatures in the range of 325-3250 MeV, as summarized in Table \ref{tab:8_ensembles}.

Using these eight gauge ensembles, we computed the meson $z$-correlators for 
the complete set of Dirac bilinears (scalar, pseudoscalar, vector, axial vector, tensor vector,
and axial-tensor vector), and each for ten combinations of quark flavors
($\bar u d$, $\bar u s$, $\bar s s$, 
 $\bar u c$, $\bar s c$, $\bar u b$, $\bar s b$, 
 $\bar c c$, $\bar c b$, $\bar b b$).
Then, we used (\ref{eq:Tc_epsilon}) and (\ref{eq:T1_epsilon})  
to determine $T_c$ and $T_1$ for each flavor combination, and obtain the 
hierarchy of restoration of chiral symmetry, in the order of
\BAN
T_{c1}^{\bar u d} < T_{c1}^{\bar u s} < T_{c1}^{\bar s s} < 
T_{c1}^{\bar u c} < T_{c1}^{\bar s c} < T_{c1}^{\bar u b} < T_{c1}^{\bar s b} < 
T_{c1}^{\bar c c} < T_{c1}^{\bar c b} < T_{c1}^{\bar b b},  \hspace{4mm} T_{c1} \equiv \max(T_c, T_1),  
\EAN 
which immediately gives the hierarchical restoration of chiral symmetry in $N_f=2+1+1+1$ QCD, i.e.,  
from the restoration of $SU(2)_L \times SU(2)_R \times U(1)_A$ chiral symmetry 
of $(u, d)$ quarks at $ T_{c1}^{\bar u d}$ 
to the $SU(3)_L \times SU(3)_R \times U(1)_A$ chiral symmetry 
of $(u, d, s)$ quarks at $ T_{c1}^{\bar s s} $, then to  
the $SU(4)_L \times SU(4)_R \times U(1)_A$ chiral symmetry 
of $(u, d, s, c)$ quarks at $ T_{c1}^{\bar c c}$, 
and finally to the $SU(5)_L \times SU(5)_R \times U(1)_A$ chiral symmetry 
of $(u, d, s, c, b)$ quarks at $ T_{c1}^{\bar b b}$.  

One of the key phenomenological outcomes of the hierarchical restoration of chiral symmetry 
is the sequential pattern of hadron dissolution as the temperature is increased successively,  
resulting in a hierarchy in both the dissolution of hadrons and their suppression 
within the quark-gluon plasma. This can be seen as follows. 
Theoretically, a meson with quark content $\bar{q} Q$ 
dissolves entirely when $\bar{q}$ and $Q$ become deconfined. 
This occurs when the screening mass of the meson exceeds that of the corresponding noninteracting theory 
with free quarks of the same masses of $q$ and $Q$. It is expected that 
$m_{\text{scr}}^{\bar{q} Q} \ge m_{\text{scr}}^{\bar{q} Q(\text{free})} $ 
at a temperature $ T_d^{\bar{q} Q} \gtrsim T_{c1}^{\bar{q} Q} $, 
where the chiral symmetry $ SU(2)_L \times SU(2)_R \times U(1)_A $ of $ \bar{q} \Gamma Q $ 
has been effectively restored.
For $N_f=2+1+1+1$ lattice QCD, this implies that the hierarchy of meson dissolution is exactly 
the same as that of chiral symmetry restoration (\ref{eq:Tc1_order}), i.e.,  
\bea
T_d^{\bar{u} d} < T_d^{\bar{u} s} < T_d^{\bar{s} s} < T_d^{\bar{u} c} < T_d^{\bar{s} c} < T_d^{\bar{u} b} 
< T_d^{\bar{s} b} < T_d^{\bar{c} c} < T_d^{\bar{c} b} < T_d^{\bar{b} b}.
\eea
This hierarchy predicts the gradual suppression of mesons within the quark-gluon plasma, 
potentially observable in relativistic heavy-ion collision experiments, 
such as those conducted at the LHC and RHIC. 
This notion builds on the pioneering work \cite{Matsui:1986dk}, 
which proposed that the dissolution of $J/\psi$ mesons in the quark-gluon plasma would 
manifest as suppressed production in heavy-ion collision experiments.

Regarding the emergent $SU(2)_{CS}$ chiral-spin symmetry, 
it is intriguing to observe that the temperature windows meeting 
the criterion (\ref{eq:SU2_CS_crit_z}) are dominated by the channels of 
heavy vector mesons with flavor contents $\bar{u} b$ and $\bar{s} b$, 
as indicated by Tables~\ref{tab:TCS_all_zT10000}--\ref{tab:TCS_all_zT30000}. 
These results represent the first findings in lattice QCD 
and suggest that, within their respective temperature windows, 
hadronlike states—especially $\bar u b$ and $\bar s b$ vector mesons-are likely bound 
into color singlets by chromoelectric interactions. 
This is notable because neither the chromomagnetic part of the quark-gluon interaction
nor the noninteracting theory with free quarks possesses any $ SU(2)_{CS} $ symmetry.
Furthermore, these findings offer valuable insights for exploring the emergent  
$SU(2)_{CS}$ symmetry in relativistic heavy-ion collision experiments, 
such as those conducted at the LHC and RHIC, by focusing on meson channels with 
$\bar{u} b$, $\bar{d} b$ and $\bar{s} b$ quark contents. 

By comparing the $T$-windows for the emergent $SU(2)_{CS}$ symmetry across different lattice QCD setups, 
we obtained a universal feature of chiral-spin symmetry in any QCD system. Specifically, we analyzed the following: \\
\noindent 1. The $\bar{u}d$ sector in $N_f=2$ lattice QCD near the physical point \cite{Rohrhofer:2019qwq} 
versus $N_f=2+1+1$ lattice QCD at the physical point \cite{Chiu:2024jyz}. \\ 
\noindent 2. The $(\bar{u}d, \bar{u}s, \bar{u}c, \bar{s}s, \bar{s}c, \bar{c}c)$ sectors 
in $N_f=2+1+1$ lattice QCD at the physical point \cite{Chiu:2024jyz} versus 
$N_f=2+1+1+1$ lattice QCD in this work. \\  
From these comparisons, we deduce the following universal feature of chiral-spin symmetry 
in any QCD system:
\textit{Increasing the number of dynamical heavy quarks shifts the $T$-windows for $SU(2)_{CS}$ symmetry 
to higher temperature ranges, with these windows being primarily dominated by the sectors involving 
the heaviest quark and the light quarks of the system.}
However, given the different lattice spacings in the $N_f=2/2+1+1/2+1+1+1$ ensembles
and the associated discretization errors---as well as the unphysical $u/d$ quark masses
in the $N_f=2+1+1+1$ lattices---it remains an open question whether this feature persists
in the physical limit; that is, with $(u,d,s,c,b)$ quarks at the physical point across all
lattice setups, in the continuum and infinite volume limits.  

To understand the nature of mesonlike states in the $J=1$ channels 
(i.e., $V_k$, $A_k$, $T_k$, and $X_k$) which are relevant to the emergent 
$SU(2)_{CS}$ symmetry, it is essential to analyze the behavior of their spectral functions 
as the temperature increases. If bound-state peaks are found within the temperature ranges 
where the $SU(2)_{CS}$ symmetry holds, and these peaks gradually broaden 
and eventually vanish as $T$ rises beyond these ranges, it would suggest that the degrees of freedom 
in these mesonlike objects correspond to color-singlet mesons, 
as opposed to deconfined quarks and gluons.
To investigate this, one could consider extending the method used in 
refs.~\cite{Bros:1992ey,Bros:2001zs,Lowdon:2022xcl} for $J=0$ mesons to the $J=1$ mesons. 
Additionally, it is necessary to compute the spatial $z$-correlators of vector mesons 
to high precisions, free of the contribution of unphysical meson states even at large distances, 
in order to reliably extract the damping factor 
for each $J=1$ meson channel. The prescription used in ref.~\cite{Chiu:2023hnm} 
(to compute two sets of quark propagators with periodic and antiperiodic boundary conditions 
in the $z$ direction) provides an effective way to eliminate the contribution of unphysical meson states 
to the $z$-correlators and offers a promising way to achieve this goal.

\appendix
\appendixsetup

\section{Notations and conventions}
\label{app:A}

This appendix summarizes the notations and conventions in this paper, which have been used 
in refs.~\cite{Chiu:2023hnm,Chiu:2024jyz}.

The correlation function of meson interpolator $ \bar q_1 \Gamma q_2 $ 
on a lattice with $(N_x, N_y, N_z, N_t)$ sites is measured according to the formula
\bea
\label{eq:C_Gamma}
C_\Gamma(x) = \left< (\bar q_1 \Gamma q_2)_x (\bar q_1 \Gamma q_2)_0^{\dagger}  \right >
= \left< \tr\left[ \Gamma (D_c + m_1)^{-1}_{0,x} \Gamma (D_c + m_2)^{-1}_{x,0} \right]
  \right >_{\text{confs}},
\eea
where $(D_c + m_q)^{-1} $ denotes the valence quark propagator with quark mass $ m_q $ in lattice QCD
with exact chiral symmetry, tr denotes the trace over the color and Dirac indices, and
the brackets $\left< \cdots \right>_{\text{confs}} $ denote averaging over the gauge configurations.
Here the label of a lattice site $x$ is understood to stand for
$ (x_1, x_2, x_3, x_4) = (x, y, z, t)$, and the overall $\pm$ sign due to
$ \gamma_4 \Gamma^\dagger \gamma_4 = \pm \Gamma $ has been suppressed. 
The $z$-correlator of the meson interpolator $ \bar q_1 \Gamma q_2 $ is defined as
\bea
\label{eq:C_Gamma_t}
C_\Gamma(z,T) = \sum_{x_1, x_2, x_4} C_{\Gamma}(x), 
\eea
where $T=1/(N_t a)$ is the temperature. 
In general, the meson $z$-correlator is expressed as a function of the dimensionless variable
\bea
zT = \frac{n_z a}{N_t a} = \frac{n_z}{N_t}
\eea
and is denoted by $C_\Gamma(zT)$.

The meson interpolators are classified according to their transformation properties, 
as listed in Table \ref{tab:bilinear}.
Due to the degeneracy (the $S_2$ symmetry) between the $k=1$ and $k=2$ components of 
the $z$-correlator for any vector meson, only the $k=1$ component is explicitly presented in this paper. 
In other words, the $k=2$ components of all meson $z$-correlators, 
as well as symmetry-breaking parameters involving $k=2$ components, are omitted.

\begin{table}[ht]
\centering
\caption{The classification of meson interpolators $\bar q_1 \Gamma q_2 $ 
and their names and notations.}
\setlength{\tabcolsep}{4pt}
\vspace{2mm}
\begin{tabular}{c c}
\hline
\hline
Name and notation & $\Gamma$ (for $z$ correlators) \\
\hline
\multicolumn{1}{l}{Scalar ($S$)}
&  $\Id$                         \\
\multicolumn{1}{l}{Pseudocalar ($P$)}
&  $\gamma_5$                    \\
\multicolumn{1}{l}{Vector ($V_k$)}
&  $\gamma_k \ (k=1,2,4)$                    \\
\multicolumn{1}{l}{Axial vector ($A_k$)}
&  $\gamma_5 \gamma_k \ (k=1,2,4)$           \\
\multicolumn{1}{l}{Tensor vector ($T_k$)}
&  $\gamma_3 \gamma_k \ (k=1,2,4)$           \\
\multicolumn{1}{l}{Axial-tensor vector ($X_k$)}
&  $\gamma_5 \gamma_3 \gamma_k \ (k=1,2,4)$  \\
\hline
\hline
\end{tabular}
\label{tab:bilinear}
\end{table}

\section{Relationship between $\sqrt{t_0}$ and $M_\pi$}
\label{app:B}

In this appendix, we estimate the variation of $ \sqrt{t_0} $ as $ M_\pi $ 
changes from 700 MeV to the chiral limit 
in lattice QCD with $ N_f=2 $ optimal domain-wall quarks. 
This serves as an estimate for the variation of $ \sqrt{t_0} $ under the same conditions 
for lattice QCD with $ N_f=2+1+1+1 $ optimal domain-wall quarks, 
where the masses of the $ s $, $ c $, and $ b $ quarks remain fixed. 
Note that performing this analysis directly for the latter case would require simulations 
on lattices with lattice size $ \gtrsim 180^4 $ and lattice spacing $ a \sim 0.03 $ fm, 
which is infeasible for the lattice community in the near future. 

For this analysis, we use results from eight $N_f = 2$ gauge ensembles with $M_\pi \sim 228$–$565$ MeV, 
as reported in ref.~\cite{Chiu:2011bm}, along with a newly generated ensemble with 
$M_\pi \sim 700$ MeV.  Table \ref{tab:Nf2} summarizes the relevant results, 
including $m_q$, $a$, $M_\pi$, $\sqrt{t_0}/a$, and $\sqrt{t_0}$. 
The first eight rows for $a$ and $M_\pi$ are taken from Tables 1 and 3 of ref.~\cite{Chiu:2011bm}, 
while the corresponding values for $\sqrt{t_0}/a$ and $\sqrt{t_0}$ are obtained 
in ref.~\cite{Chen:2014hva} as part of the determination of $\sqrt{t_0} = 0.1415(9)$ fm 
for lattice QCD with $N_f = 2$ optimal domain-wall quarks in the chiral limit. 
The data in the last row of Table \ref{tab:Nf2} are obtained in the present work.  

As detailed in ref.~\cite{Chiu:2011bm}, the lattice spacings listed 
in the second column of Table \ref{tab:Nf2} are determined using the heavy quark potential 
with the Sommer parameter $ r_0 = 0.49 $ fm. 
The pion masses in the third column are extracted from the ground state of the 
pseudoscalar time-correlation function. The values in the fourth column are obtained 
by the Wilson flow with the condition $ \langle t^2 E(t) \rangle |_{t=t_0} = 0.3 $,  
and the fifth column combines the inputs from the second and fourth columns.

\begin{table}[H]
\caption{The relationship between $\sqrt{t_0}$ and $M_\pi$ in 
         $N_f=2$ lattice QCD with optimal domain-wall quarks. See text for details.}
\vspace{2mm}
\centering
\begin{tabular}{|ccccc|}
\hline
$m_q a$  &  $a$[fm]     &  $M_\pi$[GeV]  & $\sqrt{t_0}/a$  &  $\sqrt{t_0}$[fm]  \\
\hline
0.0100 \  &  0.1045(13) \ &  0.2275(76)  \  & 1.3533(53) \	&	0.1414(18) \\	
0.0200 \  &  0.1051(10) \ &  0.3089(49)  \  & 1.3461(38) \ 	&	0.1415(14) \\	
0.0300 \  &  0.1060(12) \ &  0.3672(56)  \  & 1.3295(48) \	&	0.1409(17) \\	
0.0400 \  &  0.1071(16) \ &  0.4135(93)  \  & 1.3253(31) \	&	0.1418(24) \\	
0.0500 \  &  0.1078(16) \ &  0.4586(100) \  & 1.3041(35) \	&	0.1406(21) \\	
0.0600 \  &  0.1089(11) \ &  0.4976(59)  \  & 1.3004(33) \	&	0.1416(15) \\	
0.0700 \  &  0.1097(10) \ &  0.5327(74)  \  & 1.2945(41) \	&	0.1420(14) \\	
0.0800 \  &  0.1105(14) \ &  0.5654(78)  \  & 1.2875(35) \	&	0.1423(18) \\	
0.1226 \  &  0.1144(10) \ &  0.7007(57)  \  & 1.2427(43) \	&	0.1422(13) \\	
\hline
\end{tabular}
\label{tab:Nf2}
\end{table}


From the last column of Table \ref{tab:Nf2}, it is evident that $\sqrt{t_0}$ remains approximately 
constant for $M_\pi \sim 228–700$~MeV, with variations well within the error bars. 
Consequencely, the value of $\sqrt{t_0}$ at $M_\pi \sim 140$ MeV is expected to fall between its values 
at $M_\pi \sim 228$ MeV and in the chiral limit \cite{Chen:2014hva},   
i.e., $0.1414(18)$ fm $\leq \sqrt{t_0} \leq 0.1415(9)$ fm.  
Thus, the difference in $\sqrt{t_0}$ between $\sqrt{t_0} = 0.1422(13)$ fm at $M_\pi \simeq 700$ MeV 
and its estimated value at $M_\pi \sim 140$ MeV is 0.0008(16) fm, 
which lies within their respective error margins. 

Similarly, the difference between 
$\sqrt{t_0} = 0.1422(13)$ fm at $M_\pi \simeq 700$ MeV for $N_f = 2$ lattice QCD 
with optimal domain-wall quarks and $\sqrt{t_0} = 0.1416(8)$ fm, as determined by the MILC Collaboration 
for $N_f = 2+1+1$ lattice QCD with highly improved staggered quarks at the physical point 
and in the continuum limit, is 0.0006(15) fm, also within their error margins. 
Therefore, it is reasonable to use $\sqrt{t_0} = 0.1416(8)$ fm, as determined by the MILC Collaboration, 
as an input parameter to set the lattice spacing for the gauge ensemble generated 
by lattice QCD with $N_f = 2+1+1+1$ optimal domain-wall quarks at $M_\pi \simeq 700$ MeV, 
as detailed in ref.~\cite{Chiu:2020tml}.

\section{Symmetry breaking parameters of $N_f=2+1+1+1$ lattice QCD}
\label{app:C} 

In this appendix, the numerical values of   
$\kappa_{VA}$, $\kappa_{TX}$, $\kappa$, and $\kappa_{CS}$ are tabulated  
for $zT=1$, 2, and 3, and for each flavor sector respectively.  
For the $z$ correlators, the possible values of $zT$ at $T=1/(N_t a)$ are
$\{n_z/N_t, n_z = 1, 2, \cdots, N_z/2 \}$. 
Thus for $N_z=40$ and $N_t$ = (20, 16, 12, 10, 8, 6, 4, 2), 
the number of available temperatures are $(8, 5, 3)$ for $zT=(1,2,3)$, as shown in Tables
\ref{tab:Kappa_ud}--\ref{tab:Kappa_bb}. 

The error in the parenthesis of each entry is statistical, which    
is estimated by the jackknife method with the binsize of 10-15 configurations
of which the statistical error saturates. Due to the single spatial volume and one lattice spacing of 
this study, the systematic errors due to finite lattice spacing and finite volume cannot be estimated.
Similarly, systematics due to the unphysically heavy $u/d$ quark masses also cannot be estimated.
In other words, the precise values of      
$\kappa_{VA}$, $\kappa_{TX}$, $\kappa$, and $\kappa_{CS}$ at each temperature 
have not been determined in this study. Nevertheless, the patterns of hierarchical restoration 
of chiral symmetry as well as the emergence of approximate chiral-spin symmetry 
in high temperature QCD can be unveiled from these data.   

\vfill  
\newpage

\begin{table}[h!]
\centering
\caption{The symmetry breaking parameters of the $\bar u d $ sector in $N_f=2+1+1+1$ lattice QCD.}
\label{tab:Kappa_ud}
\setlength{\tabcolsep}{6pt}
\vspace{2mm}
\begin{tabular}{lccccccr}
\toprule
$T$	& $N_t$	& $zT$  &	$k_{VA}$ &	$k_{TX}$ & $\kappa$	&	$k_{CS}$  \\    
\midrule
325	& 20 & 1 &	
$3.363(56) \times 10^{-3}$ & $3.64(7) \times 10^{-3}$ & 0.1282(49) &	0.2847(83) \\
406 & 16 & 1 &	
$1.808(46) \times 10^{-3}$ & $2.01(5) \times 10^{-3}$ & 0.1399(33) &	0.2493(73) \\
542 & 12 & 1 &	
$7.76(15) \times 10^{-4}$ & $8.79(16) \times 10^{-4}$ &	0.1719(39) & 0.2076(42) \\
650 & 10 & 1 &	
$4.38(6) \times 10^{-4}$ & $5.003(65) \times 10^{-4}$ & 0.2051(42) & 0.1900(44)   \\
813	&  8 & 1 &	
$2.37(3) \times 10^{-4}$ & $2.721(32) \times 10^{-4}$ & 0.2267(34) & 0.1738(36)  \\
1084&  6 & 1 &	
$1.0771(53) \times 10^{-4}$ & $1.2521(74) \times 10^{-4}$ &	0.285(3) & 0.1573(17) \\
1626&  4 & 1 &	
$3.476(12) \times 10^{-5}$ & $3.889(15) \times 10^{-5}$	& 0.3945(14) & 0.1334(6)  \\
3252&  2 & 1 &  
$3.818(72) \times 10^{-6}$ & $2.900(68) \times 10^{-6}$ & 0.5953(3) & 0.0749(1)  \\
\midrule
650 & 10 & 2 &	
$5.218(96) \times 10^{-4}$ & $5.51(11) \times 10^{-4}$ & 0.1327(47) & 0.195(9) \\
813	&  8 & 2 &	
$2.791(41) \times 10^{-4}$ & $2.946(44) \times 10^{-4}$ & 0.1542(47) & 0.1817(61) \\
1084&  6 & 2 &	
$1.2418(79) \times 10^{-4}$ & $1.3292(94) \times 10^{-4}$ & 0.1895(39) & 0.1568(37) \\
1626&  4 & 2 &	
$3.937(16) \times 10^{-5}$ & $3.957(18) \times 10^{-5}$ & 0.2570(24) & 0.1241(16) \\
3252&  2 & 2 &  
$5.411(65) \times 10^{-6}$ & $2.583(84) \times 10^{-6}$ & 0.4841(8) & 0.0770(2)	\\
\midrule
1084&  6 & 3 &	
$1.35(1) \times 10^{-4}$ & $1.392(11) \times 10^{-4}$ & 0.1448(34) & 0.1694(45) \\
1626&  4 & 3 &	
$4.105(17) \times 10^{-5}$ & $3.887(22) \times 10^{-5}$ & 0.1982(28) & 0.1290(22) \\
3252&  2 & 3 &  
$7.901(65) \times 10^{-6}$ & $3.67(11) \times 10^{-6}$ & 0.4009(12) & 0.0798(3) \\	
\bottomrule
\end{tabular}
\end{table}

\vfill  
\newpage

\begin{table}[h!]
\centering
\caption{The symmetry breaking parameters of the $\bar u s $ sector in $N_f=2+1+1+1$ lattice QCD.}
\label{tab:Kappa_us}
\setlength{\tabcolsep}{6pt}
\vspace{2mm}
\begin{tabular}{lccccccr}
\toprule
$T$	& $N_t$	& $zT$  &	$k_{VA}$ &	$k_{TX}$ & $\kappa$	&	$k_{CS}$  \\    
\midrule
325	& 20 & 1 &	
$5.031(83) \times 10^{-3}$ & $5.5(1) \times 10^{-3}$ & 0.1281(49) & 0.2844(83) \\
406 & 16 & 1 &	
$2.708(69) \times 10^{-3}$ & $3.015(75) \times 10^{-3}$ & 0.1398(33) & 0.2491(73) \\
542 & 12 & 1 &	
$1.163(23) \times 10^{-3}$ & $1.317(24) \times 10^{-3}$ & 0.1718(39) & 0.2075(42) \\
650 & 10 & 1 &	
$6.573(90) \times 10^{-4}$ & $7.502(97) \times 10^{-4}$ & 0.2051(42) & 0.1900(44) \\
813	&  8 & 1 &	
$3.558(45) \times 10^{-4}$ & $4.082(48) \times 10^{-4}$ & 0.2267(34) & 0.1738(36) \\
1084&  6 & 1 &	
$1.616(8) \times 10^{-4}$ & $1.878(11) \times 10^{-4}$ & 0.285(3) & 0.1574(17) \\
1626&  4 & 1 &	
$5.212(20) \times 10^{-5}$ & $5.832(25) \times 10^{-5}$ & $0.3945(14)$ & $0.1334(6)$ \\
3252&  2 & 1 &  
$5.727(87) \times 10^{-6}$ & $4.350(79) \times 10^{-6}$ & $0.5953(3)$ & $0.0749(1)$ \\
\midrule
650 & 10 & 2 &	
$7.83(14) \times 10^{-4}$ & $8.26(16) \times 10^{-4}$ & $0.1327(47)$ & $0.1948(90)$ \\
813	&  8 & 2 &	
$4.187(62) \times 10^{-4}$ & $4.419(65) \times 10^{-4}$ & $0.1542(47)$ & $0.1817(61)$ \\
1084&  6 & 2 &	
$1.863(12) \times 10^{-4}$ & $1.994(14) \times 10^{-4}$ & $0.1895(39)$ & $0.1568(37)$ \\
1626&  4 & 2 &	
$5.906(24) \times 10^{-5}$ & $5.936(29) \times 10^{-5}$ & $0.2570(24)$ & $0.1241(16)$ \\
3252&  2 & 2 &  
$8.109(95) \times 10^{-6}$ & $3.877(95) \times 10^{-6}$ & $0.4841(8)$ & $0.0770(2)$ \\
\midrule
1084&  6 & 3 &	
$2.017(16) \times 10^{-4}$ & $2.088(17) \times 10^{-4}$ & $0.1448(34)$ & $0.1694(45)$ \\
1626&  4 & 3 &	
$6.159(26) \times 10^{-5}$ & $5.833(32) \times 10^{-5}$ & $0.1982(28)$ & $0.1290(22)$ \\
3252&  2 & 3 &  
$1.1757(79) \times 10^{-5}$ & $5.415(96) \times 10^{-6}$ & $0.4009(11)$ & $0.0798(3)$ \\
\bottomrule
\end{tabular}
\end{table}

\vfill  
\newpage

\begin{table}[h!]
\centering
\caption{The symmetry breaking parameters of the $\bar s s $ sector in $N_f=2+1+1+1$ lattice QCD.}
\label{tab:Kappa_ss}
\setlength{\tabcolsep}{6pt}
\vspace{2mm}
\begin{tabular}{lccccccr}
\toprule
$T$	& $N_t$	& $zT$  &	$k_{VA}$ &	$k_{TX}$ & $\kappa$	&	$k_{CS}$  \\    
\midrule
325	& 20 & 1 &	
$7.53(12) \times 10^{-3}$ & $8.15(16) \times 10^{-3}$ & $1.279(49) \times 10^{-1}$ & $2.841(83) \times 10^{-1}$ \\
406 & 16 & 1 &	
$4.1(1) \times 10^{-3}$ & $4.52(11) \times 10^{-3}$ & 0.1398(33) & 0.2490(73) \\
542 & 12 & 1 &	
$1.743(34) \times 10^{-3}$ & $1.975(37) \times 10^{-3}$ & 0.1718(39) & 0.2075(42) \\
650 & 10 & 1 &	
$9.86(13) \times 10^{-4}$ & $1.125(15) \times 10^{-3}$ & 0.2051(42)  & 0.1900(44) \\
813	&  8 & 1 &	
$5.337(68) \times 10^{-4}$ & $6.122(71) \times 10^{-4}$ & 0.2267(34) & 0.1738(36) \\
1084&  6 & 1 &	
$2.424(12) \times 10^{-4}$ & $2.817(17) \times 10^{-4}$ & 0.285(3) & 0.1574(17) \\
1626&  4 & 1 &	
$7.816(28) \times 10^{-5}$ & $8.745(34) \times 10^{-5}$ & 0.3945(14) & 0.1334(6)  \\
3252&  2 & 1 &  
$8.59(6) \times 10^{-6}$ & $6.527(81) \times 10^{-6}$ & 0.5952(3) & 0.0749(1)  \\
\midrule
650 & 10 & 2 &	
$1.174(22) \times 10^{-3}$ & $1.239(24) \times 10^{-3}$ & 0.1326(47) & 0.195(9) \\
813	&  8 & 2 &	
$6.280(92) \times 10^{-4}$ & $6.628(98) \times 10^{-4}$ & 0.1542(47) & 0.1816(61)  \\
1084&  6 & 2 &	
$2.795(18) \times 10^{-4}$ & $2.991(21) \times 10^{-4}$ & 0.1895(39) & 0.1568(38) \\
1626&  4 & 2 &	
$8.861(39) \times 10^{-5}$ & $8.907(43) \times 10^{-5}$ & 0.2570(24) & 0.1241(16) \\
3252&  2 & 2 &  
$1.216(9) \times 10^{-5}$ & $5.828(54) \times 10^{-6}$ & 0.4841(8) & 0.0770(2) \\
\midrule
1084&  6 & 3 &	
$3.026(24) \times 10^{-4}$ & $3.133(26) \times 10^{-4}$ & 0.1448(34) & 0.1694(45) \\
1626&  4 & 3 &	
$9.239(39) \times 10^{-5}$ & $8.75(5) \times 10^{-5}$ & 0.1982(28) & 0.1290(22)  \\
3252&  2 & 3 &  
$1.752(11) \times 10^{-5}$ & $8.00(13) \times 10^{-6}$ & 0.4009(11) & 0.798(3)  \\
\bottomrule
\end{tabular}
\end{table}

\vfill  
\newpage

\begin{table}[h!]
\centering
\caption{The symmetry breaking parameters of the $\bar u c $ sector in $N_f=2+1+1+1$ lattice QCD.}
\label{tab:Kappa_uc}
\setlength{\tabcolsep}{6pt}
\vspace{2mm}
\begin{tabular}{lccccccr}
\toprule
$T$	& $N_t$	& $zT$  &	$k_{VA}$ &	$k_{TX}$ & $\kappa$	&	$k_{CS}$  \\    
\midrule
325	& 20 & 1 &	
0.0432(5) & 0.0475(7) & 0.1382(42) & 0.2289(57) \\
406 & 16 & 1 &	
0.0263(5) & 0.0296(6) & 0.1520(36) & 0.2170(59) \\
542 & 12 & 1 &	
0.0131(2) & 0.0150(3) & 0.1769(39) & 0.1943(38) \\
650 & 10 & 1 &	
$7.9(1) \times 10^{-3}$ & $9.05(11) \times 10^{-3}$ & $0.2079(42)$ & $0.1829(41)$ \\
813	&  8 & 1 &	
$4.466(53) \times 10^{-3}$ & $5.137(56) \times 10^{-3}$ & $0.2282(33)$ & $0.1705(33)$ \\
1084&  6 & 1 &	
$2.09(1) \times 10^{-3}$ & $2.438(14) \times 10^{-3}$ & $0.285(3)$ & $0.1568(17)$ \\
1626&  4 & 1 &	
$6.856(23) \times 10^{-4}$ & $7.67(3) \times 10^{-4}$ & $0.3946(14)$ & $0.1336(6)$ \\
3252&  2 & 1 &  
$7.614(27) \times 10^{-5}$ & $5.776(38) \times 10^{-5}$ & $0.5949(3)$ & $0.0748(1)$ \\
\midrule
650 & 10 & 2 &	
$9.40(16) \times 10^{-3}$ & $0.0100(2)$ & $0.134(5)$ & $0.1824(84)$ \\
813	&  8 & 2 &	
$5.253(72) \times 10^{-3}$ & $5.556(77) \times 10^{-3}$ & $0.1547(47)$ & $0.1753(58)$ \\
1084&  6 & 2 &	
$2.412(15) \times 10^{-3}$ & $2.583(18) \times 10^{-3}$ & $0.1893(39)$ & $0.1551(37)$ \\
1626&  4 & 2 &	
$7.786(33) \times 10^{-4}$ & $7.817(37) \times 10^{-4}$ & $0.2564(24)$ & $0.1243(16)$ \\
3252&  2 & 2 &  
$1.078(5) \times 10^{-4}$ & $5.160(58) \times 10^{-5}$ & $0.4835(8)$ & $0.0771(2)$ \\
\midrule
1084&  6 & 3 &	
$2.61(2) \times 10^{-3}$ & $2.705(22) \times 10^{-3}$ & $0.1446(34)$ & $0.1670(45)$ \\
1626&  4 & 3 &	
$8.121(34) \times 10^{-4}$ & $7.680(43) \times 10^{-4}$ & $0.1974(28)$ & $0.1292(22)$ \\
3252&  2 & 3 &  
$1.5441(91) \times 10^{-4}$ & $6.981(91) \times 10^{-5}$ & $0.4004(12)$ & $0.0800(3)$ \\
\bottomrule
\end{tabular}
\end{table}

\vfill  
\newpage
 
\begin{table}[h!]
\centering
\caption{The symmetry breaking parameters of the $\bar s c $ sector in $N_f=2+1+1+1$ lattice QCD.}
\label{tab:Kappa_sc}
\setlength{\tabcolsep}{6pt}
\vspace{2mm}
\begin{tabular}{lccccccr}
\toprule
$T$	& $N_t$	& $zT$  &	$k_{VA}$ &	$k_{TX}$ & $\kappa$	&	$k_{CS}$  \\    
\midrule
325	& 20 & 1 &	
0.0646(7) & 0.0710(11) & 0.1356(41) & 0.2298(57) \\
406 & 16 & 1 &	
0.0395(7) & 0.0443(8) & 0.1504(36) & 0.2177(59) \\
542 & 12 & 1 &	
0.0197(3) & 0.0224(4) & 0.1761(39) & 0.1947(38) \\
650 & 10 & 1 &	
0.0118(1) & 0.0136(2) & 0.2073(41) & 0.1831(41) \\
813	&  8 & 1 &	
$6.70(8) \times 10^{-3}$ & $7.705(85) \times 10^{-3}$ & $0.2279(33)$ & $0.1707(33)$ \\
1084&  6 & 1 &	
$3.134(15) \times 10^{-3}$ & $3.648(21) \times 10^{-3}$ & $0.2850(30)$ & $0.1568(17)$ \\
1626&  4 & 1 &	
$1.0281(34) \times 10^{-3}$ & $1.1494(45) \times 10^{-3}$ & $0.3945(14)$ & $0.1336(6)$ \\
3252&  2 & 1 &  
$1.1424(39) \times 10^{-4}$ & $8.668(56) \times 10^{-5}$ & $0.5948(3)$ & $0.0748(1)$ \\
\midrule
650 & 10 & 2 &	
0.0141(2) & 0.0149(3) & 0.134(5) & 0.1824(84) \\
813	&  8 & 2 &	
$7.88(11) \times 10^{-3}$ & $8.33(12) \times 10^{-3}$ & $0.1544(47)$ & $0.1754(58)$ \\
1084&  6 & 2 &	
$3.618(22) \times 10^{-3}$ & $3.875(27) \times 10^{-3}$ & $0.1891(39)$ & $0.1551(37)$ \\
1626&  4 & 2 &	
$1.1681(49) \times 10^{-3}$ & $1.1729(56) \times 10^{-3}$ & $0.2563(24)$ & $0.1243(16)$ \\
3252&  2 & 2 &  
$1.619(8) \times 10^{-4}$ & $7.777(84) \times 10^{-5}$ & $0.4834(8)$ & $0.0771(2)$ \\
\midrule
1084&  6 & 3 &	
$3.917(29) \times 10^{-3}$ & $4.058(33) \times 10^{-3}$ & $0.1444(34)$ & $0.1670(45)$ \\
1626&  4 & 3 &	
$1.2184(51) \times 10^{-3}$ & $1.1525(64) \times 10^{-3}$ & $0.1973(28)$ & $0.1292(22)$ \\
3252&  2 & 3 &  
$2.306(14) \times 10^{-4}$ & $1.037(14) \times 10^{-4}$ & $0.4002(12)$ & $0.0800(3)$ \\
\bottomrule
\end{tabular}
\end{table}

\vfill  
\newpage

\begin{table}[h!]
\centering
\caption{The symmetry breaking parameters of the $\bar u b $ sector in $N_f=2+1+1+1$ lattice QCD.}
\label{tab:Kappa_ub}
\setlength{\tabcolsep}{6pt}
\vspace{2mm}
\begin{tabular}{lccccccr}
\toprule
$T$	& $N_t$	& $zT$  &	$k_{VA}$ &	$k_{TX}$ & $\kappa$	&	$k_{CS}$  \\    
\midrule
325	& 20 & 1 &	
0.0626(8) & 0.0673(9) & 0.1344(51) & 0.1196(39) \\
406 & 16 & 1 &	
0.0424(5) & 0.0465(6) & 0.160(5) & 0.1338(42) \\
542 & 12 & 1 &	
0.0256(3) & 0.0289(3) & 0.1779(44) & 0.1387(32) \\
650 & 10 & 1 &	
0.0177(2) & 0.0203(2) & 0.2065(48) & 0.1446(38) \\
813	&  8 & 1 &	
0.0116(1) & 0.0134(1) & 0.228(3) & 0.149(2) \\
1084&  6 & 1 &	
$6.29(2) \times 10^{-3}$ & $7.378(32) \times 10^{-3}$ & $0.2855(27)$ & $0.1538(17)$ \\
1626&  4 & 1 &	
$2.407(7) \times 10^{-3}$ & $2.6534(99) \times 10^{-3}$ & $0.3939(13)$ & $0.1370(6)$ \\
3252&  2 & 1 &  
$3.078(11) \times 10^{-4}$ & $2.277(15) \times 10^{-4}$ & $0.5904(4)$ & $0.0746(1)$ \\
\midrule
650 & 10 & 2 &	
0.0215(3) & 0.0228(3) & 0.1268(54) & 0.1268(67) \\
813	&  8 & 2 &	
0.0138(1) & 0.0147(1) & 0.1488(49) & 0.1409(49) \\
1084&  6 & 2 &	
$7.446(36) \times 10^{-3}$ & $8.001(47) \times 10^{-3}$ & $0.1805(39)$ & $0.1455(36)$ \\
1626&  4 & 2 &	
$2.792(12) \times 10^{-3}$ & $2.731(14) \times 10^{-3}$ & $0.2487(23)$ & $0.1318(17)$ \\
3252&  2 & 2 &  
$4.502(22) \times 10^{-4}$ & $2.093(23) \times 10^{-4}$ & $0.4772(8)$ & $0.0793(2)$ \\
\midrule
1084&  6 & 3 &	
$8.078(47) \times 10^{-3}$ & $8.36(6) \times 10^{-3}$ & $0.1358(38)$ & $0.1532(48)$ \\
1626&  4 & 3 &	
$2.939(13) \times 10^{-3}$ & $2.665(18) \times 10^{-3}$ & $0.1867(26)$ & $0.1370(23)$ \\
3252&  2 & 3 &  
$6.547(38) \times 10^{-4}$ & $2.923(39) \times 10^{-4}$ & $0.3948(12)$ & $0.0837(4)$ \\
\bottomrule
\end{tabular}
\end{table}

\vfill  
\newpage

\begin{table}[h!]
\centering
\caption{The symmetry breaking parameters of the $\bar s b $ sector in $N_f=2+1+1+1$ lattice QCD.}
\label{tab:Kappa_sb}
\setlength{\tabcolsep}{6pt}
\vspace{2mm}
\begin{tabular}{lccccccr}
\toprule
$T$	& $N_t$	& $zT$  &	$k_{VA}$ &	$k_{TX}$ & $\kappa$	&	$k_{CS}$  \\    
\midrule
325	& 20 & 1 &	
0.0935(12) & 0.1005(13) & 0.131(5) & 0.1209(39) \\
406 & 16 & 1 &	
0.0635(7) & 0.0696(8) & 0.1576(49) & 0.1349(41) \\
542 & 12 & 1 &	
0.0384(4) & 0.0433(5) & 0.1765(43) & 0.1395(35) \\
650 & 10 & 1 &	
0.0266(3) & 0.0304(3) & 0.2055(48) & 0.1453(38) \\
813	&  8 & 1 &	
0.0173(1) & 0.0201(1) & 0.227(3) & 0.150(2) \\
1084&  6 & 1 &	
$9.432(31) \times 10^{-3}$ & $0.0111(0)$ & $0.2851(26)$ & $0.1541(17)$ \\
1626&  4 & 1 &	
$3.609(11) \times 10^{-3}$ & $3.979(15) \times 10^{-3}$ & $0.3936(13)$ & $0.1371(6)$ \\
3252&  2 & 1 &  
$4.619(17) \times 10^{-4}$ & $3.417(23) \times 10^{-4}$ & $0.5902(4)$ & $0.0746(1)$ \\
\midrule
650 & 10 & 2 &	
0.0323(4) & 0.0342(4) & 0.1257(53) & 0.1271(67) \\
813	&  8 & 2 &	
0.0207(2) & 0.0221(2) & 0.1479(49) & 0.1411(49) \\
1084&  6 & 2 &	
0.0112(1) & 0.0120(1) & 0.1800(39) & 0.1457(36) \\
1626&  4 & 2 &	
$4.189(18) \times 10^{-3}$ & $4.098(21) \times 10^{-3}$ & $0.2483(23)$ & $0.1318(17)$ \\
3252&  2 & 2 &  
$6.759(32) \times 10^{-4}$ & $3.154(35) \times 10^{-4}$ & $0.4767(8)$ & $0.0793(2)$ \\
\midrule
1084&  6 & 3 &	
0.0121(1) & 0.0125(1) & 0.1353(37) & 0.1532(48) \\
1626&  4 & 3 &	
$4.41(2) \times 10^{-3}$ & $3.999(26) \times 10^{-3}$ & $0.1863(26)$ & $0.1370(23)$ \\
3252&  2 & 3 &  
$9.782(58) \times 10^{-4}$ & $4.346(59) \times 10^{-4}$ & $0.3943(12)$ & $0.0836(4)$ \\
\bottomrule
\end{tabular}
\end{table}

\vfill  
\newpage

\begin{table}[h!]
\centering
\caption{The symmetry breaking parameters of the $\bar c c $ sector in $N_f=2+1+1+1$ lattice QCD.}
\label{tab:Kappa_cc}
\setlength{\tabcolsep}{6pt}
\vspace{2mm}
\begin{tabular}{lccccccr}
\toprule
$T$	& $N_t$	& $zT$  &	$k_{VA}$ &	$k_{TX}$ & $\kappa$	&	$k_{CS}$  \\    
\midrule
325	& 20 & 1 &	
0.518(4) & 0.5701(61)	& 0.0865(24) & 0.5701(61) \\	
406 & 16 & 1 &	
0.3743(46) & 0.4205(53) & 0.1175(31) & 0.4205(53) \\
542 & 12 & 1 &	
0.2205(32) & 0.2513(35)	& 0.1551(34) & 0.2513(35) \\
650 & 10 & 1 &	
0.1417(17) & 0.1629(19)	& 0.1922(37) & 0.1877(39) \\
813	&  8 & 1 &	
0.0840(9) & 0.097(1) & 0.219(3) & 0.174(3) \\
1084&  6 & 1 &	
0.0405(2) & 0.0472(3) & 0.2799(28) & 0.1593(17) \\
1626&  4 & 1 &	
0.0135(1) & 0.0151(1) & 0.3916(14) & 0.1345(6) \\
3252&  2 & 1 &  
$1.5194(55) \times 10^{-3}$ & $1.1517(76) \times 10^{-3}$ &	0.5925(3)&	0.0747(1) \\
\midrule
650 & 10 & 2 &	
0.1683(26) & 0.1787(29) & 0.1167(42) & 0.1787(29) \\
813	&  8 & 2 &	
0.0986(13) & 0.1045(13) & 0.1429(44) & 0.1724(52) \\
1084&  6 & 2 &	
0.0468(3) & 0.0502(3) & 0.1825(37) & 0.1550(37) \\
1626&  4 & 2 &	
0.0154(1) & 0.0154(1) & 0.2523(24) & 0.1246(16) \\
3252&  2 & 2 &  
$2.16(1) \times 10^{-3}$& $1.043(11) \times 10^{-3}$&	0.4791(8)&	0.0767(2) \\
\midrule
1084&  6 & 3 &	
0.0507(4) & 0.0525(4) & 0.1382(32) & 0.1655(45) \\
1626&  4 & 3 &	
0.0161(1) & 0.0152(1) & 0.1933(28) & 0.1290(22) \\
3252&  2 & 3 &  
$3.050(18) \times 10^{-3}$ & $1.356(18)\times 10^{-3}$ & 0.3952(12)& 0.0794(3) \\
\bottomrule
\end{tabular}
\end{table}

\vfill  
\newpage

\begin{table}[h!]
\centering
\caption{The symmetry breaking parameters of the $\bar c b $ sector in $N_f=2+1+1+1$ lattice QCD.}
\label{tab:Kappa_cb}
\setlength{\tabcolsep}{6pt}
\vspace{2mm}
\begin{tabular}{lccccccr}
\toprule
$T$	& $N_t$	& $zT$  &	$k_{VA}$ &	$k_{TX}$ & $\kappa$	&	$k_{CS}$  \\    
\midrule
325	& 20 & 1 &	
0.6980(45) & 0.7473(52) & 0.0670(89) & 0.7473(52) \\
406 & 16 & 1 &	
0.5773(35) & 0.6303(38) & 0.097(11) & 0.6303(38) \\
542 & 12 & 1 &	
0.4171(32) & 0.4695(38) & 0.1342(84) & 0.4695(38) \\
650 & 10 & 1 &	
0.3119(26) & 0.3579(29) & 0.1714(77) & 0.3579(29) \\
813	&  8 & 1 &	
0.2150(16) & 0.2497(17) & 0.203(4) & 0.2497(17) \\
1084&  6 & 1 &	
0.1213(4) & 0.1424(6) & 0.2686(29) & 0.1653(16) \\
1626&  4 & 1 &	
0.0474(1) & 0.0522(2) & 0.3819(13) & 0.1396(5) \\
3252&  2 & 1 &  
$6.144(22) \times 10^{-3}$ & $4.540(30) \times 10^{-3}$ & $0.5811(4)$ & $0.0738(1)$ \\
\midrule
650 & 10 & 2 &	
0.3742(42) & 0.3971(43) & 0.0886(32) & 0.3971(43) \\
813	&  8 & 2 &	
$0.2561(24)$ & $0.2727(24)$ & $0.1176(37)$ & $0.2727(24)$ \\
1084&  6 & 2 &	
$0.1438(7)$ & $0.1545(9)$ & $0.1602(33)$ & $0.1545(9)$ \\
1626&  4 & 2 &	
$0.0552(2)$ & $0.0538(3)$ & $0.2342(22)$ & $0.1318(17)$ \\
3252&  2 & 2 &  
$9.019(43) \times 10^{-3}$ & $4.231(47) \times 10^{-3}$ & $0.4601(8)$ & $0.0772(2)$ \\
\midrule
1084&  6 & 3 &	
$0.1558(9)$ & $0.1613(11)$ & $0.1164(32)$ & $0.1613(11)$ \\
1626&  4 & 3 &	
$0.0581(3)$ & $0.0526(3)$ & $0.1726(25)$ & $0.1351(25)$ \\
3252&  2 & 3 &  
$0.0129(1)$ & $5.700(78) \times 10^{-3}$ & $0.3748(12)$ & $0.0805(4)$ \\
\bottomrule
\end{tabular}
\end{table}

\vfill  
\newpage

\begin{table}[h!]
\centering
\caption{The symmetry breaking parameters of the $\bar b b $ sector in $N_f=2+1+1+1$ lattice QCD.}
\label{tab:Kappa_bb}
\setlength{\tabcolsep}{6pt}
\vspace{2mm}
\begin{tabular}{lccc c c cr}
\toprule
$T$	& $N_t$	& $zT$  &	$k_{VA}$ &	$k_{TX}$ & $\kappa$	&	$k_{CS}$  \\    
\midrule
325	& 20 & 1 &	
0.8746(21) & 0.930(2) & 0.0378(11) & 0.930(2) \\
406 & 16 & 1 &	
0.8191(17) & 0.8908(19) & 0.0585(13) & 0.8908(19) \\
542 & 12 & 1 &	
0.715(2) & 0.8059(23) & 0.0917(14) & 0.8059(23) \\
650 & 10 & 1 &	
0.6235(22) & 0.7186(28) & 0.1245(18) & 0.7186(28) \\
813	&  8 & 1 &	
0.506(2) & 0.5925(22) & 0.1606(19) & 0.5925(22) \\
1084&  6 & 1 &	
0.3444(7) & 0.4029(12) & 0.2296(16) & 0.4029(12) \\
1626&  4 & 1 &	
0.1638(5) & 0.1758(6) & 0.3435(11) & 0.1758(6) \\
3252&  2 & 1 &  
0.0249(1) & 0.0179(1) & 0.5462(5) & 0.0709(1) \\
\midrule
650 & 10 & 2 &	
0.7381(34) & 0.781(4) & 0.0416(13) & 0.781(4) \\
813	&  8 & 2 &	
0.6085(32) & 0.6473(29) & 0.0629(19) & 0.6473(29) \\
1084&  6 & 2 &	
0.4179(13) & 0.4460(19) & 0.1072(18) & 0.4460(19) \\
1626&  4 & 2 &	
0.1952(8) & 0.181(1) & 0.1852(17) & 0.181(1) \\
3252&  2 & 2 &  
0.0378(2) & 0.0173(2) & 0.3998(9) & 0.0717(2) \\
\midrule
1084&  6 & 3 &	
0.4547(18) & 0.4632(25) & 0.0648(19) & 0.4632(25) \\
1626&  4 & 3 &	
0.210(1) & 0.1754(14) & 0.123(2) & 0.1754(14) \\
3252&  2 & 3 &  
0.0550(3)&	0.0242(3)&	0.3069(13)&	0.0726(4)	 \\
\bottomrule
\end{tabular}
\end{table}

\vfill  
\newpage

\section{Symmetry breaking parameters of $N_f=2+1+1$ lattice QCD \cite{Chiu:2024jyz}}
\label{app:D} 

For comparison of the symmetry breaking parameters between $N_f=2+1+1+1$ lattice QCD in this work
to those of $N_f=2+1+1$ lattice QCD at the physical point \cite{Chiu:2024jyz}, 
we tabulate the numerical values of $\kappa_{VA}$, $\kappa_{TX}$, $\kappa$, and $\kappa_{CS}$ 
obtaind in ref.~\cite{Chiu:2024jyz}, for $zT$=0.5, 1, and 2, and for each flavor content of 
($\bar u d, \bar u s, \bar s s, \bar u c, \bar s c$, $\bar c c$) respectively.  
For the $z$ correlators, the possible values of $zT$ at $T=1/(N_t a)$ are
$\{n_z/N_t, n_z = 1, 2, \cdots, N_z/2 \}$. 
Thus for $N_z=32$ and $N_t$ = (16, 12, 10, 8, 6, 4, 2), 
the number of available temperatures are $(7, 7, 4)$ for $zT=(0.5,1,2)$, as shown in Tables
\ref{tab:Kappa_ud_Nf2p1p1}--\ref{tab:Kappa_cc_Nf2p1p1}. 
The error in the parenthesis of each entry is statistical, which    
is estimated by the jackknife method with the binsize of 10-15 configurations
of which the statistical error saturates. Due to the single spatial volume and one lattice spacing of 
the study in ref.~\cite{Chiu:2024jyz}, 
the systematic errors due to finite lattice spacing and finite volume cannot be estimated.

\vfill  
\newpage

\begin{table}[h!]
\centering
\caption{The symmetry breaking parameters of the $\bar u d $ sector in $N_f=2+1+1$ lattice QCD 
at the physical point \cite{Chiu:2024jyz}.}
\label{tab:Kappa_ud_Nf2p1p1}
\setlength{\tabcolsep}{6pt}
\vspace{2mm}
\begin{tabular}{lccccccr}
\toprule
$T$	& $N_t$	& $zT$  &	$k_{VA}$ &	$k_{TX}$ & $\kappa$	&	$k_{CS}$  \\    
\midrule
 192 & 16  & 0.5 &	
$3.32(56) \times 10^{-4}$ & $1.88(92) \times 10^{-3}$ & $0.0755(26)$ & $0.391(14)$ \\
 257 & 12  & 0.5 &
$2.36(14) \times 10^{-5}$ & $1.32(39) \times 10^{-5}$ & $0.1388(41)$ & $0.3105(46)$ \\
 308 & 10  & 0.5 &	
$8.57(64) \times 10^{-6}$ & $5.0(2) \times 10^{-6}$ & $0.2014(26)$ & $0.2675(25)$ \\
 385 &  8  & 0.5 &	
$3.35(14) \times 10^{-6}$ & $3.91(21) \times 10^{-6}$ & $0.2649(37)$ & $0.2357(17)$ \\
 513 &  6  & 0.5 &	
$1.2(1) \times 10^{-6}$ & $1.506(79) \times 10^{-6}$ & $0.366(6)$ & $0.1978(11)$ \\
 770 &  4  & 0.5 &	
$3.62(76) \times 10^{-7}$ & $2.41(64) \times 10^{-7}$ & $0.5144(58)$ & $0.1423(3)$ \\
1540 &  2  & 0.5 &	
$6(2) \times 10^{-8}$ & $6(2) \times 10^{-8}$ & $0.6397(2)$ & $0.0613(1)$ \\
\midrule
 192 & 16  & 1 &  	
$8.54(55) \times 10^{-5}$ & $2.23(61) \times 10^{-3}$ & $0.0271(26)$ & $0.531(43)$ \\
 257 & 12  & 1 &  	
$4.21(52) \times 10^{-5}$ & $4.37(56) \times 10^{-5}$ & $0.0644(35)$ & $0.362(12)$ \\
 308 & 10  & 1 &  	
$1.408(68) \times 10^{-5}$ & $3.35(95) \times 10^{-5}$ & $0.1064(27)$ & $0.2877(54)$ \\
 385 &  8  & 1 &  	
$6.15(78) \times 10^{-6}$ & $1.49(27) \times 10^{-5}$ & $0.145(8)$ & $0.232(4)$ \\
 513 &  6  & 1 &  	
$2.09(25) \times 10^{-6}$ & $3.85(55) \times 10^{-6}$ & $0.21(1)$ & $0.1909(22)$ \\
 770 &  4  & 1 &  	
$4(1) \times 10^{-7}$ & $4.03(38) \times 10^{-7}$ & $0.3544(52)$ & $0.1452(5)$ \\
1540 &  2  & 1 &  	
$6(2) \times 10^{-8}$ & $4(1) \times 10^{-8}$ & $0.5888(4)$ & $0.0731(2)$ \\
\midrule
 385 &  8  & 2 &  	
$0.35(14) \times 10^{-4}$ & $0.43(15) \times 10^{-4}$ & $0.0737(44)$ & $0.300(9)$ \\
 513 &  6  & 2 & 	
$1.03(38) \times 10^{-5}$ & $1.39(44) \times 10^{-5}$ & $0.121(6)$ & $0.2263(57)$ \\
 770 &  4  & 2 &	
$5.62(57) \times 10^{-7}$ & $3.86(34) \times 10^{-7}$ & $0.216(5)$ & $0.1571(12)$ \\
1540 &  2  & 2 & 	
$1.3(3) \times 10^{-7}$ & $0.70(26) \times 10^{-7}$ & $0.4715(9)$ & $0.0774(2)$ \\
\bottomrule
\end{tabular}
\end{table}

\vfill  
\newpage

\begin{table}[h!]
\centering
\caption{The symmetry breaking parameters of the $\bar u s$ sector in $N_f=2+1+1$ lattice QCD
at the physical point \cite{Chiu:2024jyz}.}
\label{tab:Kappa_us_Nf2p1p1}
\setlength{\tabcolsep}{6pt}
\vspace{2mm}
\begin{tabular}{lccccccr}
\toprule
$T$	& $N_t$	& $zT$  &	$k_{VA}$ &	$k_{TX}$ & $\kappa$	&	$k_{CS}$  \\    
\midrule
 192 & 16  & 0.5 &	
$5.02(34) \times 10^{-3}$ & $5.87(45) \times 10^{-3}$ & $0.0822(25)$ & $0.3729(128)$ \\
 257 & 12  & 0.5 &
$6.46(27) \times 10^{-4}$ & $7.2(3) \times 10^{-4}$ & $0.140(4)$ & $0.3081(45)$ \\
 308 & 10  & 0.5 &	
$2.4(1) \times 10^{-4}$ & $2.810(55) \times 10^{-4}$ & $0.2019(26)$ & $0.2668(25)$ \\
 385 &  8  & 0.5 &	
$10.0(2) \times 10^{-5}$ & $1.267(26) \times 10^{-4}$ & $0.2651(37)$ & $0.2354(16)$ \\
 513 &  6  & 0.5 &	
$3.674(85) \times 10^{-5}$ & $4.60(12) \times 10^{-5}$ & $0.366(6)$ & $0.1977(11)$ \\
 770 &  4  & 0.5 &	
$1.033(14) \times 10^{-5}$ & $1.206(16) \times 10^{-5}$ & $0.5144(58)$ & $0.1423(3)$ \\
1540 &  2  & 0.5 &	
$2.018(92) \times 10^{-6}$ & $2.001(78) \times 10^{-6}$ & $0.6397(2)$ & $0.0613(1)$ \\
\midrule
 192 & 16  & 1 &  	
$0.87(16) \times 10^{-2}$ & $0.68(14) \times 10^{-2}$ & $0.0325(22)$ & $0.467(38)$ \\
 257 & 12  & 1 &  	
$1.150(64) \times 10^{-3}$ & $1.031(68) \times 10^{-3}$ & $0.0655(35)$ & $0.357(18)$ \\
 308 & 10  & 1 &  	
$3.95(11) \times 10^{-4}$ & $3.79(29) \times 10^{-4}$ & $0.1067(27)$ & $0.2860(54)$ \\
 385 &  8  & 1 &  	
$1.56(5) \times 10^{-4}$ & $1.84(12) \times 10^{-4}$ & $0.145(8)$ & $0.2315(39)$ \\
 513 &  6  & 1 &  	
$5.601(35) \times 10^{-5}$ & $6.54(67) \times 10^{-5}$ & $0.21(1)$ & $0.1907(22)$ \\
 770 &  4  & 1 &  	
$1.385(14) \times 10^{-5}$ & $1.4354(74) \times 10^{-5}$ & $0.3544(52)$ & $0.1451(5)$ \\
1540 &  2  & 1 &  	
$1.711(87) \times 10^{-6}$ & $1.107(89) \times 10^{-6}$ & $0.5887(4)$ & $0.0731(2)$ \\
\midrule
 385 &  8  & 2 &  	
$2.43(42) \times 10^{-4}$ & $0.24(13) \times 10^{-3}$ & $0.0738(44)$ & $0.2992(89)$ \\
 513 &  6  & 2 & 	
$0.84(11) \times 10^{-4}$ & $0.69(16) \times 10^{-4}$ & $0.121(6)$ & $0.2260(57)$ \\
 770 &  4  & 2 &	
$1.578(15) \times 10^{-5}$ & $1.393(12) \times 10^{-5}$ & $0.216(5)$ & $0.1571(12)$ \\
1540 &  2  & 2 & 	
$2.9(1) \times 10^{-6}$ & $1.158(77) \times 10^{-6}$ & $0.4715(9)$ & $0.0774(2)$ \\
\bottomrule
\end{tabular}
\end{table}

\vfill  
\newpage

\begin{table}[h!]
\centering
\caption{The symmetry breaking parameters of the $\bar s s $ sector in $N_f=2+1+1$ lattice QCD
at the physical point \cite{Chiu:2024jyz}.}
\label{tab:Kappa_ss_Nf2p1p1}
\setlength{\tabcolsep}{6pt}
\vspace{2mm}
\begin{tabular}{lccccccr}
\toprule
$T$	& $N_t$	& $zT$  &	$k_{VA}$ &	$k_{TX}$ & $\kappa$	&	$k_{CS}$  \\    
\midrule
 192 & 16  & 0.5 &	
$0.0803(33)$ & $0.0980(53)$ & $0.0678(24)$ & $0.375(11)$ \\
 257 & 12  & 0.5 &
$0.0174(6)$ & $0.0201(5)$ & $0.1336(39)$ & $0.2899(43)$ \\
 308 & 10  & 0.5 &	
$6.729(76) \times 10^{-3}$ & $8.238(86) \times 10^{-3}$ & $0.1972(26)$ & $0.2596(24)$ \\
 385 &  8  & 0.5 &	
$2.96(3) \times 10^{-3}$ & $3.752(38) \times 10^{-3}$ & $0.2618(37)$ & $0.2322(16)$ \\
 513 &  6  & 0.5 &	
$1.1058(89) \times 10^{-3}$ & $1.390(12) \times 10^{-3}$ & $0.364(6)$ & $0.1965(11)$ \\
 770 &  4  & 0.5 &	
$3.279(12) \times 10^{-4}$ & $3.802(17) \times 10^{-4}$ & $0.5132(58)$ & $0.1420(3)$ \\
1540 &  2  & 0.5 &	
$6.461(22) \times 10^{-5}$ & $6.412(28) \times 10^{-5}$ & $0.6390(2)$ & $0.0612(1)$ \\
\midrule
 192 & 16  & 1 &  	
$0.140(11)$ & $0.142(18)$ & $0.0250(23)$ & $0.440(17)$ \\
 257 & 12  & 1 &  	
$0.0296(13)$ & $0.0288(14)$ & $0.0615(32)$ & $0.328(11)$ \\
 308 & 10  & 1 &  	
$0.0106(1)$ & $0.0115(2)$ & $0.1032(27)$ & $0.2749(54)$ \\
 385 &  8  & 1 &  	
$4.383(54) \times 10^{-3}$ & $4.901(66) \times 10^{-3}$ & $0.1427(79)$ & $0.2269(39)$ \\
 513 &  6  & 1 &  	
$1.605(19) \times 10^{-3}$ & $1.789(23) \times 10^{-3}$ & $0.21(1)$ & $0.1891(22)$ \\
 770 &  4  & 1 &  	
$4.422(15) \times 10^{-4}$ & $4.597(21) \times 10^{-4}$ & $0.3533(52)$ & $0.1447(5)$ \\
1540 &  2  & 1 &  	
$5.501(21) \times 10^{-5}$ & $3.59(3) \times 10^{-5}$ & $0.5884(4)$ & $0.0730(2)$ \\
\midrule
 385 &  8  & 2 &  	
$5.90(13) \times 10^{-3}$ & $6.1(2) \times 10^{-3}$ & $0.0725(43)$ & $0.2931(88)$ \\
 513 &  6  & 2 & 	
$1.959(28) \times 10^{-3}$ & $1.967(38) \times 10^{-3}$ & $0.1204(60)$ & $0.2241(57)$ \\
 770 &  4  & 2 &	
$5.055(21) \times 10^{-4}$ & $4.471(26) \times 10^{-4}$ & $0.2153(50)$ & $0.1566(12)$ \\
1540 &  2  & 2 & 	
$9.1185(535) \times 10^{-5}$ & $3.764(51) \times 10^{-5}$ & $0.4711(9)$ & $0.0774(2)$ \\
\bottomrule
\end{tabular}
\end{table}

\vfill  
\newpage

\begin{table}[h!]
\centering
\caption{The symmetry breaking parameters of the $\bar u c $ sector in $N_f=2+1+1$ lattice QCD
at the physical point \cite{Chiu:2024jyz}.}
\label{tab:Kappa_uc_Nf2p1p1}
\setlength{\tabcolsep}{6pt}
\vspace{2mm}
\begin{tabular}{lccccccr}
\toprule
$T$	& $N_t$	& $zT$  &	$k_{VA}$ &	$k_{TX}$ & $\kappa$	&	$k_{CS}$  \\    
\midrule
 192 & 16  & 0.5 &	
$0.0118(8)$ & $0.0138(7)$ & $0.1386(31)$ & $0.2394(55)$ \\
 257 & 12  & 0.5 &
$3.229(84) \times 10^{-3}$ & $4.0(1) \times 10^{-3}$ & $0.1872(35)$ & $0.2343(26)$ \\
 308 & 10  & 0.5 &	
$1.73(7) \times 10^{-3}$ & $2.161(69) \times 10^{-3}$ & $0.2343(25)$ & $0.2281(17)$ \\
 385 &  8  & 0.5 &	
$9.09(15) \times 10^{-4}$ & $1.180(23) \times 10^{-3}$ & $0.2884(37)$ & $0.2167(13)$ \\
 513 &  6  & 0.5 &	
$4.032(88) \times 10^{-4}$ & $5.07(12) \times 10^{-4}$ & $0.3783(58)$ & $0.190(1)$ \\
 770 &  4  & 0.5 &	
$1.2897(46) \times 10^{-4}$ & $1.4777(65) \times 10^{-4}$ & $0.5152(56)$ & $0.1394(3)$ \\
1540 &  2  & 0.5 &	
$2.7319(85) \times 10^{-5}$ & $2.7034(86) \times 10^{-5}$ & $0.6385(2)$ & $0.0604(1)$ \\
\midrule
 192 & 16  & 1 &  	
$0.0221(28)$ & $0.0205(14)$ & $0.0664(36)$ & $0.2291(44)$ \\
 257 & 12  & 1 &  	
$5.59(16) \times 10^{-3}$ & $5.94(18) \times 10^{-3}$ & $0.0977(35)$ & $0.2248(29)$ \\
 308 & 10  & 1 &  	
$2.827(81) \times 10^{-3}$ & $3.18(12) \times 10^{-3}$ & $0.1248(29)$ & $0.213(2)$ \\
 385 &  8  & 1 &  	
$1.443(48) \times 10^{-3}$ & $1.671(56) \times 10^{-3}$ & $0.1550(58)$ & $0.1977(15)$ \\
 513 &  6  & 1 &  	
$6.15(33) \times 10^{-4}$ & $7.13(54) \times 10^{-4}$ & $0.216(9)$ & $0.182(1)$ \\
 770 &  4  & 1 &  	
$1.7404(55) \times 10^{-4}$ & $1.7870(82) \times 10^{-4}$ & $0.3540(48)$ & $0.1457(3)$ \\
1540 &  2  & 1 &  	
$2.3015(79) \times 10^{-5}$ & $1.467(13) \times 10^{-5}$ & $0.5865(4)$ & $0.0730(1)$ \\
\midrule
 385 &  8  & 2 &  	
$2.03(17) \times 10^{-3}$ & $2.15(15) \times 10^{-3}$ & $0.0813(35)$ & $0.2333(69)$ \\
 513 &  6  & 2 & 	
$8.24(75) \times 10^{-4}$ & $9.43(98) \times 10^{-4}$ & $0.1214(58)$ & $0.2078(46)$ \\
 770 &  4  & 2 &  	
$2.022(13) \times 10^{-4}$ & $1.721(11) \times 10^{-4}$ & $0.2122(49)$ & $0.1592(12)$ \\
1540 &  2  & 2 & 	
$3.917(24) \times 10^{-5}$ & $1.537(25) \times 10^{-5}$ & $0.4686(9)$ & $0.0784(2)$ \\
\bottomrule
\end{tabular}
\end{table}

\vfill  
\newpage

\begin{table}[h!]
\centering
\caption{The symmetry breaking parameters of the $\bar s c $ sector in $N_f=2+1+1$ lattice QCD
at the physical point \cite{Chiu:2024jyz}.}
\label{tab:Kappa_sc_Nf2p1p1}
\setlength{\tabcolsep}{6pt}
\vspace{2mm}
\begin{tabular}{lccccccr}
\toprule
$T$	& $N_t$	& $zT$  &	$k_{VA}$ &	$k_{TX}$ & $\kappa$	&	$k_{CS}$  \\    
\midrule
 192 & 16  & 0.5 &	
$0.2068(52)$ & $0.2548(71)$ & $0.1273(31)$ & $0.2650(52)$ \\
 257 & 12  & 0.5 &
$0.0888(17)$ & $0.1107(18)$ & $0.1802(32)$ & $0.2487(29)$ \\
 308 & 10  & 0.5 &	
$0.0490(3)$ & $0.0622(4)$ & $0.2301(21)$ & $0.2363(16)$ \\
 385 &  8  & 0.5 &	
$0.0271(2)$ & $0.0348(2)$ & $0.2861(36)$ & $0.2214(12)$ \\
 513 &  6  & 0.5 &	
$0.0122(1)$ & $0.0153(1)$ & $0.3771(59)$ & $0.1921(9)$ \\
 770 &  4  & 0.5 &	
$4.114(15) \times 10^{-3}$ & $4.718(21) \times 10^{-3}$ & $0.5146(89)$ & $0.1399(9)$ \\
1540 &  2  & 0.5 &	
$8.751(24) \times 10^{-4}$ & $8.661(33) \times 10^{-4}$ & $0.6296(2)$ & $0.0596(1)$ \\
\midrule
 192 & 16  & 1 &  	
$0.361(13)$ & $0.389(15)$ & $0.0499(45)$ & $0.389(15)$ \\
 257 & 12  & 1 &  	
$0.1464(32)$ & $0.1596(35)$ & $0.0879(29)$ & $0.2367(63)$ \\
 308 & 10  & 1 &  	
$0.0771(7)$ & $0.0860(9)$ & $0.1189(27)$ & $0.2197(41)$ \\
 385 &  8  & 1 &  	
$0.0405(4)$ & $0.0461(4)$ & $0.1517(56)$ & $0.2018(29)$ \\
 513 &  6  & 1 &  	
$0.0178(2)$ & $0.0200(2)$ & $0.2139(89)$ & $0.1836(19)$ \\
 770 &  4  & 1 &  	
$5.569(20) \times 10^{-3}$ & $5.713(27) \times 10^{-3}$ & $0.3526(99)$ & $0.1460(18)$ \\
1540 &  2  & 1 &  	
$7.418(26) \times 10^{-4}$ & $4.767(39) \times 10^{-4}$ & $0.5817(4)$ & $0.0724(1)$ \\
\midrule
 385 &  8  & 2 &  	
$0.0535(8)$ & $0.0568(12)$ & $0.0781(34)$ & $0.2364(67)$ \\
 513 &  6  & 2 & 	
$0.0216(2)$ & $0.0220(3)$ & $0.1194(57)$ & $0.2087(46)$ \\
 770 &  4  & 2 &	
$6.433(27) \times 10^{-3}$ & $5.537(37) \times 10^{-3}$ & $0.211(10)$ & $0.1590(38)$ \\
1540 &  2  & 2 & 	
$1.2495(72) \times 10^{-3}$ & $5.129(69) \times 10^{-4}$ & $0.4634(9)$ & $0.0776(2)$ \\
\bottomrule
\end{tabular}
\end{table}

\vfill  
\newpage

\begin{table}[h!]
\centering
\caption{The symmetry breaking parameters of the $\bar c c $ sector in $N_f=2+1+1$ lattice QCD
at the physical point \cite{Chiu:2024jyz}.}
\label{tab:Kappa_cc_Nf2p1p1}
\setlength{\tabcolsep}{6pt}
\vspace{2mm}
\begin{tabular}{lccccccr}
\toprule
$T$	& $N_t$	& $zT$  &	$k_{VA}$ &	$k_{TX}$ & $\kappa$	&	$k_{CS}$  \\    
\midrule
 192 & 16  & 0.5 &	
0.6161(48) & 0.7452(57) &  0.1008(51) & 0.7452(57)  \\
 257 & 12  & 0.5 &
0.4705(28) & 0.590(3) &  0.1557(36) &	0.590(3) \\
 308 & 10  & 0.5 &	
0.3594(12) & 0.4576(19) & 0.2062(22) &	0.4576(19) \\	
 385 &  8  & 0.5 &	
0.2467(11) & 0.3168(16) & 0.270(4) &	0.3168(16)  \\
 513 &  6  & 0.5 &	
0.1335(7) &	0.1660(9)&	0.3647(66) & 0.2033(9)  \\	
 770 &  4  & 0.5 &	
0.0516(2) &	0.0585(2) & 0.5011(59) & 0.1405(2)	\\
1540 &  2  & 0.5 &	
0.0119(0) &	0.0117(0)&	0.6295(2)&	0.0595(1)	\\
\midrule
 192 & 16  & 1 &  	
0.8373(73) & 0.8956(75) & 0.0222(68) & 0.8956(75) \\
 257 & 12  & 1 &  	
0.6774(51)&	0.7486(57) & 0.0491(54) & 0.7486(57) \\	
 308 & 10  & 1 &  	
0.5305(25) & 0.5961(34) & 0.0780(31) &	0.5961(34) \\ 
 385 &  8  & 1 &  	
0.3647(22) & 0.4164(26) & 0.1186(37) & 0.4164(26) \\
 513 &  6  & 1 &  	
0.1955(14) & 0.2196(16) & 0.188(7) & 0.2196(16) \\	
 770 &  4  & 1 &  	
0.0700(2) &	0.0707(3) &	0.3299(46) & 0.1474(5) \\
1540 &  2  & 1 &  	
0.0100(0) & $6.340(52) \times 10^{-3} $ &	0.5651(4)&	0.0711(1)  \\	
\midrule
 385 &  8  & 2 &  	
0.4621(49) & 0.4894(68) & 0.0459(18) &	0.4894(68)	\\
 513 &  6  & 2 & 	
0.2361(19) & 0.2389(22) & 0.0910(46) &	0.2389(22) \\	
 770 &  4  & 2 &	
0.0819(3) &	0.0682(4) &	0.1840(44) & 0.1558(13) \\
1540 &  2  & 2 & 	
0.0171(1) & $7.000(95) \times 10^{-3}$ & 0.431(1) & 0.0745(2) \\
\bottomrule
\end{tabular}
\end{table}

\vfill  
\newpage

\noindent\textbf{Funding:} This research was funded by National Science and Technology Council (Taiwan),
grant number 108-2112-M-003-005, 109-2112-M-003-006, 110-2112-M-003-009 and
Academia Sinica Grid Computing Centre (Taiwan), grant number AS-CFII-112-103.

\noindent\textbf{Data Availability Statement:} The original contributions presented in this study are
included in the article. Further inquiries can be directed to the corresponding author.

\noindent\textbf{Acknowledgments:} The author is grateful to Academia Sinica Grid Computing Center
and National Center for High Performance Computing for the computer time and facilities.
This work is supported by the National Science and Technology Council
(Grants No.~108-2112-M-003-005, No.~109-2112-M-003-006, No.~110-2112-M-003-009),
and Academia Sinica Grid Computing Centre (Grant No.~AS-CFII-112-103).

\noindent\textbf{Conflicts of Interest:} The author declares no conflict of interest. The funders had no role
in the design of the study; in the collection, analyses, or interpretation of data; in the writing of
the manuscript, or in the decision to publish the results.

\noindent\textit{Note added.}---This version corresponds to the published version
[Symmetry \textbf{17}, 700 (2025)] with its Correction incorporated: the six entries of
$\kappa_{VA}$ and $\kappa_{TX}$ of the $\bar sc$ sector at $T=770$~MeV in Table~\ref{tab:Kappa_sc_Nf2p1p1}
and the corresponding $N_f=2+1+1$ data points in the $\bar sc$ panels of Figure~\ref{fig:kVA_kTX_ss_sc_cc_compare},
which in the original publication were evaluated with the definition $|1-C_{\Gamma_B}/C_{\Gamma_A}|$
of ref.~\cite{Chiu:2023hnm} instead of (\ref{eq:kappa_AB}), as in the Erratum of
ref.~\cite{Chiu:2024jyz}. All other results and conclusions are unchanged.

\end{document}